\documentclass[amsfonts, amssymb, amsmath, showkeys, nofootinbib, reprint,floatfix, prx, superscriptaddress]{revtex4-2}

\usepackage[english]{babel}
\usepackage[colorinlistoftodos, color=green!40,prependcaption]{todonotes}
\usepackage{graphicx}
\usepackage{dcolumn}
\usepackage{bm}
\usepackage{float}
\usepackage{gensymb}
\usepackage{xcolor}
\usepackage{amsmath}
\usepackage{amssymb}
\usepackage{units}
\usepackage{placeins}
\usepackage{ulem}
\usepackage{hyperref}
\usepackage{bbm}
\usepackage{multirow}
\usepackage{amsfonts}
\usepackage{amsmath}
\usepackage{amssymb}

\usepackage{lipsum}
\def\eia{EuIn$_2$As$_2$}
\usepackage{comment}
\def\Neel{ N\'{e}el}

\begin{document}

\title{Symmetry-breaking pathway towards the unpinned broken helix}

\author{E. Donoway}
\author{T. V. Trevisan}
\author{A. Liebman - Pel\'{a}ez}
\author{R. P. Day}
\author{K. Yamakawa}
\author{Y. Sun}
\affiliation {Department of Physics, University of California, Berkeley, California 94720, USA}
\affiliation {Materials Science Division, Lawrence Berkeley National Laboratory, Berkeley, California 94720, USA}
\author{J. R. Soh}
\affiliation {Institute of Physics, École Polytechnique Fédérale de Lausanne (EPFL), Lausanne, Switzerland}

\author{D. Prabhakaran}
\author{A. T. Boothroyd}
\affiliation {Department of Physics, University of Oxford, Clarendon Laboratory, Oxford, OX1 3PU, UK}

\author{R. M. Fernandes}
\affiliation {School of Physics and Astronomy, University of Minnesota, Minneapolis, MN, 55455, USA}

\author{J. G. Analytis}
\author{J. E. Moore}
\author{J. Orenstein}

\author{V. Sunko}
\email{vsunko@berkeley.edu}
\affiliation {Department of Physics, University of California, Berkeley, California 94720, USA}
\affiliation {Materials Science Division, Lawrence Berkeley National Laboratory, Berkeley, California 94720, USA}

\begin{abstract}
One of the prime material candidates to host the axion insulator state is \eia. Initial first-principles calculations predicted the emergence of this exotic topological phase based on the assumption of a simple collinear antiferromagnetic structure. Recently, however, neutron scattering measurements revealed a much more intricate magnetic ground state, characterized by two coexisting magnetic wavevectors, reached by successive thermal phase transitions. The proposed high and low temperature phases were a spin helix and a state with interpenetrating helical and \Neel~antiferromagnetic order, termed a `broken helix,' respectively. Despite its complexity, the broken helix would still protect the axion state because the product of time-reversal and a rotational symmetry is preserved. Here we unambiguously identify the magnetic structure associated with these two phases of \eia~using a multimodal approach that combines symmetry-sensitive optical probes, scattering, and group theoretical analysis. We find that the higher temperature phase is characterized by a variation of the magnetic moment amplitude from layer to layer, with the moment vanishing entirely in every third Eu layer. The lower temperature structure is similar to the `broken helix', with one important difference: due to local strain the relative orientation of the magnetic structure and the lattice is not fixed, resulting in an `unpinned broken helix'.  As a result of the consequent breaking of rotational symmetry, the axion phase is not generically protected in \eia. Nevertheless, we show that it can be restored if the magnetic structure is tuned with uniaxial strain. Finally, we present a spin Hamiltonian that identifies the spin interactions that account for the complex magnetic order in \eia. Our work highlights the importance of a multimodal approach in determining the symmetry of complex order-parameters.
\end{abstract}

\maketitle

\section{Introduction}

The search for materials exhibiting novel emergent properties relies on the identification of their characteristic symmetries. Examples of such properties include the anomalous and topological Hall effects, arising from Berry curvature in momentum\cite{nagaosa_anomalous_2010} and real space \cite{nagaosa_topological_2013}, respectively, momentum-dependent Zeeman splitting of electronic bands \cite{smejkal_altermagnetism_2022}, and quantized response functions in topological systems~\cite{deng_quantum_2020, serlin_intrinsic_2020}. These phenomena all depend on the underlying magnetic order and present the exciting prospect of tuning by manipulating the symmetry of the magnetic state.

\begin{figure}[t]
 \centering
 \includegraphics[width=0.95\linewidth]{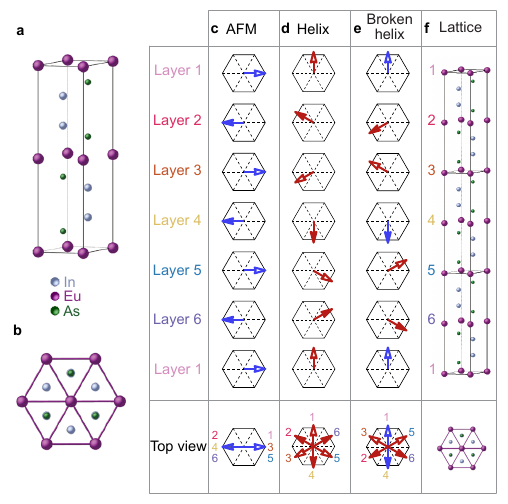}
 \caption{ (a) The side view and (b) the top view of the crystal structure of \eia. (c-e) A table showing the three magnetic structures discussed for \eia~in previous work: the theoretically assumed A-type antiferromagnet (AFM), the helix proposed in Phase I, and the broken helix proposed in Phase II. In all the structures the moments lie in Eu planes, and individual layers are ferromagnetically aligned. For each structure we show the layer-resolved spin-orientation, with the hexagon denoting the lattice orientation, as well as a top view (last row), with spins labeled by the layer number. (f) The magnetic unit cell, corresponding to three crystallographic unit cells. We mark odd and even layers with white and colored arrowheads, respectively.}
 \label{fig:EIA_Fig1}
\end{figure}

Promising material candidates for observing these phenomena can be identified by \textit{ab initio} calculations~\cite{frey_high-throughput_2020, xu_high_2020}. Of necessity, such calculations are based on an assumed magnetic structure, as it is difficult to reliably predict the magnetic ground state. An effective search strategy is to combine \textit{ab initio} calculations with the magnetic order deduced from scattering measurements. However, in some of the most interesting material systems the interpretation of scattering data can lead to ambiguities of critical importance. For example, states that exhibit order characterized by multiple symmetry-related wavevectors ($\textbf{Q}$) pose a challenge, since the diffraction pattern does not readily distinguish phase-sensitive mixed order from an equal population of domains characterized by a single $\textbf{Q}$. Similarly, domains complicate determining the orientation of magnetic moments even in a single-$\textbf{Q}$ structure. The distinct scenarios consistent with a given diffraction pattern are radically different from a symmetry perspective, and have distinct consequences for response functions and topological properties, motivating specialized scattering experiments~\cite{choi_unveiling_2022, bluschke_imaging_2022, Kim_Imaging_2005}, and comparisons with complementary experimental techniques~\cite{takagi_spontaneous_2023, park_tetrahedral_2023}.

\eia, whose structure is shown in Fig.~\ref{fig:EIA_Fig1}(a,b), is a perfect example of the scenario outlined above. \textit{Ab initio} calculations, based on an assumed antiferromagnetic structure (AFM, Fig.~\ref{fig:EIA_Fig1}c), predicted that it hosts the elusive axion insulator state~\cite{xu_higher-order_2019}, exhibiting quantized responses to electromagnetic fields~\cite{essin_agnetoelectric_2009, turner_quantized_2012}. However, a more complex magnetic behavior was uncovered by neutron scattering measurements, with two close-by transitions at $T_{N1}\approx\unit[17.5]{K}$ and $T_{N2}\approx\unit[16]{K}$~\cite{riberolles_magnetic_2021}, in which two distinct propagation vectors appeared sequentially. Neither of the two phases appearing at $T_{N1}$ and $T_{N2}$ are consistent with the previously assumed AFM structure. Nonetheless, the magnetic structures assigned to the two phases (Fig.~\ref{fig:EIA_Fig1}d,e) were shown to host the axion state~\cite{riberolles_magnetic_2021}.

Here we introduce a multimodal approach that reveals a different picture of how the  magnetic order of \eia~evolves with temperature. We combine information from scattering experiments and symmetry-sensitive optical probes with group theory analysis to identify two magnetic states that would have remained hidden to the application of any single technique. A higher-temperature Phase I ($T_{N1}>T>T_{N2}$) is characterized by the onset of a single wavevector, $\textbf{Q}_1$, which we identify as a `nodal amplitude-modulated state', in which the expectation value of the magnetic moment vanishes on every third Eu layer. In the lower-temperature Phase II ($T<T_{N2}$) an additional  wavevector $\textbf{Q}_2$ appears,  forming an `unpinned broken helix', in which the orientation of the magnetic moments with respect to the crystal axes varies continuously with location on the sample. Only special orientations, with moments aligned to high-symmetry directions of the lattice, maintain symmetries that protect the topological phase. 

Although the experimentally-informed group theory analysis, mentioned above, is well-suited to identifying the magnetic phases that emerge in \eia, it does not reveal their microscopic origin. To address this point, we propose a minimal spin model that captures the unpinned broken helix as the ground state. We show that this state requires the exchange interactions, $J(\textbf{Q})$, to peak sharply at two values of the wavevector $\textbf{Q}$. This requires long-ranged interactions, naturally arising from the coupling of itinerant electrons to the magnetic degrees of freedom. We further show that such exchange interactions alone cannot account for the higher-temperature amplitude modulated state. Whether it can be stabilized by thermal fluctuations close to the two transitions is an interesting question posed by our work.

This paper is organized as follows. We summarize the current understanding of \eia~in Sec.~\ref{sec:currentEia}. In Sec.~\ref{sec:Experiment} we introduce two optical techniques, sensitive to rotational and time-reversal symmetries, and show the results of these experiments as a function of temperature, position on the sample, and cooling protocol. We demonstrate that the symmetries revealed by these measurements are inconsistent with the previous understanding of magnetic phases in \eia, and therefore motivate a new analysis of the order parameters. In Sec~\ref{sec:Symmetry} we perform a systematic analysis of symmetry-allowed magnetic structures by combining scattering and optical data with group theory. We uniquely identify Phases I and II as a nodal amplitude-modulated state and an unpinned broken helix, respectively. We complement this analysis with a phenomenological Landau free energy model (Sec~\ref{sec:freeE}), demonstrating the symmetry-breaking pathway towards the ground state. Furthermore, we show experimentally and theoretically how the symmetry of the magnetic ground state and electronic topology can be tuned by uniaxial strain (Sec.~\ref{Sec:StrainTuning}). Finally, in Sec~\ref{sec:SpinHam} we introduce a microscopic spin-Hamiltonian, pointing to the importance of the coupling between electronic and magnetic degrees of freedom. We conclude by emphasizing that our measurements and analyses show that \eia~hosts a remarkably rich and tunable system of coupled electrons and localized moments, opening doors for future efforts to control topological phases. 

\section{Current understanding of $\mathrm{EuIn}_2\mathrm{As}_2$}\label{sec:currentEia}

\eia~is a rare-earth-based magnetic material with triangular layers of Eu stacked along the crystallographic $\bm{c}$ direction and separated by blocks of In$_2$As$_2$. Hereafter, we set $\bm{c}$ parallel to $\hat{z}$ (Fig.~\ref{fig:EIA_Fig1}a,b). Magnetism arises from the localized moments of the Eu$^{2+}$ ions ($S=7/2$, $L=0$). The moments in each Eu layer are ferromagnetically aligned with respect to each other, and localized in the Eu-Eu planes. This easy-plane anisotropy, confirmed by magnetization and neutron scattering experiments \cite{riberolles_magnetic_2021}, is implicitly assumed throughout this paper; when we refer to `moment orientation', we are referring to the orientation within the Eu planes. 

The low-energy electronic bands, dominated by the 5s orbitals of In and 4p orbitals of As, were predicted to host topological properties. Combined with the Eu-based magnetism, this makes \eia~a promising platform to explore the interplay between topology and magnetism, as has been proposed and debated in the context of several Eu-based magnetic materials~\cite{wang_single_2019, ma_emergence_2020, wang_magnetic_2021, li_engineering_2021, pierantozzi_evidence_2022, santos-cottin_eucd_2as_2_2023, cuono_ab_2023, valadkhani_influence_2023}. More specifically, \eia~was predicted to host the axion insulator state~\cite{xu_higher-order_2019,riberolles_magnetic_2021}, a phase characterized by half-quantized magneto-electric coupling in the bulk and half-quantized Hall surface conductivity. The axion phase requires $\mathcal{T}$ to be broken in order to gap the surface states, while the quantization is protected by another symmetry that reverses the sign of the magnetoelectric coupling constant, such as spatial inversion ($\mathcal{P}$), $\mathcal{T}$ combined with half-translations, or the product of $\mathcal{T}$ and two-fold rotations ($\mathcal{T} C_{2}$).

The original prediction of an axion insulator state in \eia~was based on first-principles calculations that showed an insulating bulk and an A-type antiferromagnetic order illustrated in Fig.~\ref{fig:EIA_Fig1}c. Note that this N\'eel-like magnetic order does not change the periodicity of the lattice, since the paramagnetic unit cell already contains two Eu layers; as a result, the magnetic ordering vector coincides with the Bragg wave-vector $(0,0,1)$. Importantly, this magnetic configuration preserves spatial inversion, which protects the axion phase~\cite{xu_higher-order_2019}. However, experiments uncovered a different, but intriguing, picture: \eia~shows metallic DC~\cite{regmi_temperature-dependent_2020, yan_field-induced_2022} and optical~\cite{xu_infrared_2021} transport, it has a Fermi surface~\cite{sato_signature_2020, regmi_temperature-dependent_2020}, and the magnetic structure is considerably more complex. Both neutron~\cite{riberolles_magnetic_2021} and resonant x-ray~\cite{soh_understanding_2023} scattering experiments found two consecutive magnetic transitions ($T_{N1}\approx\unit[17.5]{K}$, $T_{N2}\approx\unit[16]{K}$), corresponding to the onsets of two distinct wave vectors. At $T_{N1}$ a single propagation vector, $\bm{Q}_{1} = (0,0,1/3)$, is observed, while at $T_{N2}$ an additional propagation vector, $\bm{Q}_{2}=(0,0,1)$, emerges. The higher and lower temperature phases that we refer to as Phase I and Phase II, respectively, were interpreted as a 60\degree-helix (Fig.~\ref{fig:EIA_Fig1}d) and a novel magnetic state exhibiting interpenetrating Type A-AFM and spin-helical orders, respectively. The latter phase was termed `broken helix' (Fig.~\ref{fig:EIA_Fig1}e). Despite this complexity, both phases still host an axion insulator state that, in this case, is protected by a magnetic symmetry $\mathcal{T} C_{2}$ that the product of time-reversal and two-fold rotation  \cite{riberolles_magnetic_2021}. In the following, we demonstrate that the findings of our optical experiments challenge this established picture. 

\section{Optical probes}\label{sec:Experiment}

We measure optical reflectance to extract information on the three-fold rotational symmetry around $\hat{z}$ ($C_{3z}$) and  time-reversal symmetry ($\mathcal{T}$); the latter we probe in two complementary ways. In this section we briefly describe the physical principles behind these measurements, and the relevant experimental considerations. We then show the findings of these measurements performed on~\eia~as a function of temperature, position on the sample and cooling protocol. We discuss how the unique capability of these probes reveals information incompatible with the current understanding of \eia.

 \begin{figure}
 \centering
 \includegraphics[width=1\linewidth]{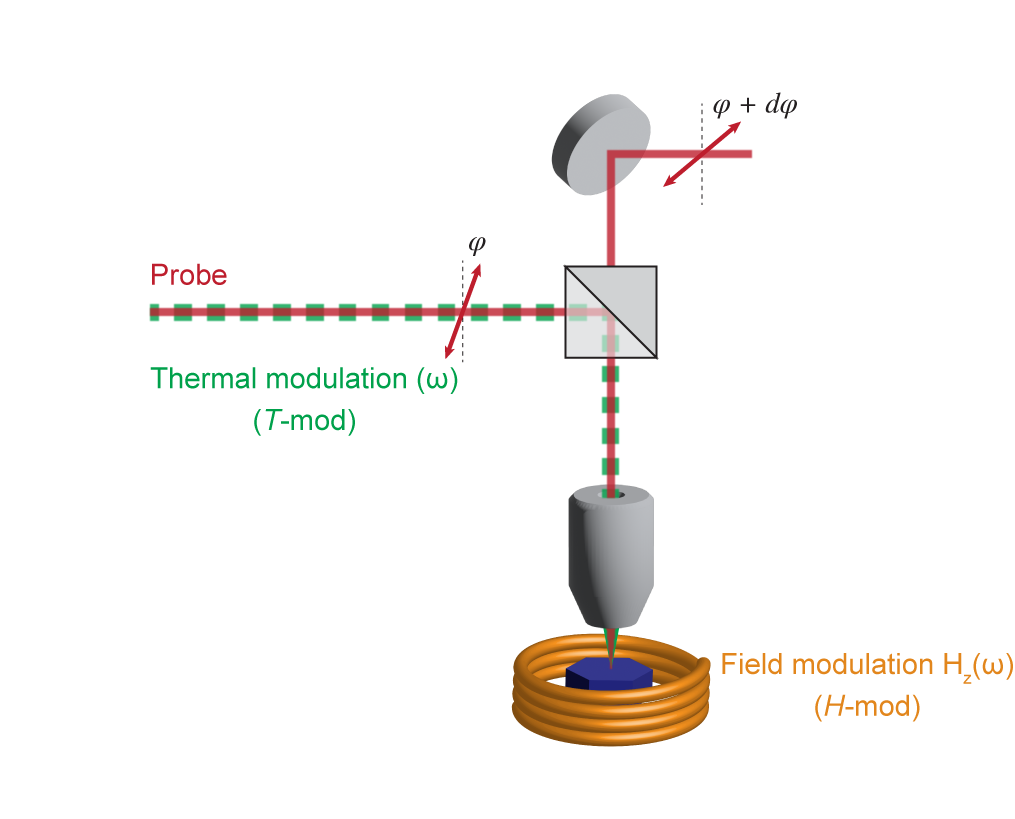}
 \caption{Schematic of the optical setup used for measurements of the polarization rotation as a function of incident polarization. For thermal modulation (\textit{T}-mod), an optically chopped second (pump) beam is spatially overlapped with the probe beam on the sample surface and used to modulate the temperature of the sample by heating it. During field modulation (\textit{H}-mod), an AC magnetic field is applied along the c-axis of the sample with a coil. During thermal (field) modulation, the field (pump beam) is turned off.}
 \label{fig:EIA_Fig2}
\end{figure}

\subsection{Symmetry sensitive experiment}

The reflectance at normal incidence is described by a $2\times2$ matrix, which can be parameterized as:
\begin{equation} \label{eq:reflectivity}
\bm{r} = r_0 \mathbbm{1}+ \delta r_s
\left(\begin{array}{cc}
\cos2\theta_0 & \sin2\theta_0 \\
\sin2\theta_0 &-\cos2\theta_0 
\end{array}\right)+\delta r_a
\left(\begin{array}{cc}
0 & 1 \\
-1 & 0 
\end{array}\right),
\end{equation} 
where $\mathbbm{1}$ denotes the unit matrix, and $\theta_0$ and $\theta_0+\pi/2$ correspond to the principal optical axes. $r_0$ is the isotropic contribution allowed in all materials, whereas $\delta r_s$ and $\delta r_a$, known as birefringence and the Kerr effect, respectively, contain distinct symmetry information. The crucial difference between them is their behavior with respect to the exchange of indices: $\delta r_s$ is symmetric and $\delta r_a$ anti-symmetric. The general reciprocity relations~\cite{shelankov_reciprocity_1992, halperin_hunt_1992} enforce $\delta r_a=0$ in $\mathcal{T}$-invariant systems; thus $\delta r_a\neq 0$ is an unambiguous probe of time-reversal symmetry breaking. In a reflectivity experiment the Kerr effect manifests as a change of polarization orientation upon reflection by an angle of $\delta r_a/r_0 \unit[~]{rad}$.

In contrast, $\delta r_s$ in not sensitive to $\mathcal{T}$, but is forbidden by rotational symmetry $C_{nz}$, if $n\geq3$. A non-zero $\delta r_s$ means that the reflectance depends on the relative orientation of the light polarization and the lattice. Like $\delta r_a$, $\delta r_s$ also induces a polarization rotation upon reflection, however the magnitude of the change depends on the incident polarization. In particular, the rotation vanishes when the incident polarization is aligned to the principal optical axes. As we show below, this fact is the basis for our detection of rotational symmetry breaking. 

Further symmetry-sensitive information can be gleaned from the change of reflectance in an applied magnetic field $H_z$. To linear order the symmetric and antisymmetric components of the field-induced change to reflectance, $r_H$, can be parameterized as: 

\begin{equation} \label{eq:reflectivityH}
\bm{r}_H = \delta \alpha_s H_z
\left(\begin{array}{cc}
\cos2\theta_H & \sin2\theta_H \\
\sin2\theta_H &-\cos2\theta_H 
\end{array}\right)+ \delta\alpha_a H_z
\left(\begin{array}{cc}
0 & 1 \\
-1 & 0 
\end{array}\right).
\end{equation} 
With the additional $\mathcal{T}$-odd factor, $H_z$, the reciprocity relations now allow the anti-symmetric term ($\delta\alpha_a$) in all materials, while the symmetric term ($\delta\alpha_s$) indicates breaking of time reversal symmetry. A nonzero ($\delta\alpha_s$) is usually referred to as the linear magneto-optic effect~\cite{eremenko_magneto-optics_1987}, or linear magneto-birefringence (LMB). The principal axes of the LMB response are given by $\theta_H$ and $\theta_H+\pi/2$. Although both $\delta r_a$ and $\delta\alpha_s$ require time-reversal to be broken, the symmetry conditions that enforce them to be non-zero  are different. This follows from the first term of Eq.~\ref{eq:reflectivityH}, which shows that $C_{3z}$ must be broken in addition to $\mathcal{T}$ for $\delta\alpha_s$ to be non-zero.

The three distinct symmetry-sensitive quantities that we obtain through reflectance measurements are therefore $\delta r_s$, $\delta r_a$, and $\delta \alpha_s$. All of them induce changes of light polarization upon reflection, and can be distinguished from each other by our optical techniques~~\cite{sunko_spin-carrier_2023, lee_observation_2022, little_three-state_2020}. The fundamental observable is the rotation of the angle of linear polarization ($d\phi$) about the optical axis at normal incidence, as a function of sample orientation. Since we cannot physically rotate the sample, we access the same information by rotating the incoming light polarization ($\phi$). The change of polarization upon reflection at temperature $T$ and field $H_z$ is given by:
\begin{equation}
\label{eq:dphi}
d\phi = A\left(T, H_z\right)\sin\left[2\left(\phi-\theta\left(T, H_z\right)\right)\right]+B\left(T, H_z\right).
\end{equation}
The amplitude of the sinusoidal variation, $A\left(T, H_z\right)$, and the principal axis orientation, $\theta\left(T, H_z\right)$, are determined by the symmetric part of the reflectance tensors ($\delta r_s$, $\delta \alpha_s$), while the constant offset $B\left(T, H\right)$ originates from the anti-symmetric $\delta r_a$ and $\delta \alpha_a$. It is now clear how the quantities of interest can be experimentally distinguished: in a $H_z = 0$ measurement the sinusoidally-varying and constant $d\phi$ originate from $\delta r_s$ and $\delta r_a$, respectively, while $\delta \alpha_s$ is captured by a sinusoidal variation proportional to a magnetic field.

In practice, we perform measurements in two modes (Fig.~\ref{fig:EIA_Fig2}; see Appendix~\ref{sec:Optical} for details). In the temperature modulated mode ($T$-mod) we modulate temperature at a frequency $f\approx\unit[2]{kHz}$ using a second laser beam as a heater, therefore measuring the temperature derivative of Eq.~\ref{eq:dphi}. This experimental procedure reveals the same information about symmetry as the unmodulated experiment at $H_z = 0$ ($\delta r_s$, $\delta r_a$), while enhancing sensitivity and rejecting contributions from inevitable setup imperfections. 

In the field-modulated mode ($H$-mod) we modulate the magnetic field supplied by a copper coil ($H_z\approx\unit[3]{mT}, f\approx\unit[100]{Hz}$), and measure the field derivative of Eq.~\ref{eq:dphi}. This experiment is sensitive to $\delta \alpha_s$. Therefore, the symmetry constraints to observe non-zero temperature and field derivatives of $A$ ($\partial_TA$ and $\partial_HA$, respectively), differ. As we will demonstrate below, the ability to simultaneously measure these quantities is crucial to determine the symmetry of the two ordered phases in \eia. 

\subsection{Onset of the two phases}

\begin{figure}[b]
 \centering
 \includegraphics[width=0.9\linewidth]{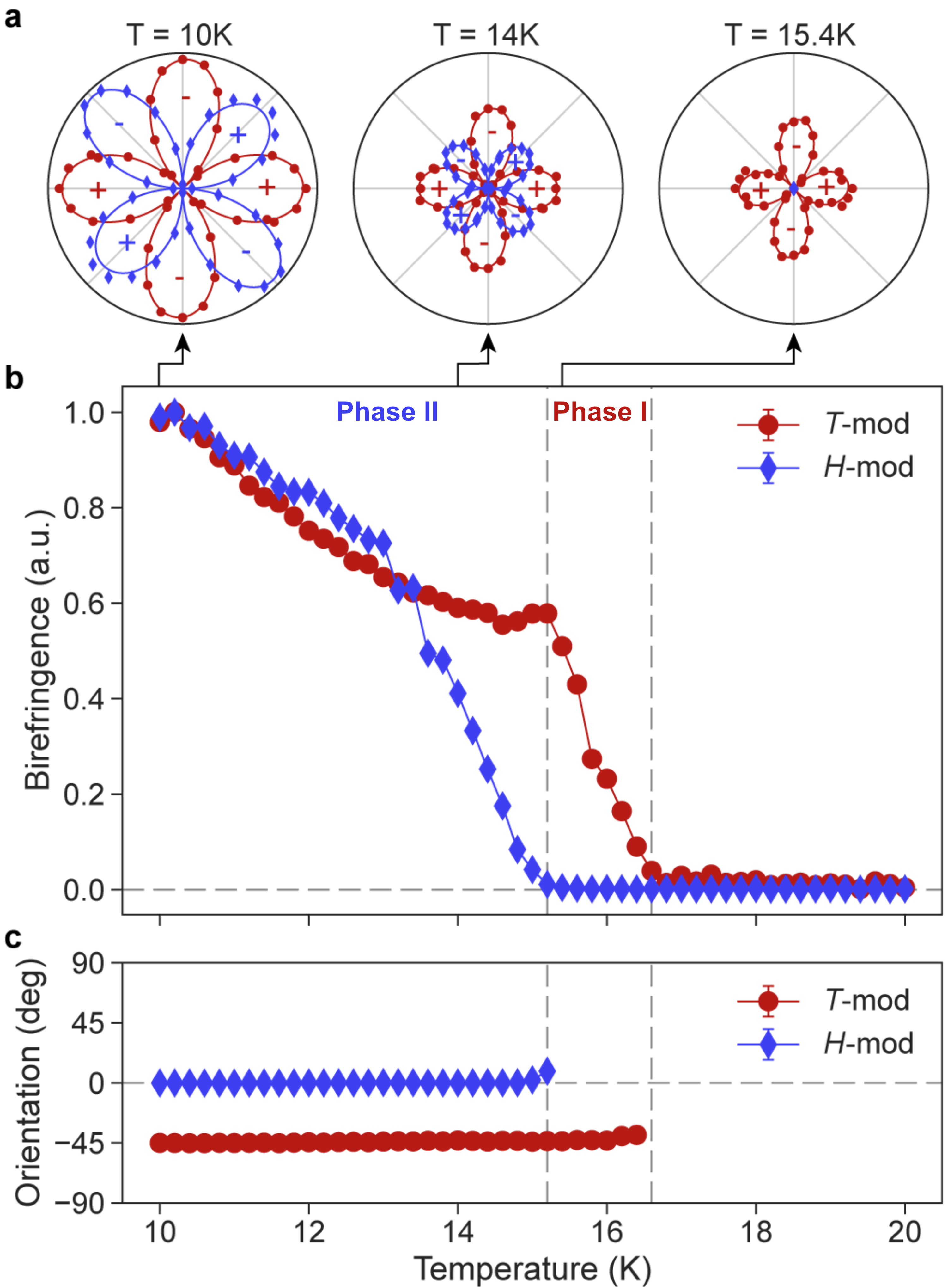}
 \caption{(a) Polar plots of thermally modulated (\textit{T}--mod, red circles) and field modulated (\textit{H}--mod, blue diamonds) polarization rotation $\delta\varphi$ at different temperatures reveal distinct signatures of broken rotational symmetry in each of the two magnetic phases. To illustrate the remnant 180$\degree$ rotational symmetry, $\delta\varphi$ measured as a function of the incident polarization angle from $-90\degree\rightarrow 90\degree$ is replicated for $90\degree\rightarrow -90\degree$. (b) Temperature dependence of \textit{T}--mod and \textit{H}--mod amplitude. The \textit{T}--mod signal onsets at the transition temperature associated with the higher temperature phase, whereas the \textit{H}--mod signal onsets at the lower transition temperature. The amplitudes are normalized at $\unit[10]{K}$. (c) Temperature dependence of the principal axes orientations corresponding to the \textit{T}--mod and \textit{H}--mod signals. The principal axes associated with the two signals are oriented $45\degree$ relative to each other and remain constant with temperature.}
 \label{fig:EIA_Fig3}
\end{figure}

Fig.~\ref{fig:EIA_Fig3} illustrates the onset with decreasing temperature of the optical signatures of broken symmetry. The polar plots in Fig.~\ref{fig:EIA_Fig3}a show the $T$-mod and $H$-mod rotation of polarization as red and blue symbols, respectively, measured at three temperatures. The orientations of the principal axes obtained by the two modulation modes differ by $\theta_H-\theta_0 =\unit[45]{\degree}$, a point we return to later. The temperature dependence of the $T$-mod and $H$-mod amplitude and principal axis orientation are plotted in Fig.~\ref{fig:EIA_Fig3}b and Fig.~\ref{fig:EIA_Fig3}c, respectively. Both amplitudes exhibit a sharp onset, separated by $\unit[1.4]{K}$. The onset of the $T$-mod birefringence coincides with the appearance of the $\bm{Q}_{1}$ peak (Phase I) and the $H$-mod signal to the appearance of antiferromagnetic $\bm{Q}_{2}$ peak (Phase II), as detected by x-ray scattering measurements performed on the same crystal~\cite{soh_understanding_2023}.

The two optical experiments, which probe two distinct symmetries,  revealed the two transitions, and unambiguously showed that the symmetry-breaking pathway proposed on the basis of scattering experiments cannot be correct. The $\unit[60]{\degree}$-helix structure previously assumed to describe Phase I preserves the $C_{3z}$ symmetry of the lattice, and is thus incompatible with the birefringence onset at $T_{N1}$. In fact, birefringence was expected to arise only at $T_{N2}$, in contrast to the observation. Furthermore, spatially-resolved measurements, discussed in the following section,  reveal additional information, which is incompatible with the past understanding of both phases in \eia.

\subsection{Spatial distribution of optical signals}\label{sec:spatial_distribution}

To investigate the spatial dependence of the two optical signals, we raster scanned the sample under the beam focus and repeated the polarization rotation measurements across a $240 \times 260 \unit{\mu m^2}$ sample region, with measurements taken every $\unit[20]{\mu m}$. These measurements revealed two surprising facts: (a) the principal optical axes in \eia~can assume any orientation with respect to the lattice, and (b) despite that breadth of orientations, the fundamental $H$--mod signal is the same in every position on the sample.

\subsubsection{$T$-mod signal}

\begin{figure}[t]
 \centering
 \includegraphics[width=1\linewidth]{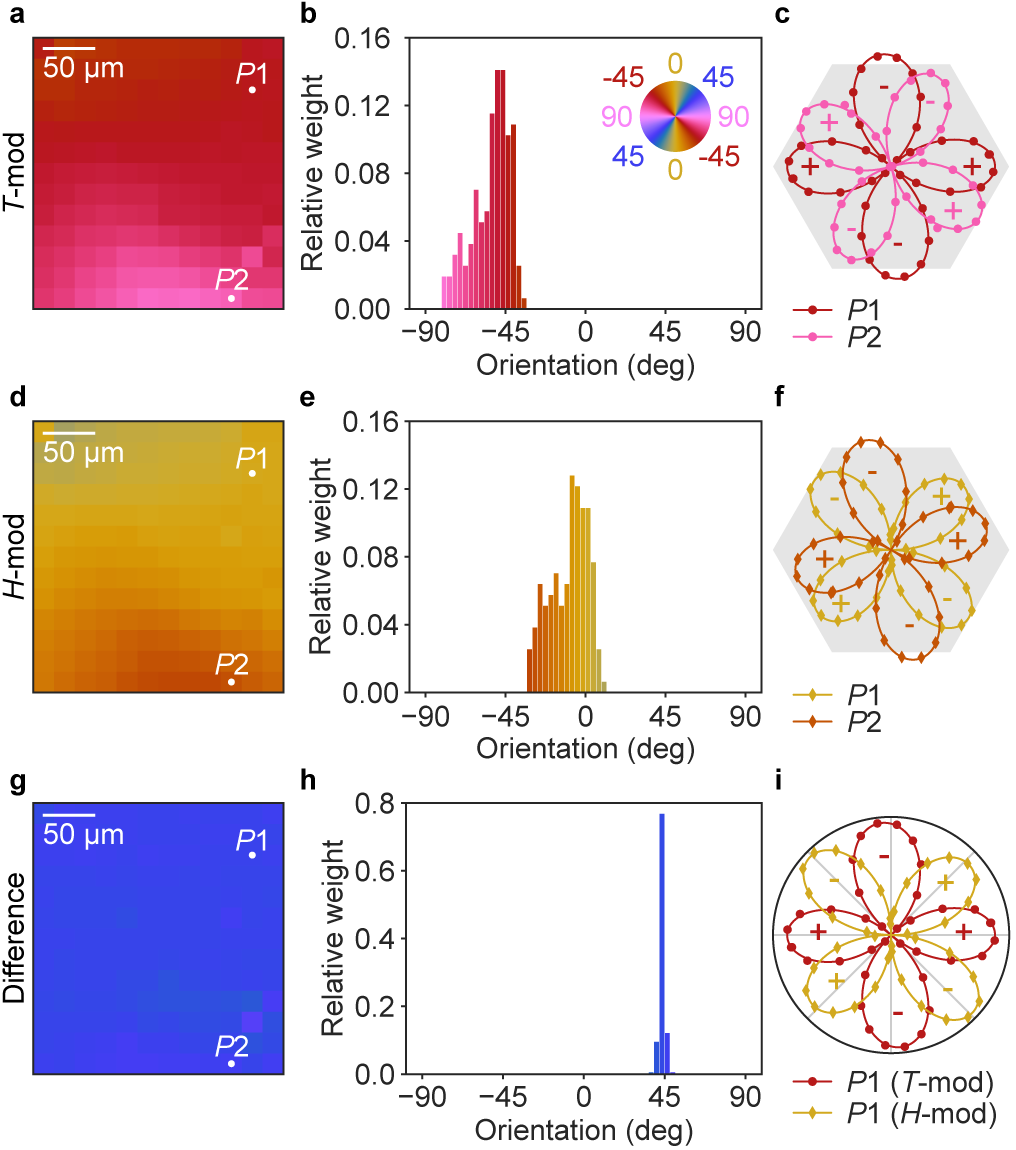}
 \caption{Principal axis orientations associated with \textit{T}--mod and \textit{H}--mod signals: (a), (d), (g) Maps of the spatial distributions of the principal axis orientations for \textit{T}--mod ($\theta_0$), \textit{H}--mod ($\theta_H$), and their difference ($\theta_H - \theta_0$), respectively. (b), (e), (h) Histograms of the distributions in (a), (d), (g), respectively. The orientations of the \textit{T}--mod and \textit{H}--mod signals are broadly distributed and continuously varying, whereas the difference between the orientations, $\theta_H - \theta_0$, is sharply peaked at $45\degree$. (c), (f) Polar plots of the polarization rotation at points labeled $P1$ and $P2$ in (a), (d), respectively, overlaid on hexagons to demonstrate the lack of registration to the crystalline axes. The principal axes of the two points are oriented $\approx 30\degree$ from each other in each of the \textit{T}--mod and \textit{H}--mod maps. (i) Polar plot of the \textit{T}--mod and \textit{H}--mod polarization rotations at point $P1$, demonstrating the principal axes of the \textit{T}--mod and \textit{H}--mod signals to be oriented $45\degree$ relative to each other.}
 \label{fig:EIA_Fig4}
\end{figure}

First we focus on the $H=0$ signal. In Figs.~\ref{fig:EIA_Fig4}(a,b) we show a map of the spatial distribution of the principal axis orientation $\theta_0$, and the corresponding histogram. We find a single broad peak in the distribution of principal axis orientations, spanning more than $\unit[30]{\degree}$ (Fig.~\ref{fig:EIA_Fig4}b), as emphasized by the polar plot showing the polarization dependence of the $T$-mod signal taken at two sample positions (Fig.~\ref{fig:EIA_Fig4}c). Since the two high-symmetry directions in the Eu planes are separated by $\unit[30]{\degree}$, these measurements show that the principal optical axes can take \textit{any} orientation with respect to the lattice. 

The observation of a broad distribution of principal axes is surprising, since breaking of the discrete $C_{3z}$ symmetry is expected to yield three domains related by $C_{3z}$, which would manifest as three narrow peaks in the histogram, separated by $\unit[60]{\degree}$~\cite{little_three-state_2020,xu_three-state_2022,sunko_spin-carrier_2023,ye_elastocaloric_2023}. This observation could be interpreted in two ways, which are important to distinguish. Either the mapped region is dominated by a single domain, and the broadening of the histogram is caused by contributions of the other two domains which cannot be resolved with our diffraction-limited resolution of 1 $\mu$m (micro-domain scenario), or the magnetic order microscopically assumes a continuum of orientations (orientation continuum scenario). We note that the broken helix structure, proposed for Phase II (Fig.~\ref{fig:EIA_Fig1}e), protects the axion state only for one orientation of the structure with respect to the lattice; the orientation continuum scenario would therefore drastically change the topological properties. Below, we show that only the orientation continuum scenario is consistent with the data, requiring a re-evaluation of symmetry and topology in~\eia.

\begin{figure}
 \centering
 \includegraphics[width=1\linewidth]{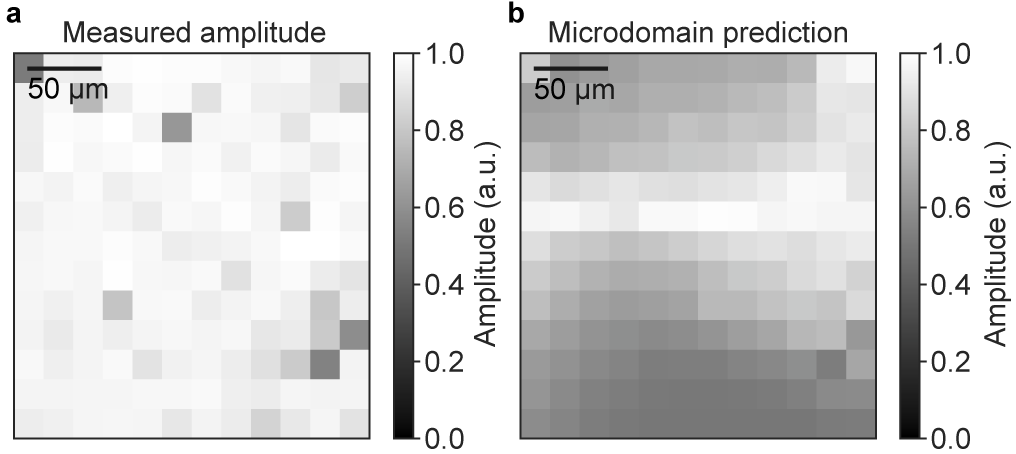}
 \caption{Measured and simulated birefringence amplitude (\textit{T}--mod): (a) Thermally modulated birefringence amplitude in the same region as the maps in Fig.~\ref{fig:EIA_Fig4}. The amplitude is nearly constant over the region. (b) Birefringence amplitude simulated using the microdomain model prediction for birefringence amplitude from orientation. The microdomain model predicted amplitude is expected to vary as the measured birefringence angle is continuously tuned, in contrast with the observed experimental amplitude.}
 \label{fig:EIA_Fig5}
\end{figure}

The orientation continuum scenario is proven by analyzing the amplitude of birefringence as a function of position. In the micro-domain scenario the amplitude is the largest when the signal is dominated by a single domain, and is reduced by averaging over micro-domains, yielding a well-defined prediction for the relationship between the birefringence amplitude and orientation (see Appendix~\ref{sec:Microdomains} for more details). In contrast, the amplitude is independent of the angle in the orientation continuum scenario. As evident in Fig.~\ref{fig:EIA_Fig5}a, the amplitude is uniform across the sample region in which the principal axis orientation spans an angle larger than $\unit[30]{\degree}$, i.e. the full range between the high-symmetry directions of the lattice. In contrast, in the microdomain scenario the amplitude would systematically vary across this region (simulated in Fig.~\ref{fig:EIA_Fig5}b).

The confirmation of the orientation continuum scenario suggests that the local principal axes are chosen by a built-in strain, and that the distributions in Figs.~\ref{fig:EIA_Fig4}(a,b) reflect the distribution of built-in strain axes. A somewhat similar scenario was proposed to explain the rotation of the nematic director in the superconducting nematic phase of twisted bilayer graphene \cite{fernandes_nematicity_2020, cao_nematicity_2021}. However, it is unusual to observe an orientation continuum in a magnetic system, since magneto-crystalline anisotropy (MCA) tends to pin moments to the high-symmetry direction of the lattice.   Our observation therefore suggests a very weak MCA, such that an inevitable built-in strain dominates, and selects an arbitrary orientation at a generic sample position. This hypothesis is proven below by deliberate application of uniaxial strain (Sec.~\ref{Sec:StrainTuning}), showing that the magnetic space group can be manipulated, with important consequences for topology.

\subsubsection{$H$-mod signal}

The $H-$mod signal measured over the same sample region reveals an angle distribution $\theta_H$ of principal optical axes of similar width (Figs.~\ref{fig:EIA_Fig4}(d-f)). However, it is quite striking that the distribution of the \textit{difference} of the two angles ($\theta_H-\theta_0$) is remarkably narrow, and centered at $\unit[45]{\degree}$ (Figs.~\ref{fig:EIA_Fig4}(g-i)). 

This finding raises two questions, which we address in the remainder of this section: why the distribution is so narrow, and why time-reversed domains of $\theta_H-\theta_0=\unit[-45]{\degree}$ are not observed anywhere on the sample.

The narrow distribution of the difference $\theta_H-\theta_0=\unit[45]{\degree}$ suggests that this relationship is enforced by symmetry, as has been observed in some other magnetic systems. For example, in Ref.~\cite{sunko_spin-carrier_2023} it was shown that the magnetic point groups $2/m$ and $2'/m'$ allow only for $\theta_H-\theta_0 = \pm 45 \degree$ and $\theta_H-\theta_0 = 0, 90 \degree$, respectively. However, further consideration reveals that this cannot be the case in \eia: since the principal axis orientation generically does not coincide with the high-symmetry directions of the crystal, no in-plane rotational axes or vertical mirror planes remain valid symmetries at a generic sample position, alone or in combination with time reversal. Therefore, there is no symmetry that can enforce a relationship between $\theta_H$ and $\theta_0$, and a microscopic mechanism must instead underlay the sharp angle distribution in Fig.~\ref{fig:EIA_Fig4}h. Identifying why $\theta_H-\theta_0 = 0, 90 \degree$ is not observed, although it is allowed by symmetry, is beyond the scope of our work, and we hope that our findings motivate further \textit{ab initio} investigation to identify the origin of the observation reported in Figs.~\ref{fig:EIA_Fig4}(g-i).

\begin{figure}[t]
 \centering
 \includegraphics[width=1\linewidth]{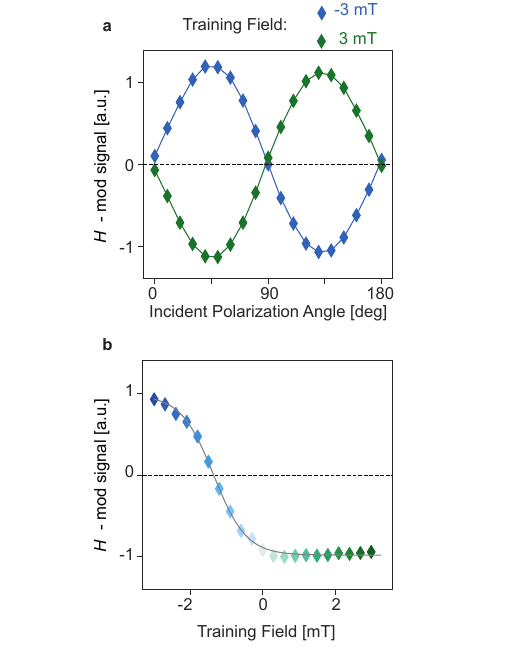}
 \caption{(a) $H$-mod signal as a function of incident polarization angle measured at $\unit[5]{K}$ in a sample cooled in a training field of $\unit[-3]{mT}$ and $\unit[3]{mT}$ (blue and green symbols, respectively). (b) $H$-mod signal measured using incident polarization of $\unit[50]{\degree}$, as a function of a training DC field applied during cooling. Both panels clearly demonstrate that the signal can be trained by a magnetic field, proving its TR-odd nature. }
 \label{fig:EIA_Fig6}
\end{figure}

The remaining puzzle is the absence of the time-reversed domain  ($\theta_H-\theta_0=\unit[-45]{\degree}$). A closely related observation is that the same $\theta_H-\theta_0=\unit[45]{\degree}$ domain was observed on each cool-down through the transition, which is not expected of a $\mathcal{T}$-breaking order parameter that condensed in the absence of a  magnetic field. To explore this, we measured $H$-mod birefringence at $\unit[5]{K}$ following cooling in a DC magnetic field (Fig.~\ref{fig:EIA_Fig6}a). The sign of the signal can be trained by the field, confirming its $\mathcal{T}$-odd nature. The two time-reversed states are characterized by values of $\theta_H$ that differ by $\unit[90]{\degree}$, and correspond to $\theta_H=\theta_0\pm\unit[45]{\degree}$.

We further monitor the maximum signal, at $\unit[50]{\degree}$ of incident polarization, as a function of the training field (Fig.~\ref{fig:EIA_Fig6}b), and find that even at a vanishing applied training field the signal reaches a value close to saturation. This indicates that the Earth's field is sufficient to train the domains, consistent with the observation that the sign of the signal does not change with sample position, or between different cooldowns. We conclude that the $\mathcal{T}$-breaking domains in \eia~are easily switched, and their sign can be detected through LMB.

\section{Determining the magnetic structures}\label{sec:Symmetry}

In this section, we use a multimodal approach to determine the magnetic structures in \eia. Both optical and scattering experiments show evidence of two magnetic transitions as a function of temperature, yielding two magnetic phases. These experiments impose complementary constraints on the corresponding magnetic structures. Our symmetry-sensitive optical measurements show that: (i) $C_{3z}$ is broken in Phases I and II; (ii) $\mathcal{PT}$ is broken in Phase II, but preserved in Phase I; (iii) the in-plane magneto-crystalline anisotropy is negligible. Additionally, scattering experiments~\cite{riberolles_magnetic_2021, soh_understanding_2023} revealed the propagation vectors: (iv) $\mathbf{Q}_{1}=(0,0,1/3)$ in Phase I and (v) $\mathbf{Q}_{1}=(0,0,1/3)$ and $\mathbf{Q}_{2}=(0,0,1)$ in Phase II. While observation (i) is inconsistent with the 60\degree-helix previously associated with Phase I~\cite{riberolles_magnetic_2021, soh_understanding_2023}, we show in this section that it is possible to reconcile all the experimental results for both phases. 

In Phase I we find a collinear state with varying amplitude of the magnetic moments that we hereafter refer to as \textit{nodal amplitude-modulated collinear order}, while in Phase II we find a state similar to the previously proposed broken-helix~\cite{riberolles_magnetic_2021}, but with one key difference: the moments in the experimentally observed structure are not pinned to the crystalline axes. We refer to this state as the \textit{unpinned broken helix}, and show that it can be manipulated with uniaxial strain (Sec.~\ref{Sec:StrainTuning}), raising the possibility of on-demand control of electronic topology in \eia. This tunability is a direct consequence of weak magneto-crystalline anisotropy, which can be overpowered even by modest built-in strain.

In the remainder of this section, we highlight the main steps involved in our experimentally guided symmetry analysis, while more details are given in Sec.~\ref{secSM:symmetry} and Sec.~\ref{secSM:op} of the Appendix. In Sec.~\ref{sec:freeE} we complement the symmetry analysis with a phenomenological free energy model, which captures the sequence of broken symmetries, and the corresponding evolution of magnetic structure. Finally, we summarize the main results of the symmetry and free energy analyses; a reader more interested in those results than in the reasoning that led to them may immediately proceed to Sec.~\ref{sec:resultsFE}. Illustrations of the magnetic structures associated with the two phases are shown in Fig.~\ref{fig:EIA_Fig7}d and Fig.~\ref{fig:EIA_Fig8}(c,e), respectively.  

\subsection{Multimodal Approach}

Scattering experiments reveal the successive onset of two propagation vectors, $\mathbf{Q}_{1}$ and $\mathbf{Q}_{2}$, resulting in a non-zero expected value of Eu magnetic moment  that can be expressed as the sum over Fourier components:

\begin{equation} \label{eq:magneticMoment}
 \mathbf{M}_\alpha(\mathbf{r}_i)=\sum_{\mathbf{Q}}e^{i\mathbf{Q}\cdot\mathbf{r}_i}\mathbf{M}_\alpha(\mathbf{Q}),
\end{equation} 
where $\mathbf{r}_i$ denotes the position of a Eu site, and $\alpha=1, 2$ is the Eu sublattice index. $\mathbf{Q}=\pm\mathbf{Q}_1$ in Phase I and $\mathbf{Q}=\pm\mathbf{Q}_1,\pm\mathbf{Q}_2$ in Phase II. However, the scattering experiments are not sufficient to uniquely determine $\mathbf{M}_\alpha(\mathbf{Q})$; removing this ambiguity is the objective of our analysis. We note that since both propagation vectors are perpendicular to the Eu layers ($\mathbf{Q}_1 || \mathbf{Q}_2 ||\mathbf{\hat{z}}$), it is sufficient to consider the average magnetic moment in each Eu layer, $\mathbf{M}_\alpha(z_i) = \left\langle\mathbf{M}_\alpha(\mathbf{r}_i)\right\rangle_{z=z_i}$, rather than the magnetic moment at each site. 

The first step is to identify all magnetic structures $\mathbf{M}_\alpha(z_i)$ consistent with the scattering experiments, and with the fact that the magnetic moments lie in Eu planes. The next step is to identify the magnetic space groups associated with each of those magnetic configurations. The symmetry-sensitive information obtained from the optical experiments can then be used to rule out a vast majority of structures obtained in the previous step. In Phase I of \eia~this symmetry-based approach has proven sufficient to uniquely identify the magnetic structure, while in Phase II it needs to be complemented with further considerations, as explained in Sec.\ref{sec:freeE}.

\subsection{Phase I: Amplitude modulated collinear order}\label{sec:PhaseI}

Each magnetic structure lowers the system symmetry from the parent paramagnetic group $P6_3/mmc1'$ (No. 194.264) to a magnetic space group (MSG) that is a subgroup of the paramagnetic group, and has the symmetry properties of one or more of its irreducible representations (Irreps). We find that all magnetic configurations $\mathbf{M}_\alpha(z_i)$ that are consistent with the observed magnetic Bragg peaks and in-plane magnetic moments transform according to the $m\Delta_6$ Irrep (see Appendices~\ref{secSM:symmetry} and~\ref{secSM:op} for a detailed analysis). One of them is the 60\degree-helix proposed in earlier works~\cite{riberolles_magnetic_2021, soh_understanding_2023}, which is inconsistent with the observed birefringence. We now show that there is only one magnetic structure for Phase I consistent with all the experimental results (i)-(iii) outlined above.

\begin{figure}[t]
 \centering
 \includegraphics[width=1\linewidth]{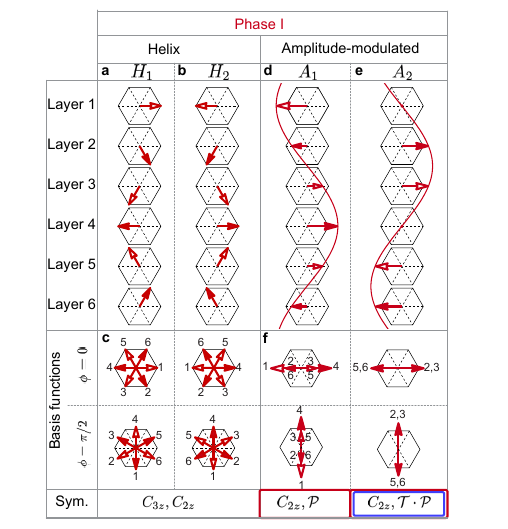}
 \caption{Top row: The layer-resolved moment orientation in (a, b) the two helical structures of opposite helicities ($H_1$ and $H_2$); (d,e) the two symmetry-distinct amplitude-modulated structures ($A_1$ and $A_2$, respectively). Middle row: Two equivalent choices for the basis function for the $m\Delta_6$ Irrep: (c) helical and (f) amplitude-modulated states. Bottom row: symmetries associated with the four different states: helical, $A_1$ and $A_2$. We note only the symmetries that are present regardless of the relative orientation of the lattice and the moments (for more details, see Table~\ref{tabSM:table1} in the Appendix). We mark in red and blue the symmetries consistent with the $T$-mod and $H$-mod measurements, respectively.}
 \label{fig:EIA_Fig7}
\end{figure}

It is useful to rewrite the order parameter of Phase I, $\mathbf{M}_\alpha(\mathbf{Q}_1)$ as a vector in the four-dimensional representation space of $m\Delta_6$, whose basis functions are any four linearly independent arrangements of the magnetic moments that transform as the $m\Delta_6$ Irrep. One possible choice of the basis functions is the four 60\degree-helix states of opposite helicities and orthogonal orientations, corresponding to the states $H_1$, $H_2$ and $\phi=0$, $\phi=\pi/2$ in Fig.~\ref{fig:EIA_Fig7}(a-c). For our purposes, however, it is more convenient to consider the amplitude-modulated collinear magnetic states of different phase, shown in Fig.~\ref{fig:EIA_Fig7}(d-f) and labeled by $A_1$, $A_2$ and $\phi=0$, $\phi=\pi/2$. In these structures, all magnetic moments point in the same direction, while their amplitude varies sinusoidally from layer to layer. 

There are two classes of amplitude-modulated structures, hereafter called \textit{nodeless} (denoted by $A_1$) and \textit{nodal} (denoted by $A_2$). In the nodeless structure, $\mathbf{M}_\alpha(z_i)$ is non-zero in all Eu layers, while in the nodal structure it vanishes in every third Eu layer (Fig.~\ref{fig:EIA_Fig7}(d,e)). In principle, the magnetic moments associated with each of the orders $A_{1,2}$ could assume any orientation. Therefore, the four structures obtained by $A_{1,2}$ with moments pointing along two orthogonal directions form a valid basis of the $m\Delta_6$ representation space (Fig.~\ref{fig:EIA_Fig7}f). Written in this basis, the order parameter for Phase I takes the form 
\begin{equation}
\bm{\Psi}_I=
\left(
a_1 \cos{\phi_1}, a_1 \sin{\phi_1}, a_2 \cos{\phi_2} , a_2 \sin{\phi_2}
\right),
\label{eq:psiI}
\end{equation}

\noindent where $\phi_{1,2}$ and $a_{1,2}$ set the orientation and amplitude, respectively, of the moments associated with $A_{1,2}$. Any magnetic structure consistent with the scattering pattern observed in Phase I can be obtained by specific choices of $a_{1,2}$ and $\phi_{1,2}$, and the relationship between $\bm{\Psi}_I$ and $\mathbf{M}_\alpha(\mathbf{Q}_1)$ is derived in Appendix~\ref{secSM:op}. For instance, while the amplitude-modulated phases $A_1$ and $A_2$ in Fig.~\ref{fig:EIA_Fig7}(d-e) are described by $a_1 \neq 0$, $a_2 = 0$ and $a_2 \neq 0$, $a_1 = 0$, respectively, the 60\degree-helix phases $H_1$ and $H_2$ are parametrized by $a_1=a_2$ and $\phi_2=\phi_1 + \pi/2$, with $\phi_1=\pi$ for the $H_1$ phase and $\phi_2=\phi_1 - \pi/2$ with $\phi_1 = 0$ for the $H_2$ phase.  

While $\bm{\Psi_I}$ is compatible with the scattering experiments for any choice of $a_{1,2}$ and $\phi_{1,2}$, optical measurements impose further constraints. Birefringence is non-zero only in structures that break the three-fold rotation symmetry along $z$ ($C_{3z}$), which can be re-cast as a non-zero value of the composite three-state Potts nematic order parameter~\cite{little_three-state_2020}, 
\begin{equation} \label{eq:nematic_psi} 
\bm{\eta}= \frac{1}{2} \left(\begin{matrix}
a_1^2 \cos(2\phi_1) + a_2^2 \cos(2\phi_2)\\
a_1^2 \sin(2\phi_1) + a_2^2 \sin(2\phi_2)
\end{matrix}\right).
\end{equation}

\noindent The condition $\bm{\eta}\neq 0$ for Phase I excludes the 60\degree-helical states, and allows any of the amplitude-modulated states. This is because, as explained above, the 60\degree-helical states have $a_1 = a_2$ and $\phi_2 = \phi_1 \pm \pi/2$, which makes $\bm{\eta} = 0$.

The absence of an $H$-mod signal in Phase I enables us to distinguish between the two amplitude-modulated states, as this signal is forbidden by the product of inversion and time reversal symmetry ($\mathcal{PT}$). Therefore, the state $A_1$ is ruled out, as the Eu sites remain centers of inversion, as they were in the parent space group, and $\mathcal{P}\mathcal{T}$ is broken.  In contrast, in $A_2$ inversion symmetry is broken, while $\mathcal{P}\mathcal{T}$ is preserved. In fact, this nodal amplitude-modulated structure is the only one belonging to the $m\Delta_6$ Irrep that allows for non-zero $\bm{\eta}$ and vanishing $H$-mod signal. Therefore, the only magnetic structure that is consistent with all experimental findings in Phase I is $A_2$, characterized by $a_1=0$ and $a_2 \neq 0$ in Eq.~\ref{eq:psiI}.

One remaining degree of freedom is the direction of the magnetic moments within the $ab$-plane, set by $\phi_2$. Setting $a_1 = 0$ in Eq. (\ref{eq:nematic_psi}), we note that the direction of the magnetic moment coincides with the orientation of the nematic director. Therefore, the principal axis extracted from our optical birefringence measurements corresponds to the direction of the magnetic moments. Although the broad distribution of principal axis orientations shown in Fig.~\ref{fig:EIA_Fig4} was measured in Phase II, we have demonstrated that the orientation does not change as a function of temperature (Fig.~\ref{fig:EIA_Fig3}e). Therefore, the distribution of orientations is equally broad in Phase I, indicating that $\phi_2$ is not constrained. This observation is surprising from a symmetry perspective, since there are always symmetry-allowed terms in the free energy,  which can be thought of as magneto-crystalline anisotropy, that favor $\phi_2$ pointing along the high-symmetry directions. The experimental observation therefore indicates that the magneto-crystalline anisotropy is weak compared to built-in strain.

Our work is the first to suggest an amplitude modulated phase in \eia, and our conclusion is supported by the results of M{\"o}ssbauer spectroscopy (reported in the Supplementary Fig.~10 of Ref.~\cite{riberolles_magnetic_2021}). These measurements also confirm that the valence Eu$^{2+}$ in \eia~is temperature independent, fixing spins on individual sites to $S=7/2$ in all the phases. The amplitude modulation therefore arises from the sinusoidal variation of the thermal average of the magnetic moments. M{\"o}ssbauer measurements are not sensitive to the phase of amplitude modulation and, therefore, cannot distinguish  between $A_1$ and $A_2$. We note that a  nodal amplitude-modulated phase had previously been observed in ground state of the Na-doped SrFe$_2$As$_2$ \cite{allred_double_2016}, where it originates from itinerant magnetism, in contrast to the local moment magnetism of \eia.

\subsection{Phase II: Mixed order} \label{sec:PhaseII}

\begin{figure}
 \centering
 \includegraphics[width=1\linewidth]{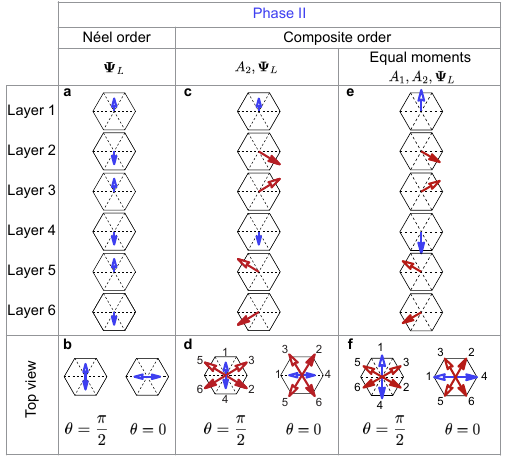}
 \caption{Top row: The layer-resolved spin orientation in (a) the \Neel~state; (c) in the mixed structure obtained by combining $A_2$ order with the \Neel~order; the blue arrows indicate the orientation of $\bm{\Psi}_L$; (e) The mixed structure obeying the equal moment condition (Eq.~\ref{eq:condition}). Bottom row: (b) the basis functions of the $m\Gamma_5^+$ Irrep, corresponding to the two perpendicular orientations of the \Neel~order. (d,f) The top view of two orthogonal orientations of the mixed order shown in (c,e), respectively. $\theta$ denotes the orientation of the \Neel~order. The structure shown in panel f with $\theta=\pi/2$ corresponds to the broken helix considered in previous work~\cite{riberolles_magnetic_2021}. }
 \label{fig:EIA_Fig8}
\end{figure}

We now turn to the analysis of Phase II, whose onset is characterized by the appearance of an additional ordering vector $\mathbf{Q}_2=(0,0,1)$, corresponding to the emergence of \Neel~order (illustrated in Fig.~\ref{fig:EIA_Fig8}a). In-plane \Neel~order transforms as the $m\Gamma_5^+$ Irrep of the paramagnetic group of \eia~and its order parameter takes the form of a two-component vector,
\begin{equation}
\label{eq:Neel}
\bm{\Psi}_{L}=l\left(\cos{\theta}, \sin{\theta}\right),
\end{equation}

\noindent in the space of $m\Gamma_5^+$. Here, $l$ denotes the amplitude of the moments in the \Neel~state and $\theta$ determines their orientation with respect to the in-plane crystal axes (Fig.~\ref{fig:EIA_Fig8}(a,b)). 

Phase II is, therefore, parameterized by both $\bm{\Psi}_{I}$ defined in Eq.(\ref{eq:psiI}) and $\bm{\Psi}_{L}$, and all candidate magnetic states for this phase can be described by different choices of $a_{1,2}$, $\phi_{1,2}$, $l$ and $\theta$. One example with $\bm{\Psi}_{I}=(0,0,a_2,0)$ and $\bm{\Psi}_{L}=(0,l)$ is shown in Fig.~\ref{fig:EIA_Fig8}(c,d). The blue arrows indicate the direction of the Néel component. A complete list of states compatible with Phase II can be found in Table~\ref{tabSM:table2} in the Appendix, where the relationship between $\mathbf{M}_{\alpha}(\mathbf{Q}_2)$ and $\bm{\Psi}_{L}$ is also derived.

To analyze the optical responses in Phase II, we first note that any non-zero $\bm{\Psi}_L$ also introduces a Potts-nematic order parameter:
\begin{equation}\label{eq:nematic_xi}
 \bm{\xi}=\frac{l^2}{2}\left(\begin{matrix}
\cos2\theta\\
\sin2\theta
\end{matrix}\right) \text{ .}
\end{equation}
Therefore, in the lowest temperature phase, both $\bm{\xi}$ and $\bm{\eta}$ [Eq.(\ref{eq:nematic_psi})] contribute to the observed birefringence via the combination $\bm{\eta} + \bm{\xi}$ (see Appendix \ref{secSM:nematic} for details). Furthermore, the Néel component breaks $\mathcal{PT}$, accounting for the onset of the $H$-mod signal in Phase II. 

From a symmetry perspective only, we cannot completely constrain the magnitudes of $a_{1,2}$ and $l$, nor their relative orientations. Many different structures with $\bm{\Psi}_{I}$ and $\bm{\Psi}_L$ both non-zero break $C_{3z}$ symmetry and $\mathcal{PT}$ symmetry, and are therefore consistent with the optical measurements. Indeed, Table \ref{tabSM:table2} shows various phases characterized by different combinations of the order parameters that result in different magnetic space groups.

However, by taking into account the local moment nature of the \eia~magnetism, we can narrow down the candidates for the magnetic ground state. The magnetism originates from the Eu$^{2+}$ localized magnetic moments ($S=7/2$), with no valence mixing with Eu$^{+3}$\cite{riberolles_magnetic_2021}. At the lowest temperatures, where fluctuations associated with the magnetic phase transitions are completely suppressed, the moments in each layer should have the same magnitude. Note that this statement is not in contradiction with our proposed amplitude-modulated Phase I, where the magnetic moments are small and fluctuate strongly because of proximity to magnetic phase transitions. Indeed, as we show below, the amplitude-modulated phase can smoothly evolve from a configuration near $T_{N2}$ in which the moments in each layer are not the same to a configuration at lower temperatures in which the moments all have the same magnitude, consistent with the aforementioned M{\"o}ssbauer spectroscopy data. 

Enforcing the moments to have the same magnitude corresponds to the additional conditions on the components of the $\bm{\Psi}_{I}$ and $\bm{\Psi}_L$ order parameters: 
\begin{equation}\label{eq:condition}
 a_{1} =  \sqrt{a_{2}^2 + 2l^2}-\sqrt{2}l , \hspace{5pt} \phi_1 = \phi_2\pm\frac{\pi}{2},\hspace{5pt} \theta=\phi_2\mp\frac{\pi}{2}. 
\end{equation}

\noindent We note that the broken helix state, proposed in Refs.~\cite{riberolles_magnetic_2021,soh_understanding_2023}, is a special combination of $\bm{\Psi}_{I}$ and $\bm{\Psi}_L$ which fulfills the equal-moment condition (Fig.~\ref{fig:EIA_Fig8}(e,f)). 
Although the equal-moment condition cannot be obeyed throughout Phase II (in contrast to assumptions in the previous work), the fact that there are no phase transitions between $T_{N2}$ and the lowest measured temperature (2K) indicates that the symmetry at $T \lesssim T_{N2}$ is the same as in the equal-moment ground state. The free energy model presented in the following section will provide a perspective on the evolution of the magnetic structure from the amplitude modulated state of Phase I to the equal-moment ground state.

In addition to the moment-length variation, the key difference between the broken helix considered previously and our findings is that the nematic order in Phase II  is not constrained by the underlying lattice, and we therefore refer to this structure as the \textit{unpinned broken helix}. As we discussed in the context of Phase I, this observation indicates that the magneto-crystalline terms in the free energy are weak compared even to the inevitable built-in strain. 

\section{Landau free-energy expansion}\label{sec:freeE}

In the previous section, we identified the symmetries of the two magnetic phases in \eia. Here, we introduce a phenomenological description of the evolution of the magnetic structure from the nodal amplitude-modulated state to the unpinned broken helix. The key finding is that it is not necessary to include two separate instabilities to describe the two phase transitions. Instead, the order $\Psi_{L}$ naturally arises from coupling to the order $\Psi_{I}$ of Phase I.

\subsection{Phenomenological description of the two transitions}

Our starting point is the Landau free energy
\begin{equation}\label{eq:FreeEn}
F\left(\bm{\Psi}_{I},\bm{\Psi}_{L}\right)=F_A\left(\bm{\Psi}_{I}\right)+F_L\left(\bm{\Psi}_{L}\right) + F_{AL}\left(\bm{\Psi}_{I},\bm{\Psi}_{L}\right).
\end{equation}

\noindent It has three classes of terms: $F_A$ involves only the order parameter $\bm{\Psi}_{I}$ of Phase I (Eq.~\ref{eq:psiI}), which transforms as the $m\Delta_6$ Irrep; $F_L$ depends only on the \Neel~component $\bm{\Psi}_{L}$ of the order parameter of Phase II, which transforms as the $m\Gamma_5^+$ Irrep; and $F_{AL}$ accounts for the coupling between them. The most general expressions for these three terms are given in the Appendix \ref{secSM:freeE}. For the discussion in this Section, it suffices to know that only combinations of the components of $\bm{\Psi}_{I}$ and $\bm{\Psi}_{L}$ that are invariant under all symmetries of the parent paramagnetic group, including time-reversal symmetry, are allowed. In particular, $F_{A}$ has a quadratic term,

\begin{equation}\label{eq:quadraticF_A}
F_A\left(\bm{\Psi}_{I}\right) = \alpha_I\left(T-T_{N1}\right)\left|\bm{\Psi}_{I}\right|^2+(...)\end{equation}
whose coefficient changes sign to induce the transition into Phase I at $T_{N1}$. The remaining parameters of $F_{A}$ are temperature independent, and chosen to favor the experimentally-observed nodal amplitude-modulated state. 

Similarly, $F_{L}$ has a quadratic term 
\begin{equation}\label{eq:quadraticF_L}
F_L\left(\bm{\Psi}_{L}\right) =\alpha_L\left|\bm{\Psi}_{L}\right|^2+(...),\end{equation}
and the second transition, at $T_{N2}$, could in principle be induced by a change of its sign. However, obtaining two transitions so close in temperature would require fine-tuning of model parameters. Instead, $\bm{\Psi}_{L}$ can be induced by the coupling term $F_{AL}$. The form of $F_{AL}$ determines whether these two transitions take place simultaneously or at different temperatures. Because $\bm{\Psi}_{I}$ and $\bm{\Psi}_{L}$ transform as different Irreps, there can be no bilinear coupling between them. This, combined with the fact that the free energy is time-reversal-even, implies that all couplings $\bm{\Psi}_{I}$ and $\bm{\Psi}_{L}$ happen at quartic order. As we show in Appendix \ref{secSM:freeE}, there are two types of quartic coupling: a linear-cubic coupling between $\bm{\Psi}_{L}$ and $\bm{\Psi}_{I}$ and a biquadratic coupling. The former implies that the onset of $\bm{\Psi}_{I}$  necessarily triggers $\bm{\Psi}_{L}$, i.e. that the two transitions are simultaneous. Since two transition temperatures are observed in \eia, linear-cubic coupling has to vanish in Phase I. Indeed, as we show in Appendix \ref{secSM:freeE}, the linear-cubic term generically vanishes if the condition $a_1 = 0$ is imposed. Therefore, the observation of two transitions, combined with the analysis of the Landau free energy, provides independent evidence that $A_2$ is the state realized in Phase I. 

The subsequent transition to Phase II can arise from the biquadratic coupling between $\bm{\Psi}_{I}$ and $\bm{\Psi}_{L}$, which is allowed even if $a_1 = 0$. As the temperature is lowered and $\left|\bm{\Psi}_I\right|^2$ increases, the biquadratic coupling renormalizes the quadratic term of the N\'eel order parameter according to:
\begin{equation}\label{eq:EffQuad}
F\left(\bm{\Psi}_{I},\bm{\Psi}_{L}\right)=(\alpha_L-\beta \left|\bm{\Psi}_I\right|^2)\left|\bm{\Psi}_L\right|^2 + (...),
\end{equation}such that it eventually changes sign if $\beta>0$, triggering the second phase transition at $T_{N2}$ (see Appendix, Sec.~\ref{sec:SM_FAL} for a detailed discussion).  

We have therefore shown that a symmetry-based Landau approach can reproduce the observed sequence of phase transitions.  However, this approach does not uniquely predict the evolution of the magnetic structure, as it does not capture the tendency towards equal moment magnitude.  To develop a description consistent with this tendency, we introduce constraints on $F$ beyond those that follow from symmetry alone.   In Appendix \ref{secSM:equal} we demonstrate that fixing the ratios between some of the coefficients of $F_{A}$ and $F_{AL}$ results in a term whose minimization gives precisely the equal-moment condition. We introduce a temperature dependence of this term to model the increase in stiffness with respect to variation in moment amplitude with reduced temperature, and show that it is sufficient to induce the transition into Phase II. 

\subsection{The symmetry-breaking pathway}\label{sec:resultsFE}

In the previous two sections we identified the order parameters in the two phases of \eia, derived the most general form of the free energy, and constrained its parameters based on experimental findings and physical arguments. We now analyze the magnetic structures obtained by minimization of the free energy as a function of temperature, and compute quantities that can be compared with experimental results.

\begin{figure}[t]
\centering
\includegraphics[width=1\linewidth]{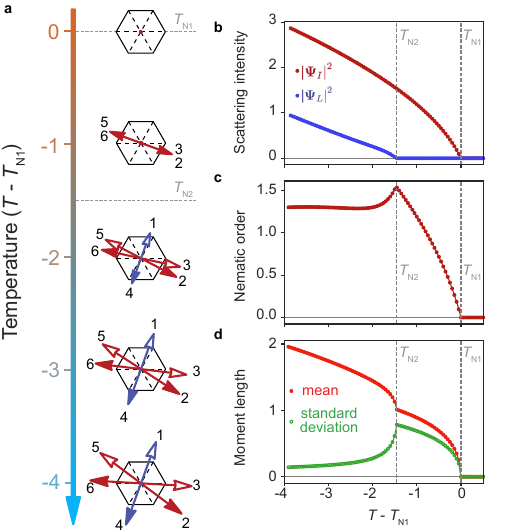} \label{fig:EIA_Fig9}
\caption{Results of the free energy minimization of the constrained free energy $\tilde{F}$ as a function of temperature: (a) Magnetic structure; (b) Scattering intensity at $\mathbf{Q}_{1}$ (red) and $\mathbf{Q}_{2}$ (blue), computed as $\left|\bm{\Psi}_I\right|^2$ and $\left|\bm{\Psi}_L\right|^2$, respectively; (c) The magnitude of the nematic order parameter, computed as $\left|\bm{\eta}+\bm{\xi}\right|$; (d) The mean (red) and standard deviation (green) of the magnetic moment length, extracted from the magnetic structures obtained by energy minimization. The values for the parameters of the Landau free-energy are given in Appendix \ref{secSM:freeE}. For illustrative purposes, the coefficients of the magneto-crystalline anisotropy terms in the Landau functional were set to zero (see Appendix, Sec.\ref{secSM:freeE}), so the magnetic moments in the resulting structure do not point along high-symmetry directions, consistent with the experiment. Of course, in the experiments, the magnetic moments are ultimately pinned by the local strain. All the axes are in arbitrary units.} 
\end{figure}

In Fig.~\ref{fig:EIA_Fig9}a we show the evolution of the magnetic structure with temperature. The transition between the paramagnet and Phase I at $T_{N1}$ is characterized by breaking of the rotational $C_{3z}$ symmetry. The magnetic structure in Phase I is identified as the nodal amplitude modulated state, in which the average magnetic moment vanishes in every third Eu layer ($A_2$, Fig.~\ref{fig:EIA_Fig7}e). This moment length variation is enabled by strong fluctuations in the vicinity of the transition, but it becomes increasingly unfavorable as the temperature is lowered. The non-linear coupling between the nodal amplitude modulated state and the N\'eel order ($\bm{\Psi}_{L}$; Fig.~\ref{fig:EIA_Fig8}a) triggers the second transition, and both the N\'eel order and the nodeless $A_1$ component  (Fig.~\ref{fig:EIA_Fig7}d) develop at $T_{N2}$. At this point, the moments are neither of the same magnitude nor point along the same direction, as can be verified in Fig.~\ref{fig:EIA_Fig9}a. As the temperature is lowered, the minimization of the constrained free energy ($\tilde{F}$, Eq.~\ref{eqSM:constF}), gives the results of Fig.~\ref{fig:EIA_Fig9}a, revealing indeed that the moments tend to acquire the same magnitude far away from the transition.

To compare the results of energy minimization with experimental findings, we compute $\left|\bm{\Psi}_I\right|^2$ and $\left|\bm{\Psi}_L\right|^2$, (Fig.~\ref{fig:EIA_Fig9}b), the nematic order parameter ($\left|\bm{\eta}+\bm{\xi}\right|$; Fig.~\ref{fig:EIA_Fig9}c), and the mean magnetic moment length, as well as its standard deviation (Fig.~\ref{fig:EIA_Fig9}d). The results of minimization of the constrained Landau free energy as a function of temperature capture the main features of the experimental findings remarkably well: (a) $\left|\bm{\Psi}_L\right|^2$ onsets at a slightly lower temperature than $\left|\bm{\Psi}_I\right|^2$, and shows a steeper temperature dependence, consistent with scattering experiments~\cite{riberolles_magnetic_2021, soh_understanding_2023}; (b) nematic order ($\left|\bm{\eta}+\bm{\xi}\right|$) onsets at $T_{N1}$, consistent with the observed birefringence reported in this work; (c) the expected value of the magnetic moment varies substantially close to the transition, but the variation is suppressed by the onset of Phase II, consistent with M{\"o}ssbauer spectroscopy~\cite{riberolles_magnetic_2021}. We conclude that our free energy model captures the essential aspects of all the optical and spectroscopic measurements on \eia~to date. 

\section{Strain-tuning the magnetic symmetry}\label{Sec:StrainTuning}

The sharp onset of birefringence at $T_{N1}$ proves its magnetic origin. On the other hand, the experimental observation of the broad distribution of the principal axis orientation (Fig.~\ref{fig:EIA_Fig4}) implies that the orientation of magnetic moments at each sample location is determined by built-in strain, which locally breaks the $C_{3z}$ symmetry (Fig.~\ref{fig:EIA_Fig4}). This in turn suggests that uniaxial strain can be used to control the principal axis orientation. As we demonstrate below, this is indeed possible, with implications for the tuning of magnetic symmetry. 

\begin{figure}[t]
 \centering
 \includegraphics[width=0.9\linewidth]{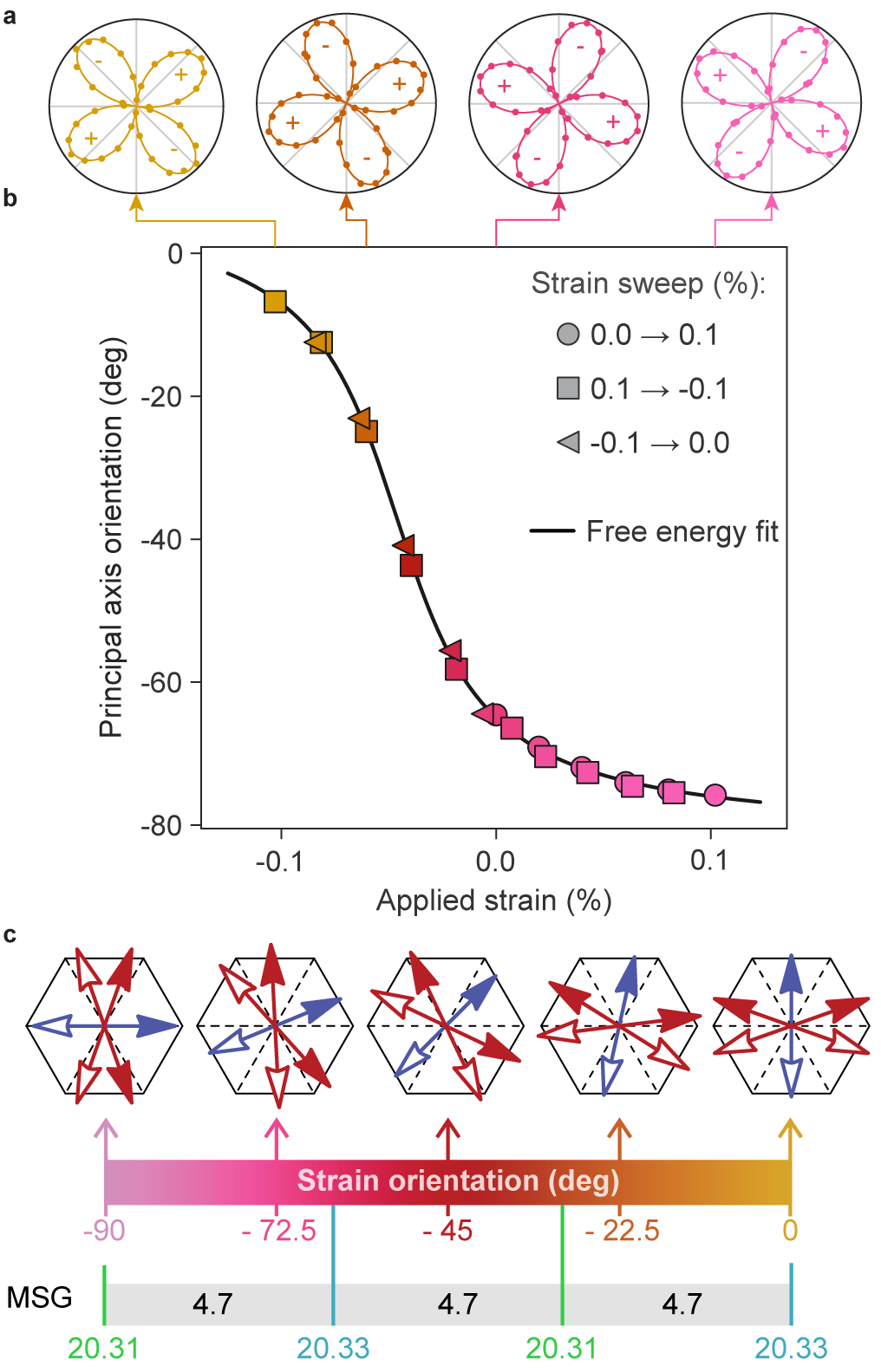}
 \caption{ (a) The polarization rotation as a function of incident polarization measured at four different values of applied strain. (b) Symbols: Principal axis orientation as a function of applied strain (positive values denote tensile strain). Agreement between the data taken while increasing and decreasing strain indicate that the tuning is reproducible, with no plastic deformations. The line is calculated by minimizing the free energy (see text for more detail; parameters are the same as in Fig.~\ref{fig:EIA_Fig9}, and $T-T_{N1}=-3.5$, $\epsilon=0.1$), assuming a built-in strain, whose magnitude and orientation ($0.056\%$, oriented at -65\degree) were extracted from the fit. The applied strain is oriented at -82\degree, indicating a slight misalignment ( 8\degree) between the applied strain axis and the principal optical axes of the laboratory. (c) The magnetic structures found by free energy minimization for different orientations of \textit{total} strain. The rotation of the magnetic structure has stark consequences for symmetry, with the vast majority of orientations belonging to MSG 4.7; only when $L$, denoted by the blue arrows, is oriented along the Eu-Eu bond direction, or perpendicular to it, are different MSGs found: 20.31 and 20.33, respectively. }
 \label{fig:EIA_Fig10}
\end{figure}

To test the hypothesis of axis tunability, we applied compressive and tensile uniaxial strain to a bar-shaped sample using a commercial strain device (see Appendix~\ref{Sec:SI_Strain} for experimental details). It is immediately clear from the $T$-mod birefringence measured at several strain values that the principal axes orientation can indeed be rotated by strain (Fig.~\ref{fig:EIA_Fig10}a). Systematic strain-dependent measurements (Fig.~\ref{fig:EIA_Fig10}b) reveal a smooth strain dependence of the orientation, spanning the range of $\sim 0-90\degree$, corresponding to the exchange of fast and slow optical axes.

To show that this is a consequence of weak magneto-crystalline anisotropy, we include strain $\bm{e}$ in our free energy model through coupling to the nematic order parameters $\bm{\eta}$ and $\bm{\xi}$ (Eqs.~\ref{eq:nematic_psi} and \ref{eq:nematic_xi}):

\begin{equation}
 F_{\epsilon} = - \epsilon \bm{e} \cdot \left(\bm{\eta}+\bm{\xi}\right),
\end{equation}
where $ \epsilon$ is the coupling constant, $\bm{e}$ the total strain assumed to be the same for both $\bm{\eta}$ and $\bm{\xi}$, since they both originate from the Eu$^{2+}$ moments (Appendix \ref{secSM:nematic}). The total strain is a sum of built-in strain $\bm{e_0}$ and applied strain $\bm{e_a}$. To obtain the curve in Fig.~\ref{fig:EIA_Fig10}b we vary the applied strain $\bm{e_a}$, and minimize the total free energy (other parameters are the same as in Fig.~\ref{fig:EIA_Fig9}, and $T-T_{N1}=-3.5$, $\epsilon=0.1$).We find excellent agreement with the measured strain-dependence of principal axis orientation for $\left|\bm{e_0}\right|=0.056\%$ (Fig.~\ref{fig:EIA_Fig10}b). The principal axis rotates smoothly, in agreement with the expectations of a 3-state Potts order parameter coupled to a conjugate field in the limit of weak magneto-crystalline anisotropy \cite{chakraborty_strain-tuned_2023}. 

To explore the consequences of this tunability, we calculate the magnetic structure that minimizes the free energy as a function of the orientation of total strain, while keeping its magnitude fixed (Fig.~\ref{fig:EIA_Fig10}c). As expected, we find that the relative orientation of the magnetic structure and the lattice can be freely tuned. Furthermore, we analyze the symmetry of all of the structures, and we find that for most orientations the magnetic point group is P2$_1$ (No. 4.7), which contains $C_{2z}$ combined with a translation by one Eu layer and identity as the only symmetries. However, when the \Neel~vector is aligned to one of the high symmetry directions of the lattice, additional symmetries are restored. In particular, when the \Neel~vector is perpendicular to the Eu-Eu bond direction the magnetic space group identified for Phase II in Ref.~\cite{riberolles_magnetic_2021} is restored, protecting the axion insulator phase. We therefore find that it is possible to use uniaxial strain to switch on and off the axion insulator state in \eia.

\section{Spin Hamiltonian}\label{sec:SpinHam}

So far we have shown that our measurements and symmetry analysis lead to a unique identification of the two magnetic phases, as well as describing the two phase transitions. A question that remains is the microscopic origin of this behavior, which we address by developing an effective spin Hamiltonian.

As we show in the following sections, there are two prerequisites for forming the broken helix order: (i) a long-range Heisenberg interaction $\mathcal{J}_{ij}$ and (ii) fourth-order exchange, $\mathcal{J}_{ijkl}$. Our proposed spin Hamiltonian,
\begin{equation}\label{eq:hamilt}
 H=\sum_{ij} \mathcal{J}_{ij}\bm{S}_i\cdot \bm{S}_j + \sum_{ijkl} \mathcal{J}_{ijkl}\left(\bm{S}_i\cdot \bm{S}_j\right) \left(\bm{S}_k\cdot \bm{S}_l\right),
\end{equation}
includes these terms, and neglects terms that depend on the spin orientation with respect to the lattice. To capture the effects of the easy-plane anisotropy, and the fact moment length is the same for every Eu$^{2+}$ ion in the ground state, we parameterize each spin by an angle, $\bm{S}_i=(\cos{\theta_i}, \sin{\theta_i})$. 

Our two theoretical approaches are complementary: the Landau free energy expansion captures the sequence of broken symmetries, and is valid in the vicinity of the transitions. One the other hand, the spin Hamiltonian describes the microscopic origin of the magnetic ground state, but provides no information about the pathway between the paramagnetic state and the ground state. In Sec.~\ref{sec:GS} we will complement the spin Hamiltonian with atomistic magnetic simulations at non-zero temperatures, hence connecting the two approaches. 

\subsection{Heisenberg term}\label{sec:Heis}
The Heisenberg term in Eq.~\ref{eq:hamilt} can induce modulated magnetic structures if coupling beyond nearest neighbors are included, with the propagation vector determined by the Fourier transform, $\mathcal{J}_{Q}$, of the spatially dependent exchange interaction, $\mathcal{J}_{ij}$. The two periodicities in the ground state of \eia~imply that $\mathcal{J}_{Q}$ is peaked at both $\bm{Q}_{1}$ and $\bm{Q}_{2}$, while the sequence of transitions indicates that $\bm{Q}_{1}$ is energetically more favorable. We capture this physical intuition with a Ruderman–Kittel–Kasuya–Yosida (RKKY)-inspired expression for the effective exchange coupling (for more details see the Appedix~\ref{sec:S_Heis}):

\begin{equation}\label{eq:interactionBoth}
 \mathcal{J}_{ij} = J_1 \frac{\cos{\left( k_1 \left(i-j\right) \right)}}{k_1 \left(i-j\right)}+ J_2 \frac{\cos{\left( k_2 \left(i-j\right) \right)}}{k_2 \left(i-j\right)}.
\end{equation}
 
\begin{figure}[t]
 \centering
 \includegraphics[width=1\linewidth]{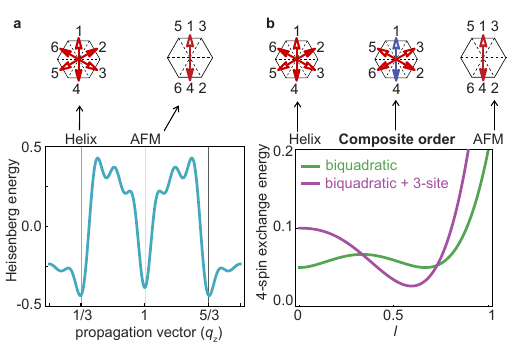}
 \caption{(a) Main plot: Heisenberg energy of a single $\bm{Q}=(0,0,q_z)$ magnetic structure, as a function of $q_z$. The distance-dependence of the Heisenberg interaction is given by Eq.~\ref{eq:interactionBoth}, with parameters ($k_1=0.4 \pi$, $k_2=\pi$, $J_1=-0.1$, $J_2=0.125$) chosen to reproduce the minima at $q_z=1/3$ and $q_z=1$, and favor the $q_z=1/3$ state.  We consider interactions up to tenth neighbor (the numerical values of the exchange parameters are given in Table \ref{tab:Heisenberg} of the Appendix~\ref{sec:SI_atom}). Top panel: magnetic structures corresponding to the helix and antiferromagnetic states. (b) Main panel: The energy of two forms of the fourth-order exchange: biquadratic ($\mathcal{J}_{1212}=0.2$, green curve) and a combination of the biquadratic and a 4-spin-3-site exchange ($\mathcal{J}_{1212}=\mathcal{J}_{1223}=0.2$, purple curve), as a function of the magnetic structure interpolated between the helical ($l=0$) and the antiferromagnetic ($l = 1$) state, while maintaining the spin normalization condition (Eq.~\ref{eq:Params}), and setting the length of each spin to 1. Biquadratic exchange exhibits degenerate minima at the helical ($l=0$) and mixed ($0<l<1$) states, and therefore is not sufficient to stabilize the mixed state. However, biquadratic exchange combined with the 4-spin-3-site exchange exhibits one minimum in the mixed state, and can stabilize it. The top row shows the magnetic structures corresponding to $l=0$ (helix), $l=0.5$ (mixed order) and $l=1$ (AFM).}
 \label{fig:EIA_Fig11}
\end{figure}

\noindent In Fig.~\ref{fig:EIA_Fig11}a we show the energy of a single $\bm{Q}=(0,0,q_z)$ state as a function of $q_z$, calculated from Eq.~\ref{eq:interactionBoth} assuming a ten-neighbor Heisenberg Hamiltonian. The parameters, $k_{1,2}$ and $J_{1,2}$ (given in the caption) are chosen so that the energy has two local minima, at $q_z=1/3$ and $q_z=1$, and the minimum at $q_z=1/3$ is slightly lower. 

An energy landscape with two well-defined minima as shown in Fig.~\ref{fig:EIA_Fig11}a requires narrow energy wells and consequently long-ranged real-space interactions. Two possible mechanisms for the long-range coupling are the RKKY interaction mediated by conduction electrons, and the dipolar interaction, which has been suggested to induce A-type antiferromagnetism in several Eu- compounds~\cite{berry_-type_2022}. A possible scenario is that the dipolar interaction promotes the AFM order, while the RKKY interaction promotes $\bm{Q}_{1}$. Regardless, the exchange parameters used here (Fig.~\ref{fig:EIA_Fig11}a) capture the tendency towards ordering at two $\bm{Q}$-values, and RKKY interaction is required for at least $\bm{Q}_1$. 

\subsection{Fourth-order terms}\label{sec:fourth}

Crucially, no parameter choice within a purely Heisenberg model will promote the coexistence of two periodicities - the system simply chooses the lower of the two energy minima in Fig.~\ref{fig:EIA_Fig11}a, i.e and antiferromagnet or a helix. To account for the coexistence of the two periodicities, it is necessary to move beyond the Heisenberg model. It has been shown in several itinerant systems that various forms of the four-spin exchange ($\mathcal{J}_{ijkl}$; Eq.~\ref{eq:hamilt}) can promote multi-$\bm{Q}$ states composed of symmetry-equivalent single-$\bm{Q}$ states~\cite{hayami_noncoplanar_2021,hayami_topological_2021, hoffmann_systematic_2020, takagi_multiple_2018}. As we show below, fourth order terms can stabilize the coexistence of states even if they belong to different Irreps, as is the case with the $\bm{Q}_{1}$ and $\bm{Q}_{2}$ orders in \eia.  We note that first principles calculations~\cite{mendive-tapia_ab_2019}, effective Hubbard models~\cite{hoffmann_systematic_2020} and perturbative expansions of Kondo models~\cite{hayami_effective_2017} have shown that coupling described by $\mathcal{J}_{ijkl}$ naturally arise in itinerant systems, and can be crucial for promoting exotic magnetic order. 

First, we find that the biquadratic term, $(\bm{S}_{i} \cdot \bm{S}_{i+1})^2$, is not sufficient to promote the mixed order. However, when combined with the symmetry-allowed three-site-four-spin term it leads to an an interaction which can stabilize the mixed ground state:

\begin{equation}\label{eq:fourthOrder}
 H_4=\mathcal{J}_{1212}(\bm{S}_{i} \cdot \bm{S}_{i+1})^2+\mathcal{J}_{1223}(\bm{S}_{i} \cdot \bm{S}_{i+1})(\bm{S}_{i+1} \cdot \bm{S}_{i+2}).
\end{equation}

We demonstrate this by showing that the energy of the fourth-order term is lower in the mixed state than in either of the single-$\bm{Q}$ states characterized by equal moments (helix and AFM). To that end we construct a continuum of magnetic structures with varying ratios between the amplitudes of the two order parameters. In these structures equal moment length in each layer ($M$) is maintained by applying the constraint expressed in Eq.~\ref{eq:condition}, leading to: 
\begin{equation}\label{eq:Params}
    a_1=\frac{M-l}{\sqrt{2}},\hspace{10pt} a_2=\frac{\sqrt{-3 l^2+2 lM+M^2}}{\sqrt{2}},
\end{equation}
with the orientations $\phi_1 = \phi_2 \pm \pi/2$ and  $\theta = \phi_2 \mp \pi/2$. For $l=0$ we find $a_1=a_2=M/\sqrt{2}$, corresponding to a helix, while for $l = M$ the structure is antiferromagnetic ($a_1 = a_2 = 0$). For each value of $0<l<M$ we calculate $a_1$ and $a_2$, determine the corresponding layer-dependent spin orientation, and compute the energy of the different Hamiltonian terms. 

 In Fig.~\ref{fig:EIA_Fig11}b we plot the energy of the fourth order term (Eq.~\ref{eq:fourthOrder}, $\mathcal{J}_{1212}=\mathcal{J}_{1223}=0.2$, purple curve) as a function of $l$, for $M=1$. The fourth-order exchange clearly exhibits a minimum for $0<l<1$, and therefore favors the mixed state over either of the single-$\bm{Q}$ states. In contrast, the biquadratic term alone exhibits two degenerate minima, at $l=0$ and $0<l<1$, and therefore is not sufficient to stabilize the mixed state ($\mathcal{J}_{1212}=0.2, \mathcal{J}_{1223}=0$, green curve).

\subsection{Evolution of the magnetic structure}\label{sec:GS}

\begin{figure}[t]
 \centering
 \includegraphics[width=1\linewidth]{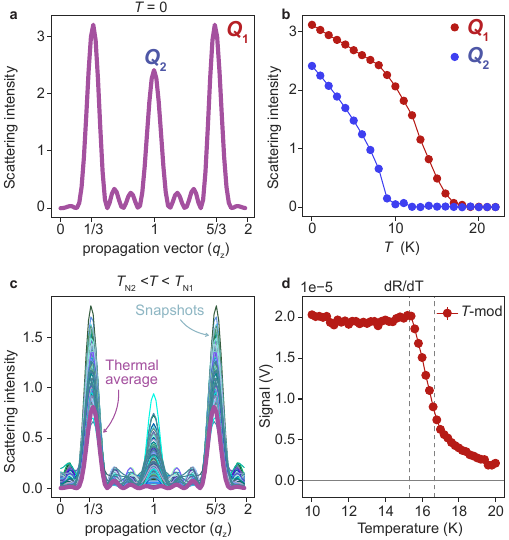}
 \caption{(a-b) Results of atomistic LLG simulations~\cite{muller_spirit_2019}, based on Hamiltonian shown in Eq.~\ref{eq:hamilt}, with the same parameters used in Fig.~\ref{fig:EIA_Fig11} (details are given in Sec.~\ref{sec:SI_atom} of the SM). (a) The Fourier transform of the spin structure found at $T = 0$, revealing two peaks, at $\bm{Q}_1$ and $\bm{Q}_2$. (b) The temperature dependence of scattering intensity at $\bm{Q}_1$ and $\bm{Q}_2$. revealing two second-order transitions, consistent with the experiment. (c) The Fourier transform of snapshots of the magnetic structure in Phase I, as well as a the Fourier transform of the structure found by averaging $10^5$ such snapshots. This demonstrates antiferromagnetic fluctuations in Phase I, despite vanishing static AFM order. (d) The temperature dependence of the temperature derivative of reflectivity measured at $\unit[633]{nm}$, showing a sharp change of electronic structure at $T_{N2}$. This indicates that the AFM fluctuations shown in panel d are accompanied by fluctuations of the electronic structure. }
 \label{fig:EIA_Fig12}
\end{figure}

The above analysis suggests that combining the long-range Heisenberg exchange with the 4-spin terms can induce the mixed structure. To verify this, we found the magnetic ground state by solving the Landau-Lifshitz-Gilbert (LLG) equation, as implemented in the Spirit framework for atomistic spin simulations~\cite{muller_spirit_2019}. The broken helix is indeed the ground state, and its Fourier transform (Fig.~\ref{fig:EIA_Fig12}a) shows good agreement with the scattering data~\cite{riberolles_magnetic_2021, soh_understanding_2023}. We have therefore identified a Hamiltonian whose ground state is the observed mixed order. 

We extend the calculation to non-zero temperature, implemented within the Spirit framework by adding a stochastic thermal field, whose distribution is determined by the fluctuation-dissipation theorem~\cite{muller_spirit_2019}. In Fig.~\ref{fig:EIA_Fig12}b we show the amplitudes of the two scattering peaks as a function of temperature, and find two second-order transitions, with $\bm{Q}_2$ vanishing at a lower temperature upon warming, in agreement with the experiments. However, the simulations reveal a helical, rather than amplitude-modulated (AM) state in Phase I, in contrast to our experimental findings. It is perhaps not surprising that the effective spin Hamiltonian cannot capture the AM phase, since neither of its terms is likely to favor it: the Heisenberg energies of the helix and AM phases are degenerate, and the four-spin exchange terms are negligible just below the second order transition at $T_{N1}$, when the AM phase is stabilized. Magneto-crystalline anisotropy, which has been suggested to promote the AM state in a few gadolinium compounds~\cite{blanco_specific_1991, bouvier_specific_1991}, is not included in the calculation because our data suggest it is not experimentally relevant. 

Although the free energy originating from the spin Hamiltonian alone does not capture the nodal amplitude modulated phase, the strong coupling of magnetic and electronic degrees of freedom can account for additional contributions to entropy. To illustrate why this is the case, we first look at individual snapshots of the magnetic structure in Phase I, obtained by the atomistic simulations, which reveal peaks at both $\bm{Q}_1$ and $\bm{Q}_2$. In contrast, the average over an ensemble of $10^4$ such snapshots, shows only the helical $\bm{Q}_1$ peaks, as is characteristic of Phase I. Therefore, the transition into Phase II is anticipated by strong AFM fluctuations. Since magnetic fluctuations are slow on electronic time scales, AFM fluctuations are accompanied by electronic ones.

Crucially, due to the strong coupling of itinerant and localized degrees of freedom in \eia, the fluctuations of the electronic structure that accompany AFM fluctuations are significant. Infrared studies~\cite{xu_infrared_2021} have shown that electronic structure dramatically reconstructs with the onset of the AFM order in Phase II. This is consistent with our own measurement of the temperature derivative of reflectivity on the same sample used for symmetry-sensitive measurements (Fig.~\ref{fig:EIA_Fig12}d), which exhibits a sharp kink at $T_{N2}$. Entropic effects in \eia~therefore cannot be correctly captured by localized spin models alone. Our findings suggest that the entropy associated with the joined system of localized spins and itinerant electrons stabilizes the nodal amplitude-modulated state, providing a strong motivation for explicit calculations considering the magnetic and electronic degrees of freedom in \eia~on equal footing. It also yields constraints for such models: a correct calculation needs to reproduce both the nodal amplitude modulated state in Phase I, and the broken helix ground state.

The key message of our spin model is that the itinerant electrons are crucial for stabilizing the magnetism of \eia: the long-range Heisenberg interaction and the fourth order term coupling three Eu layers are the simplest terms that can stabilize the magnetic ground state, but neither can arise just from direct exchange of the localized $4f$ orbitals. Instead, they are likely to be mediated by conduction electrons, with possible contributions from coupling to the lattice, as proposed in Ref.~\cite{soh_understanding_2023}. The understanding of the origin of the magnetic interactions offers a natural way to resolve a slight difference in $\bm{Q}_1$ reported in the two scattering experiments: only Soh et al.~\cite{soh_understanding_2023} found $\bm{Q}_1=(0,0,1/3)$ within experimental resolution, while Riberolles et al.~\cite{riberolles_magnetic_2021} found $\bm{Q}_1 \sim (0,0,0.303)$. Since the magnetism is mediated by conduction electrons, such differences can naturally be explained by variations in stoichiometry. In addition to explaining the existing data, this suggests that electron density could be a powerful tuning knob for magnetism in \eia.

\section{Discussion and Conclusions}

To summarize, we demonstrated a multimodal approach that combines scattering and symmetry sensitive optical measurements with group theory analysis to uniquely determine the magnetic structures in two phases of \eia. In addition to answering the open questions regarding the magnetism of \eia, we set the scene for future research in materials with strongly intertwined magnetic and electronic degrees of freedom. We have further shown that, although the axion state is not reached simply by cooling a sample through the two transitions, it can be retrieved by application of uniaxial strain. Finally, we have identified the coupling of local magnetic and itinerant electronic degrees of freedom as the origin of the rich magnetic behavior observed in \eia. 

The multimodal approach has proven invaluable to determine the nature of the broken symmetry phases, offering information beyond that available to any of the techniques alone. While scattering is unique in its ability to determine ordering wavevectors, spatially-resolved probes are crucial to distinguish systems that possess a given symmetry from those in which apparent symmetry is restored by averaging over domains. This has proven critical for recognizing that the magnetic structure in Phase I is an amplitude-modulated state, rather than a helix. More broadly, the multimodal approach to identifying broken symmetry phases can be employed regardless of the nature of the order: here it was demonstrated on a material with complex magnetism, but it is also applicable to systems exhibiting charge density or orbital ordering. 

Each of the magnetic phases that we have identified has unique aspects. Amplitude-modulated structures in systems of \textit{localized} moments are rare. They have been deduced on the basis of thermodynamic measurements in several  Gd$^{3+}$ compounds \cite{blanco_incommensurate_1990,blanco_specific_1991, bouvier_specific_1991, blanco_thermodynamical_1991}, however the phase of the amplitude modulation was not determined. Magneto-crystalline anisotropy, thought to be the driver of amplitude modulation in the Gd compounds, is experimentally irrelevant in \eia, raising the intriguing possibility that its amplitude modulated state is stabilized by entropic effects. 

A further consequence of the decoupling of magnetic states from the \eia~lattice is that the magnetic symmetry can be controlled by uniaxial strain. Thus \eia~ can be tuned into and out of a topologically non-trivial axion state. This capability allows the exploration of approximate symmetries in topological systems, addressing the following question: if a symmetry protects a topological state for a particular orientation of the magnetic structure and the lattice, how do the response functions change when the orientation is infinitesimally changed? In other words, how robust are topological responses to continuous tuning parameters? Our work identifies Eu$^{2+}$ - based compounds as ideal platforms for such studies. 

Our understanding of itinerant electrons as mediators of magnetic interactions radically changes the picture of \eia: while the initial interest in this material was spurred by the notion that a simple magnetic structure will generate exotic electronic features \cite{xu_higher-order_2019}, we have shown that electronic states instead generate an exotic magnetic ground state. Of course, this does not mean that the electronic states themselves do not possess interesting properties. In addition to the topological phase restored for specific relative orientations of the magnetic structure and the lattice, AFM order was shown to induce altermagnetic momentum-dependent spin polarization of the electronic bands in \eia~\cite{cuono_ab_2023}; the influence of such band-splitting on electron-mediated magnetic interactions is an important open question.  Our work therefore challenges \textit{ab-initio} calculations to study the interplay of magnetic and itinerant degrees of freedom in \eia~in a realistic way, in order to capture the experimental findings on both subsystems.

\begin{acknowledgments}

This research was primarily funded by the Quantum Materials program under the Director, Office of Science, Office of Basic Energy Sciences, Materials Sciences and Engineering Division, of the U.S. Department of Energy, Contract No. DE-AC02-05CH11231. E. D. received additional support from the National Science Foundation Graduate Research Fellowship Program under Grant No. DGE-2146752 and the Ford Foundation Predoctoral Fellowship. R.D. is supported by the Canadian Government under a Banting Fellowship. R.M.F. (phenomenological modeling) was supported by the U.S. Department of Energy, Office of Science, Basic Energy Sciences, under award no. DE-SC0020045. D. P. and A.T.B. would like to acknowledge the Engineering and Physical Sciences Research Council (EPSRC), UK and the Oxford-ShanghaiTech collaboration project for financial support. J.O received support from the Gordon and Betty Moore Foundation's EPiQS Initiative through Grant GBMF4537 to J.O. at UC Berkeley. V.S. is supported by the Miller Institute for Basic Research in Science, UC Berkeley.

 \end{acknowledgments}

\bibstyle{apsrev4-1}

\bibliography{EIA_main}

\appendix

\newpage

\counterwithout{equation}{section}

\renewcommand\theequation{S\arabic{equation}}
\renewcommand\thefigure{S\arabic{figure}}
\renewcommand\thetable{S\arabic{table}}
\renewcommand\thesection{S\arabic{section}}
\renewcommand\bibnumfmt[1]{[S#1]}
\setcounter{equation}{0}
\setcounter{figure}{0}
\setcounter{enumiv}{0}

\newpage

\section*{Supplementary Material for `Symmetry-breaking pathways to the broken helix state'}

\section{Crystal growth and characterization}

The optical data in Figs.~\ref{fig:EIA_Fig2}-~\ref{fig:EIA_Fig5} were taken on the same crystal used for the resonant x-ray scattering in Ref.~\cite{soh_understanding_2023}, synthesized at the University of Oxford by a method described in Ref.~\cite{soh_understanding_2023}. Since scattering and optical experiments were performed on physically the same crystal, there are no ambiguities about sample-sample variation that could complicate our multimodal approach. Samples made at UC Berkeley gave qualitatively the same optical response, with $T_{N1}$ and $T_{N2}$ revealed by the $T$-mod and $H$-mod measurements, respectively. The strain experiment (Fig.~\ref{fig:EIA_Fig10}) was performed on one of those samples, with the corresponding crystal growth method described below. 

Single crystals of \eia~were prepared by a flux method similar to \cite{riberolles_magnetic_2021}. Europium (99.9999$\%$ from Ames Lab) was cut up and combined in an alumina crucible under a dry nitrogen glove box atmosphere with arsenic pieces (99.99$\%$ from Alfa Aesar) and indium shot (99.9999$\%$ from Alfa Aesar) in a 1:3:12 molar ratio with total mass 4g. The crucible was sealed in an evacuated quartz ampule without being exposed to air.
The ampule was heated in a box furnace from room temperature to $\unit[300]{\degree C}$ over 2 hours, held for 1 hour, then heated to $\unit[580]{\degree C}$ over 3 hours, held for 2 hours, then ramped to $\unit[900]{\degree C}$ over 20 hours before soaking at $\unit[900]{\degree C}$ for 4 hours. Finally, the ampule was slowly cooled to $\unit[770]{\degree C}$ at $\unit[1.8]{\degree C}$/hour, and spun in a centrifuge to remove excess indium flux. The method is similar to the one used in Ref.~\cite{soh_understanding_2023}, with a few differences: the highest temperature in that work was $\unit[950]{\degree C}$, the centrifuging temperature at $\unit[750]{\degree C}$, and the Eu purity 99.99$\%$. Both methods used heat soaks before reaching the peak temperature to avoid any high vapor pressure of arsenic and increase the homogeneity of the flux solution.

\section{Optical experiments}\label{sec:Optical}

\begin{figure}[h]
 \centering
 \includegraphics[width=1\linewidth]{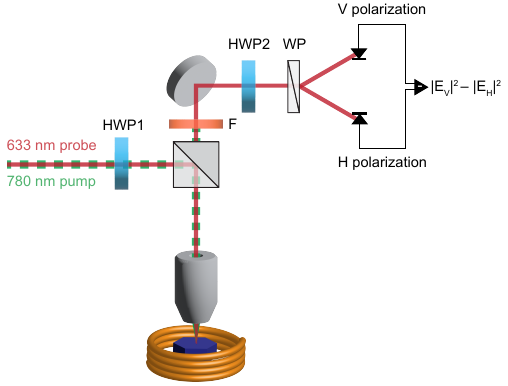}
 \caption{Schematic of optical setup for modulated birefringence measurements.}
 \label{fig:SM_setup}
\end{figure}

A schematic of the optical setup is shown in Fig. \ref{fig:SM_setup}. The incident polarization angle, $\varphi$, of the probe laser (633 nm) is set by a polarizer followed by a half-wave plate (HWP1). The measured polarization rotation from the sample is independent of the pump (780 nm) polarization. After reflecting off the sample, the pump beam is rejected by a color filter (F), and the probe beam polarization is further rotated by an angle $\varphi + 45\degree$ by a second half-wave plate (HWP2). In the case where the polarization state is not altered by the setup or the sample, the polarization of the reflected probe beam upon exiting the second half-wave plate is an equal superposition of vertical (V) and horizontal (H) linearly polarized light. The beam is then sent through a Wollaston prism (WP), which spatially separates vertical and horizontal components of the light, and the two orthogonal components are each focused onto separate, unbiased photodiodes of an balanced optical bridge detector. When the V and H components have equal intensity, which occurs when the polarization state of the light is unchanged by the sample, the net photocurrent is zero; in contrast, any measured signal indicates a change of polarization, such that the technique has a high sensitivity to detecting such changes.

While changes in the final polarization state can also be introduced by birefringence or ellipticity of the setup---resulting in artifacts---we largely mitigate this issue by performing thermally-modulated (\textit{T}--mod) and field-modulated (\textit{H}--mod) experiments, as these setup effects are independent of both temperature and field. For thermal modulation, a second (pump) laser is focused onto the same spot on the sample as the probe beam and optically chopped at kHz frequencies to locally modulate the temperature of the sample. For field modulation, the sample is placed in a coil that is driven with an alternating current to create an oscillating magnetic field. The experiment is sensitive only to effects which are proportional to the modulation parameter, which the artifacts arising from the setup are not. Despite this, small cross-coupling terms can still occur if there is more than one optical constant proportional to the modulation parameter, as discussed in detail in the Supplementary Material of Ref.~\cite{sunko_spin-carrier_2023}.

\begin{figure*}[t!]
 \centering
 \includegraphics[width=1\linewidth]{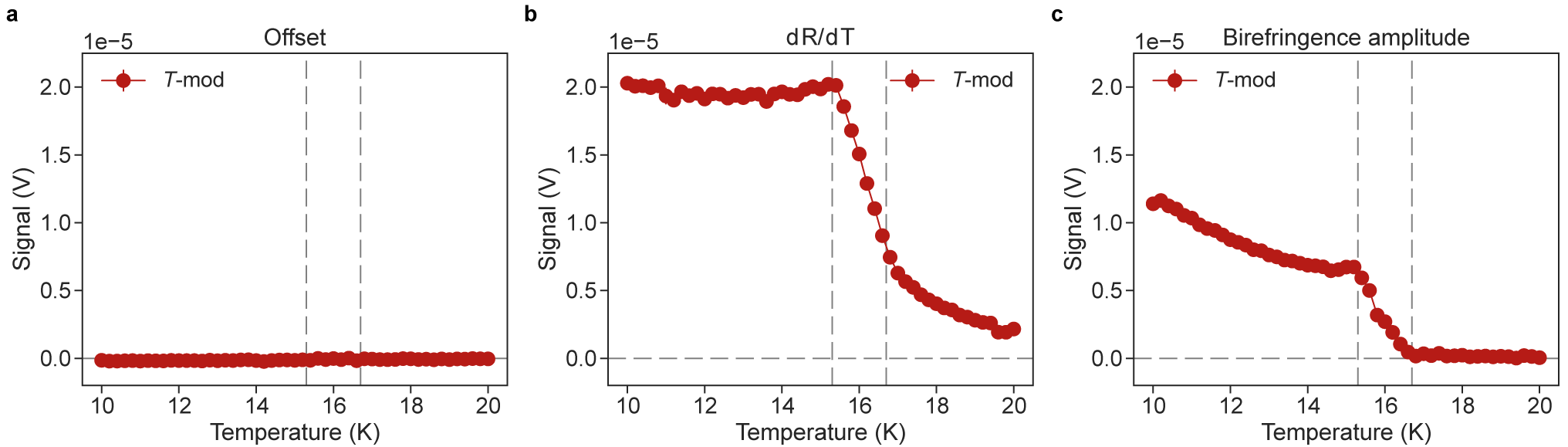}
 \caption{Thermally modulated (a) offset, (b) $d\mathrm{R}/d\mathrm{T}$, and (c) birefringence amplitude.}
 \label{fig:off_drdt_amp}
\end{figure*}

In Phase I, the sample becomes birefringent, rotating the polarization of the probe beam, which unbalances the contributions of the orthogonally polarized light components entering the balanced photodiode and causes a signal to be measured. The intensity of the light admitted to each of the photodiodes can be calculated using the Jones calculus formalism, in which the polarization state of the light is represented by a vector in the (V, H) basis and each optical component is represented by a 2x2 matrix. The Jones matrix for a half-wave plate with its fast axis rotated by $\theta$ with respect to the horizontal axis, thereby rotating the polarization by $\varphi=2\theta$, is described by:

\begin{equation}\label{eq:HWP}
    J_{hwp}\left(\theta\right)=\left(\begin{array}{cc}
\cos(2\theta) & \sin(2\theta)\\
\sin(2\theta) & -\cos(2\theta)
\end{array}\right).
\end{equation}

The Jones matrix representing the sample is:

\begin{equation}
        J_{sam}\left(r,b,k, \varphi_0\right)=R\left(-\varphi_{0}\right)\left(
\begin{array}{cc}
 r+b & k \\
 -k & r-b \\
\end{array}
\right)R\left(\varphi_{0}\right),
\end{equation}
where $r$ represents the sample reflectivity, $b$ the birefringence (difference in reflectivity between V and H polarizations), and $k$ for the polar Kerr effect (difference in reflectivity between the left- and right-circularly polarized light caused by an out-of-plane magnetization). $R\left(\varphi_{0}\right)$ is a rotation matrix, which encodes the different orientations of the sample and its domains with respect to the lab coordinate system: 

\begin{equation}
    R\left(\varphi_0\right)=\left(
\begin{array}{cc}
 \cos (\varphi_0 ) & -\sin (\varphi_0 ) \\
 \sin (\varphi_0 ) & \cos (\varphi_0 ) \\
\end{array}
\right)
\end{equation}

Modeling the experiment in this way, the final polarization state of the reflected light is given by,

\begin{widetext}
\begin{equation}
    \begin{pmatrix}
        E_V \\ E_H
    \end{pmatrix} = J_{\mathrm{HWP2}}(\varphi/2+22.5\degree)\cdot R(-\varphi_0)\cdot J_{\mathrm{sample}}\cdot R(\varphi_0) \cdot J_{\mathrm{HWP1}}(\varphi/2)\cdot \begin{pmatrix}
        1 \\ 0
    \end{pmatrix}.
\end{equation}
\end{widetext}

The balanced photodiode detector measures the intensity difference between the two orthogonal polarization components, $I = |E_V|^2-|E_H|^2$. In the experiment, we rotate the two half-wave plates, which rotates the polarization of the incident probe beam relative to the sample to simulate rotation of the sample. 

The data were fit to the following function:
\begin{equation}
    I(\varphi) = A\mathrm{sin}(2(\varphi - \varphi_0)) + B\mathrm{sin}(4(\varphi - \varphi_1)) + C.
\end{equation}

The term $A\mathrm{sin}(2(\varphi - \varphi_0))$ describes the birefringence: $A$ is the amplitude of the birefringence and $\varphi_0$ is the orientation of the sample's principal axes relative to the lab coordinate frame. The term $B\mathrm{sin}(4(\varphi - \varphi_1))$ is small and arises from second order couplings of the sample birefringence as well as weak residual birefringence and ellipticity of the setup that can cross-couple with modulated birefringence from the sample.

The offset term, $C$, typically describes the derivative of the sample magnetization with respect to the modulation parameter; this corresponds to the magneto-optical Kerr effect for the thermally modulated experiment and the magnetic susceptibility for the field modulated experiment. However, cross-couplings between the sample and setup can result in a small offset artifact that is unrelated to these derivatives and which can be resolved when they are equal to zero. Fig. \ref{fig:off_drdt_amp} shows the thermally modulated offset, reflectivity, and birefringence amplitude. In the thermally modulated measurements, $C$ is small (two orders of magnitude smaller than the birefringence amplitude $A$ and thermally-modulated, temperature-dependent reflectivity $dR/dT$). Contributions to this term arise from cross-coupling between the sample and setup birefringence, ellipticity, and $dR/dT$, resulting in a small offset (typically at least an order of magnitude smaller than the birefringence amplitude) when no magneto-optical Kerr effect is detected in the material, as is the case for \eia. See Ref.~\cite{sunko_spin-carrier_2023} for an in-depth analysis of the ways in which these cross-coupling terms arise, the material and optical properties they couple, the artifacts they can create, their influence on the measured signal, and methods for mitigating them to isolate the true signal.

\section {Strain experiment and calibration}\label{Sec:SI_Strain}

\begin{figure*}[t!]
 \centering
 \includegraphics[scale=0.8]{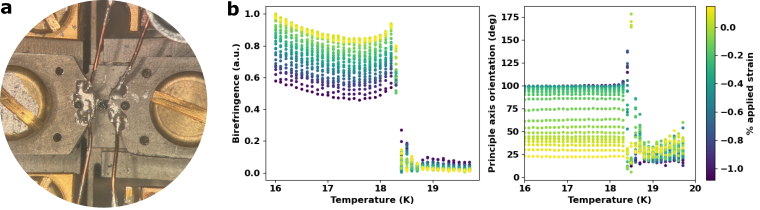}
 \caption{(a) sample loaded into strain cell. (b) strain dependence of birefringence vs temperature.}
 \label{fig:strainSI_fig1}
\end{figure*}

\subsection{Experiment}

For the measurement of $T$-mode birefringence under applied strain, a EuIn$_2$As$_2$ sample (600$\times$500 $\mu$m) was loaded into a Razorbill Instruments CS130 strain cell. The sample was secured between two titanium plates at each end with Stycast 2850FT epoxy, which were in turn bolted to the piezo-actuated jaws of the cell (Fig.\ \ref{fig:strainSI_fig1}a). A custom thermal link to the cryostat base was mounted onto the strain cell, from which two copper wires were used to heat sink the sample via EPO-TEK H20E silver epoxy. Two additional wires were routed from the sample to a Cernox thermometer and resistive heater respectively, both of which were mounted onto the strain cell body and isolated from the thermal link by an air gap. The CS130 is equipped with a capacitive sensor to monitor the displacement of the jaws, which we measured using a Keysight E4980AL LCR meter.

To explore the temperature-strain phase diagram of \eia, we measured the birefringence as a function of both temperature and strain across the two transitions. Fig.\ \ref{fig:strainSI_fig1}b shows the temperature dependence of the $T$-mod birefringence for different strains at a characteristic spot on the sample. For each curve, the sample was cooled to 16 K in zero applied strain, after which the voltage applied to the piezos was ramped to a fixed value at $\sim 0.5$ V/s. The birefringence was measured while warming under the applied voltage, allowing the reading on the strain cell capacitor to stabilize at each temperature. Because the voltage-strain relationship is temperature dependent, strictly speaking the strain changes with temperature for each curve, however we report the average strain and find that the variance is negligible over this small temperature range. Importantly, the transition temperature and overall shape of the birefringence with temperature is unmodified by strain, indicating that the essential character of the phase transition remains intact with strain.

Furthermore, to demonstrate strain tunability of the magnetic structure, we swept the strain at a fixed temperature of 16K (Fig.\ \ref{fig:EIA_Fig10}). After cooling the sample in zero applied strain, we swept the voltage on the piezos from 0 $\rightarrow$ 20 V $\rightarrow$ $-20$ V $\rightarrow$ 0, measuring the capacitance at each point.

\subsection{Calibration}

The methods for calculating the strain applied to a sample mounted in a CS130 strain cell are described in detail in the Razorbill Instruments application notes \cite{razorbill_app_notes}. The strain cell incorporates a capacitive displacement sensor to quantify the strain in the sample for a given applied voltage to the peizo stacks. In the ideal scenario, the displacement of the capacitor plates is equivalent to the displacement of the crystal, from which the strain can be calculated as $\epsilon = \Delta L/L$. However several caveats must be taken into account: (1) the capacitance of the sensor exhibits a temperature dependence independent of the true gap between the jaws of the cell, primarily due to a contraction of the distance between the capacitor plates at lower temperatures, (2) the differential thermal expansion between the sample and the titanium cell leads to an offset to the displacement of the sample, and (3) the finite stiffness of the epoxy and bolting plates securing the sample to cell manifests as an additional reduction to the sample displacement. Obtaining an accurate strain measurement thus requires careful consideration. Here, we describe the specific procedure used towards these corrections.

The displacement of the piezo stacks corrected for temperature dependent contractions in the capacitor can be parameterized as:

\begin{equation}\label{eq:capacitor_displacement}
 \Delta L_{\textnormal{piezo}}(T, C) = \frac{\alpha}{C(T) - C_p(T)} - d_0,
\end{equation}

\noindent where $T$ and $C$ are the temperature and measured capacitance respectively, $\alpha$ and $d_0$ are provided by Razorbill, and $C_p(T)$ represents a temperature dependent parallel capacitance. $C_p(T)$ can be obtained by loading a stiff titanium dummy sample, which isolates the temperature dependence of the measured capacitance to changes in the capacitor itself, and measuring the capacitance with no applied voltage, $C_{\textnormal{dummy}}(T)$, as a function of temperature:

\begin{equation}\label{eq:capacitor_temp_dep}
 C_p(T) = C_{\textnormal{dummy}}(T) - \frac{\alpha}{d_0}.
\end{equation}

\begin{figure}
 \centering
 \includegraphics[scale=0.5]{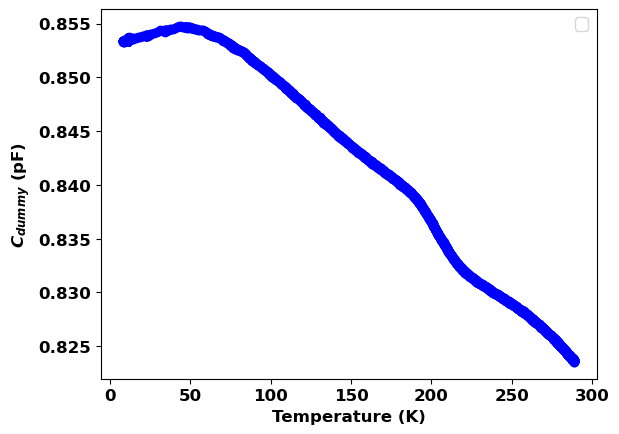}
 \caption{Measured $C_\textnormal{dummy}(T)$ vs.\ $T$.}
 \label{fig:strainSI_fig2}
\end{figure}

\noindent The experimental values for $C_{\textnormal{dummy}}(T)$ measured in our system are shown in Fig.\ \ref{fig:strainSI_fig2}.

As mentioned above, $\Delta L_{\textnormal{piezo}}$ ought to be corrected to obtain the true sample displacement $\Delta L$. However, as neither the thermal expansion nor the Young's modulus are available for EuIn$_2$As$_2$, we have chosen to disregard these corrections in reporting strain with the understanding that the values are relative and should be taken as an upper bound to the true strain. These need not be small corrections; simulations for the case of the iron-pnictide supercondctors show strain transmission of 70\% due to the pliability of the epoxy, and the thermal expansion between titanium and a test sample can differ by several multiples~\cite{ikeda_symmetric_2018, straquadine_evidence_2022} . However, the qualitative observation that the optical axis is tunable under strain is independent of these corrections.

\section{Microdomain scenario}\label{sec:Microdomains}

\begin{figure*}[t!]
 \centering
 \includegraphics[scale=0.4]{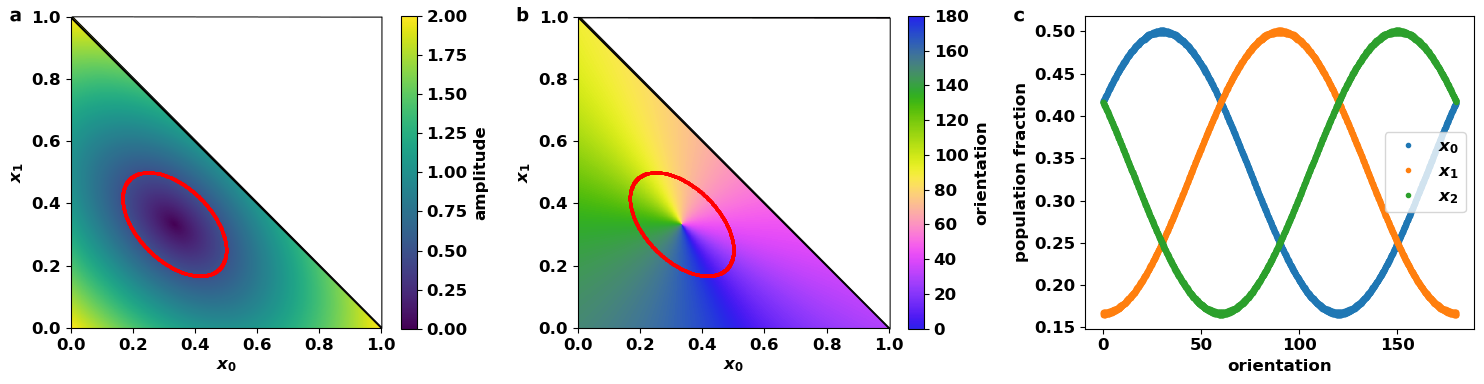}
 \caption{(a) Birefringence amplitude and (b) principle axis orientation as a function of $x_0$ and $x_1$ with $x_2$ set by the normalization condition. The ellipsoid indicates a possible phase-space trajectory that yields a constant amplitude and changing orientation. (c) Population fraction phase-space trajectories around the constant amplitude ellipsoid versus expected principal axis.}
 \label{fig:microdomain_fig}
\end{figure*}

In any symmetry-breaking transition, energetically equivalent but distinguishable configurations of the order parameter, i.e. domains, can be obtained by the application of the broken symmetry operators. In EuIn$_2$As$_2$, the transition into Phase I breaks the $C_{3z}$ symmetry of the paramagnetic state, so three domains related by $C_{3z}$ rotation are generally expected. However, our experiment does not show the expected three domains, and reveals a broad orientation continuum instead (Fig.~\ref{fig:EIA_Fig4}).

One possible explanation of this observation is that the size of such domains is smaller than our resolution ($\sim$ 5 $\mu$m), so that the measured birefringence results from averaging over many domains. Here we model this microdomain scenario to illustrate that it in fact cannot naturally account for the smoothly varying principle axis orientation and constant birefringence amplitude observed in EuIn$_2$As$_2$ as shown in Figs.\ \ref{fig:EIA_Fig4} and \ref{fig:EIA_Fig5}.

We consider a collection of small domains within the area of our laser spot. Each domain is described by a nematic vector with its principle axis oriented along one of the three $C_{3z}$-equivalent directions of the crystal, which we label $i = 0,1,2$; taking orientation 0 to point along the $x$-axis, these correspond to the nematic vector pointing along $i\pi/3$. We denote the population fraction of domains at orientation $i$ as $x_i$, such that $\sum_i x_i = 1$. The reflectivity matrix for a domain at orientation $i$ can be expressed as

\begin{equation}\label{eq:reflectivity_matrix_microdomain}
 r_i = R(-i\pi/3)\begin{pmatrix} 1 + \delta r & 0 \\ 0 & 1 - \delta r \end{pmatrix}R(i\pi/3),
\end{equation}

where $R(\theta) = \begin{pmatrix} \cos(\theta) & -\sin(\theta) \\ \sin(\theta) & \cos(\theta) \end{pmatrix}$ is a typical rotation matrix.

As discussed in Sec.\ \ref{sec:Optical}, each domain at orientation $i$ contributes a polarization rotation as a function of incident polarization angle of $d\phi_i = \delta r \sin(2(\phi - i\pi/3))$ such that the total signal is $d\phi = \sum_{i} x_i d\phi_i.$ To first order in $\delta r$, this evaluates to

\begin{equation}
 d\phi = A(\mathbf{x})\sin(2(\phi + \alpha(\mathbf{x}))),
\end{equation}

\noindent where the effective birefringence amplitude $A(\mathbf{x})$ and principle axis orientation $\alpha(\mathbf{x})$ are given by

\begin{align}
 A(\mathbf{x}) &= \delta r \sqrt{3(x_1 - x_2)^2 + (x_1 + x_2 - 2x_0)^2} \label{eq:microdomain_amp} \\
 \alpha(\mathbf{x}) &= \frac{1}{2}\tan^{-1}\left( \frac{\sqrt{3}(x_1 - x_2)}{x_1 + x_2 - 2x_0} \right) \label{eq:microdomain_ang}\\
 \mathbf{x} &= (x_0, x_1, x_2).
\end{align}

Having developed a model for the microdomain scenario, we want to determine whether a continuously changing principle axis orientation can be consistent with a constant amplitude as observed in the experiments. To this end, we explore both quantities in the $\mathbf{x}$ phase space, which represents the possible spread of domains at different spots on the sample. The normalization constraint $\sum_i x_i = 1$ allows us to rewrite Eqns.\ \ref{eq:microdomain_amp} and \ref{eq:microdomain_ang} in terms of just two components of $\mathbf{x}$, arbitrarily taken as $x_0$ and $x_1$. Fig.\ \ref{fig:microdomain_fig}a-b shows the effective birefringence amplitude and principle axis within this reduced parameter space. As indicated by the red ellipses, we find that the only trajectories that are constant in amplitude with changing principle axis are very fine-tuned, involving contrived changes in populations of all three domains (Fig.\ \ref{fig:microdomain_fig}c). While this scenario is technically feasible, the strict restriction on the phase space has seemingly no physical basis. On these grounds we discard the microdomain scenario as a viable explanation of the birefringence in EuIn$_2$As$_2$.

The predictions based on this model in the main text show the case of only two competing domain orientations. Specifically, we define the orientation with maximum weight in Fig.\ \ref{fig:EIA_Fig4}c to be $\theta_0 = 0$ with respect to the crystal with population $x_0$, and consider the broad distribution to be a consequence of mixing in some of the other two domains; for orientations clockwise to $\theta_0$ we allow a population $x_2$ of $\phi_2 = 2\pi/3$ domains, and for orientations counterclockwise to $\theta_0$ we allow a population $x_1$ of $\phi_1 = \pi/3$ domains, taking the population of the third domain to zero. For example, $x_2=0$ corresponds to the bottom edges of the phase space in Fig.\ \ref{fig:microdomain_fig}. For each point in Fig.\ \ref{fig:EIA_Fig5}b, we then use Eqns.\ \ref{eq:microdomain_amp} and \ref{eq:microdomain_ang} to uniquely specify the amplitude expected for the measured orientation.

\section{Symmetry analysis}\label{secSM:symmetry}
In this section, we provide details of the symmetry analysis and group theory methods that we applied to pinpoint the magnetic configuration for Phase I and Phase II of \eia.   

Two ingredients are important to characterize a magnetically ordered state: the propagation vector(s) $\mathbf{Q}$, which determines the periodicity of the magnetic structure, and the symmetries that they preserve. The spatial dependence of the expectation values of the magnetic moments in the ordered state is given by
\begin{equation}
\mathbf{M}_{\alpha}(\mathbf{r}_i)=\sum\limits_{\mathbf{Q}}e^{i\mathbf{Q}\cdot\mathbf{r}_i}\mathbf{M}_{\alpha}(\mathbf{Q}) \text{ .}
\label{eqSM:M}
\end{equation}

\noindent Here, $\alpha$ denotes the sublattice index and $\mathbf{r}_i$ is the position of the atom $i$ of the underlying lattice. $\mathbf{M}_{\alpha}(\mathbf{r}_i)$ transforms as one or more Irreducible representations (Irreps) of the parent paramagnetic space group (also called gray group). The gray group is formed by the crystal space group $G$ combined with time-reversal ($\mathcal{T}$), $G\oplus\mathcal{T}G$. The development of a magnetic order breaks $\mathcal{T}$ alone and certain spatial operations $g\in G$ \footnote{$g$ can be proper or improper rotations, spatial inversions, and also non-symmorphic operations in which point group operations are combined with half-lattice translations.}, but may preserve certain combinations of spatial operations and time-reversal $\mathcal{T}g$. Therefore, the magnetic order lowers $G\oplus\mathcal{T}G$ down to one of its subgroups that define the magnetic space group (MSG) $G_{M}$ of the ordered state. 

\begin{figure*}[t!]
    \centering
    \includegraphics[width=0.72\linewidth]{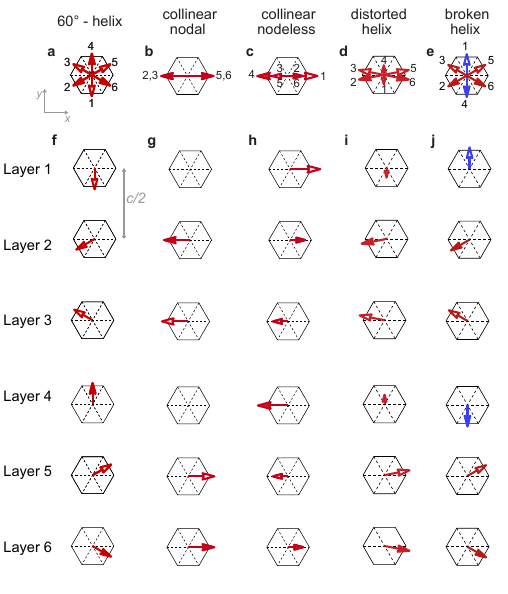}
    \caption{Representative moment configurations of the class of magnetic structures listed in Tables \ref{tabSM:table1} and \ref{tabSM:table2}. The moments in each Eu layer are in-plane and ferromagnetically aligned, pointing along the direction indicated by the arrows. The top view of each structure is shown in the top row of the figure (panels a-e).  In panels (e-f), the broken helix structure fulfills the equal moment condition in Eq.~\ref{eqSM:equalM}}.
    \label{figSM:structures}
\end{figure*}

For \eia, the gray group is $P6_3/mmc1'$ (No. 194.264), $\mathbf{Q}=\mathbf{Q}_{1}=\pm(0,0,1/3)$ for Phase I and $\mathbf{Q}=\mathbf{Q}_{1}, \mathbf{Q}_{2}=\pm(0,0,1)$ for Phase II. Besides, $\alpha=1,2$ since there are two non-equivalent Eu atoms per crystallographic unit cell. For the discussion that comes later in this section, we can focus on the dependence of $\mathbf{M}_{\alpha}(z_i)$ on the $z$-component of the position of the Eu layer $i$, since the moments are ferromagnetically aligned within each plane and the ordering vectors are perpendicular to the direction $\hat{z}$ of stacking of the Eu planes (see Fig.\ref{figSM:structures}).

Note that throughout the paper, we write $\mathbf{Q}$ in terms of the reciprocal lattice vectors, $\mathbf{Q}=(h,k,l)=h\mathbf{b}_1+k\mathbf{b}_2+l\mathbf{b}_3$, where $\mathbf{b}_1=-\frac{4\pi}{3a}\hat{x}$, $\mathbf{b}_2=\frac{2\pi}{3a}\left(-\hat{x}+\sqrt{3}\hat{y}\right)$.

Magnetic structures with the same $\mathbf{Q}$ but distinct $\mathbf{M}_{\alpha}(\mathbf{Q})$ have different symmetry properties and transform as distinct Irreps of the parent gray group. In this section, we explain how to get a list of all magnetic structures, and corresponding $\mathbf{M}_{\alpha}(\mathbf{Q})$, that are candidates for Phases I and II from a symmetry perspective only. We further narrowed it down based on experimental evidence. Our multimodal protocol can be summarized as follows:  
\begin{itemize}
    \item \textit{Step 1}- given the paramagnetic space group and the ordering vector(s) for the phase in case, the list of all magnetic space groups that are subgroups of the paramegnetic groups and can be reached by a magnetic phase transition can be readily obtained using the software Isotropy~\cite{stokes2017isotropy}. The corresponding Irreps and order parameters are also listed. 
    \item \textit{Step 2}- For each Irrep, there are multiple magnetic structures belonging to different magnetic space groups. Coupled Irreps are also possible. We constrained which Irrep to focus on by requiring that the magnetic structure has magnetic moments lying on the Eu planes and magnetic Bragg peaks consistent with scattering experiments\cite{riberolles_magnetic_2021, soh_understanding_2023}. These constraints singled out Irrep $m\Delta_6$ for Phase I and $m\Delta_6$ coupled with $m\Gamma_5^+$ for Phase II.
    \item \textit{Step 3}- All the magnetic structures that transform as the Irrep(s) selected in step 2 are listed in Table \ref{tabSM:table1} for Phase I and \ref{tabSM:table2} for Phase II. 
    \item \textit{Step 4}- The symmetries of each magnetic structure from step 3 are compared with the constraints set by the optical measurements, which are used to rule out a subset of them.
\end{itemize}

\noindent For Phase I, our multimodal approach allowed us to identify three states  that are consistent with both scattering and optical experiments. They correspond to nodal amplitude-modulated states with different orientations of the magnetic moments. One of the possibilities is that the moments are parallel to $[100]=\hat{x}$, as illustrated in Fig.\ref{figSM:structures}b and characterized by an order parameter $\bm{\Psi}_{I}=(a,a,0,0)$. Due to the high-symmetry direction of the moments, this structure preserves the two-fold rotation around $\hat{y}$ ($C_{2[010]}$), as well as a non-symmorphic two-fold rotation around $\hat{z}$ ($C_{2z}$) and the product between time-reversal and inversion ($\mathcal{T}\mathcal{P}$), and belongs to the MSG $Cm'c'm'$ (No. 65.564). Note that any structure obtained from $\bm{\Psi}_{I}=(a,a,0,0)$ by a three-fold rotation around $z$ ($C_{3z}$) is a related domain and belongs to the same MSG. The second allowed nodal structure is characterized by the order parameter $\bm{\Psi}_{I}=(0,0,a,-a)$ and has moments parallel to $\hat{y}$. This is also a high-symmetry direction in the Eu plane. The moments break $C_{2[010]}$, but preserve a mirror symmetry $M_{[010]}=\mathcal{P}C_{2[010]}$. The third allowed nodal structure is parametrized by $\bm{\Psi}_{I}=(a,a,b,-b)$ and is a linear superposition of the two structures mentioned before. As a result, the magnetic moments point along a direction that is not a high-symmetry one, and neither $M_{[010]}$ nor $C_{2,[010]}$ are preserved. The only symmetries left in the generic orientation nodal structure are the non-symmetric $C_{2z}$ and $\mathcal{P}\mathcal{T}$, lowering the MSG to $P2_1/m'$ (No. 11.53). 

Note that the sixth-order terms in the Landau functional (see Sec.\ref{secSM:freeE}) enforce the moments to point along a high-symmetry direction, reflecting the underlying crystalline anisotropy. However, in the regime where built-in strain dominates over such a crystal anisotropy, the nodal structure with lower symmetry is generically favored by uniaxial strain along an arbitrary direction. Motivated by the observed broad distribution of crystal axes, we will consider this regime.      

For Phase II, on the other hand, all structures in Table \ref{tabSM:table2} are consistent with both scattering and optical experiments. Among them, the three broken-helix structures are more likely to describe Phase II since they are favored by the equal moment condition, which in turn is expected if the magnetism of \eia arises from localized $S=7/2$ Eu$^{2+}$ moments (see Sec.\ref{secSM:equal}). Similarly to the three nodal amplitude modulated structures possible for Phase I, the three broken helices in Phase II are distinguished by the orientation of the \Neel~component with respect to the crystal axis. Once again, crystal anisotropy favors moments parallel to the in-plane high-symmetry direction, but the dominant built-in strain favors a less symmetric orientation of the moments (MSG $P2_1$ (No. 4.7)). Given the broad distribution of built-in strain direction suggested by our optical experiments, the broken helix in Phase II is essentially unpinned from the crystal axes, presumably due to built-in strain.  

Representatives of each type of state appearing in Tables \ref{tabSM:table1} and \ref{tabSM:table2} are shown in Fig. \ref{figSM:structures}. Note that in Phase I there is a class of structures labeled ``distorted helix'', which is obtained by a superposition of an amplitude-modulated structure and a 60\degree-helix. The moments get distorted in comparison with the more symmetric 60-\degree counterpart, but there is still a well-defined single-handed circulation of the moments' direction around $\hat{z}$. In Phase II, an AFM component is also added to the magnetic structure, in agreement with the development of the additional $\mathbf{Q}_2=(0,0,1)$ ordering vector. When strong enough, this AFM component leads to the interchange of four moment directions, leading to a broken helix similar to that originally introduced in Ref. \onlinecite{riberolles_magnetic_2021} (compare Fig. \ref{figSM:structures}d and e). However, as mentioned in the main text, a key difference is that the broken helix proposed in this work is not pinned to a high-symmetry direction in the Eu plane due to the strong effect of the built-in strain.

\begin{table*}[t!]
\caption{Candidate magnetic structures for Phase I that are consistent with the experimentally observed Bragg peaks $\mathbf{Q}_{1}=(0,0,1/3)$ and magnetic moments lying within the Eu planes. For each structure, we specify the form of the order parameter $\bm{\Psi}_I$ written in the basis of the representation space of $m\Delta_6$ adopted by Isotropy (helical basis). The transformation matrix between this basis and the collinear basis that we adopt in this work  is shown in Sec. \ref{secSM:basis}. The magnetic space group (MSG) and its generators are also listed. $1$ denotes the identity operation, $\mathcal{P}$ denotes spatial inversion, and $\mathcal{T}$ represents time reversal. $C_{n,[uvw]}$ denotes a rotation of $2\pi/n$ around the axis $[uvw]=u\mathbf{a}_1+v\mathbf{a}_2+\mathbf{a}_3$, where $\mathbf{a}_j$ are primitive crystal axes. We set $[001]\parallel\hat{z}$. The mirror operations are $M_{[uvw]}=\mathcal{P}C_{2,[uvw]}$. When any of these operations (generically denoted by $g$) is combined with a translation $\tau=\tau_1\mathbf{a}_1+\tau_2\mathbf{a}_2+\tau_3\mathbf{a}_3$ by a fraction of the primitive unit cell, we have a non-symmorphic operation $\{\left. g\right|\tau_1\tau_2\tau_3\}$. The last column shows the two optical responses considered in this work. $\checkmark$ ($\times$) signifies that the response is symmetry allowed (forbidden).}
\label{tabSM:table1}
\resizebox{0.99\textwidth}{!}{
\begin{tabular}{|c|c|c|c|cc|}
\hline
\multirow{2}{*}{Magnetic structure}                                                               &\multirow{2}{*}{Order parameter}                                    & \multirow{2}{*}{MSG}    & \multirow{2}{*}{Generators of the MSG}                                                                                                                                                    & \multicolumn{2}{c|}{Optical responses}                                             \\  \cline{5-6} 
                                                                                                  & \multicolumn{1}{c|}{}                  &                         &
                                                                                                  & \multicolumn{1}{c|}{T-mod signal}                  & H-mod signal                  \\ \hline
\multirow{5}{*}{60\degree-helix}                                                   & 
\multicolumn{1}{c|}{$(0,0,0,a)$}                                       & P6$_1$2'2' (178.159)                 & $1$, $\{\left. C_{2z}\right|00\frac{3c}{2}\}$, $\{\left. C_{3z}\right|00c\}$, $\mathcal{T}C_{2,[010]}$ & \multicolumn{1}{c|}{$\times$}                      & $\times$                      \\ \cline{2-6} 
&
\multicolumn{1}{c|}{$(0,0,a,0)$}                                       & P6$_5$2'2' (179.165)                 & $1$, $\{\left. C_{2z}\right|00\frac{3c}{2}\}$, $\{\left. C_{3z}\right|002c\}$, $\mathcal{T}C_{2,[010]}$ & \multicolumn{1}{c|}{$\times$}                      & $\times$                      \\ \cline{2-6} 
                                                                                                  & \multicolumn{1}{c|}{$(a,0,0,0)$}                                       & P6$_5$22 (179.161)                 &          $1$, $\{\left. C_{2z}\right|00\frac{3c}{2}\}$, $\{\left. C_{3z}\right|002c\}$, $C_{2,[010]}$                                                                                                                                                                                 & \multicolumn{1}{c|}{$\times$}                      & $\times$                      \\ \cline{2-6} 
                                                                            &              \multicolumn{1}{c|}{$(0,a,0,0)$}                                       & P6$_1$22 (178.155)                 & $1$, $\{\left. C_{2z}\right|00\frac{3c}{2}\}$, $\{\left. C_{3z}\right|00c\}$, $C_{2,[010]}$ & \multicolumn{1}{c|}{$\times$}                      & $\times$                      \\ \cline{2-6}         & \multicolumn{1}{c|}{$(a,0,b,0)$}                                       & P6$_5$ (170.117)                 &                  $1$, $\{\left. C_{2z}\right|00\frac{3c}{2}\}$, $\{\left. C_{3z}\right|002c\}$                                                                                                                                                                        & \multicolumn{1}{c|}{$\times$}                      & $\times$                      \\ \hline
\multirow{2}{*}{\begin{tabular}[c]{@{}c@{}}colinear nodal\\ amplitude modulated\end{tabular}}     & \multicolumn{1}{c|}{\multirow{2}{*}{$(a,a,0,0)$}}     & \multirow{2}{*}{Cm'c'm' (63.465)} & \multirow{2}{*}{$1$, $\{\left. C_{2z}\right|00\frac{3c}{2}\}$, $C_{2,[010]}$, $\mathcal{T}\mathcal{P}$}                                                                                                                                                                         & \multicolumn{1}{c|}{\multirow{2}{*}{$\checkmark$}} & \multirow{2}{*}{$\times$}     \\
                                                                                                  & \multicolumn{1}{c|}{}                               &                   &                                                                                                                                                                                                                    & \multicolumn{1}{c|}{}                              &                               \\ \hline
\multirow{2}{*}{\begin{tabular}[c]{@{}c@{}}colinear nodeless\\ amplitude modulated\end{tabular}}  & \multicolumn{1}{c|}{\multirow{2}{*}{$(0,0,a,a)$}}  & \multirow{2}{*}{Cm'c'm (63.462)} & \multirow{2}{*}{$1$, $\{\left. C_{2z}\right|00\frac{3c}{2}\}$, $\mathcal{T}C_{2,[010]}$, $\mathcal{P}$}                                                                                                                                                                         & \multicolumn{1}{c|}{\multirow{2}{*}{$\checkmark$}} & \multirow{2}{*}{$\checkmark$} \\
                                                                                                  & \multicolumn{1}{c|}{}                               &                   &                                                                                                                                                                                                                    & \multicolumn{1}{c|}{}                              &                               \\ \hline
\begin{tabular}[c]{@{}c@{}}colinear nodal \\ amplitude modulated\end{tabular}                     & \multicolumn{1}{c|}{$(0,0,a,-a)$}                                     & Cmcm' (63.461)                  &        $1$, $\{\left. C_{2z}\right|00\frac{3c}{2}\}$, $M_{[010]}$, $\mathcal{T}\mathcal{P}$                                                                                                                                                                                   & \multicolumn{1}{c|}{$\checkmark$}                  & $\times$                      \\ \hline
\multirow{2}{*}{\begin{tabular}[c]{@{}c@{}}colinear nodeless\\ amplitude modulated\end{tabular}}  & \multicolumn{1}{c|}{\multirow{2}{*}{$(a,-a,0,0)$}}   & \multirow{2}{*}{Cmcm (63.457)} & \multirow{2}{*}{$1$, $\{\left. C_{2z}\right|00\frac{3c}{2}\}$, $C_{2,[010]}$, $\mathcal{P}$ }                                                                                                                                                                         & \multicolumn{1}{c|}{\multirow{2}{*}{$\checkmark$}} & \multirow{2}{*}{$\checkmark$} \\
                                                                                                  & \multicolumn{1}{c|}{}                                                  &                         &                                                                                                                                                                                           & \multicolumn{1}{c|}{}                              &                               \\ \hline
\multirow{2}{*}{\begin{tabular}[c]{@{}c@{}}colinear nodeless\\ amplitude modulated\end{tabular}}  & \multicolumn{1}{c|}{\multirow{2}{*}{$(a,a,b,b)$}}   & \multirow{2}{*}{Cm'c'2$_1$ (36.176)} & \multirow{2}{*}{$1$, $\{\left. C_{2z}\right|00\frac{3c}{2}\}$, $\mathcal{T}M_{[010]}$}                                                                                                                                                                         & \multicolumn{1}{c|}{\multirow{2}{*}{$\checkmark$}} & \multirow{2}{*}{$\checkmark$} \\
                                                                                                  & \multicolumn{1}{c|}{}                                                  &                         &                                                                                                                                                                                           & \multicolumn{1}{c|}{}                              &                               \\ \hline
\multirow{2}{*}{\begin{tabular}[c]{@{}c@{}}colinear nodeless\\ amplitude modulated\end{tabular}}  & \multicolumn{1}{c|}{\multirow{2}{*}{$(a,-a,b,-b)$}} & \multirow{2}{*}{Cmc2$_1$( 36.172)} & \multirow{2}{*}{$1$, $\{\left. C_{2z}\right|00\frac{3c}{2}\}$, $M_{[010]}$}                                                                                                                                                                         & \multicolumn{1}{c|}{\multirow{2}{*}{$\checkmark$}} & \multirow{2}{*}{$\checkmark$} \\
                                                                                                  & \multicolumn{1}{c|}{}                                                  &                         &                                                                                                                                                                                           & \multicolumn{1}{c|}{}                              &                               \\ \hline
distorted helix                                                                                   & \multicolumn{1}{c|}{$(0,0,a,b)$}                                       & C2'2'2$_1$ (20.33)                   &  $1$, $\{\left. C_{2z}\right|00\frac{3c}{2}\}$, $\mathcal{T}C_{2,[010]}$                                                                                                                                                                                         & \multicolumn{1}{c|}{$\checkmark$}                  & $\checkmark$                  \\ \hline
distorted helix                                                                                   & \multicolumn{1}{c|}{$(a,b,0,0)$}                                       & C222$_1$ (20.31)                   &                                                   $1$, $\{\left. C_{2z}\right|00\frac{3c}{2}\}$, $C_{2,[010]}$                                                                                                                                        & \multicolumn{1}{c|}{$\checkmark$}                  & $\checkmark$                  \\ \hline
\multirow{2}{*}{\begin{tabular}[c]{@{}c@{}}colinear nodal\\ amplitude modulated\end{tabular}}     & \multicolumn{1}{c|}{\multirow{2}{*}{$(a,a,b,-b)$}}  & \multirow{2}{*}{P2$_1$/m' (11.53)}  & \multirow{2}{*}{$1$, $\{\left. C_{2z}\right|00\frac{3c}{2}\}$, $\mathcal{T}\mathcal{P}$}                                                                                                                                                                         & \multicolumn{1}{c|}{\multirow{2}{*}{$\checkmark$}} & \multirow{2}{*}{$\times$}     \\
                                                                                                  & \multicolumn{1}{c|}{}                               &                   &                                                                                                                                                                                                                    & \multicolumn{1}{c|}{}                              &                               \\ \hline
\multirow{2}{*}{\begin{tabular}[c]{@{}c@{}}colinear nodeless \\ amplitude modulated\end{tabular}} & \multicolumn{1}{c|}{\multirow{2}{*}{$(a,-a,b,b)$}}  & \multirow{2}{*}{P2$_1$/m (11.50)}  & \multirow{2}{*}{$1$, $\{\left. C_{2z}\right|00\frac{3c}{2}\}$, $\mathcal{P}$}                                                                                                                                                                         & \multicolumn{1}{c|}{\multirow{2}{*}{$\checkmark$}} & \multirow{2}{*}{$\checkmark$} \\
                                                                                                  & \multicolumn{1}{c|}{}                                                  &                         &                                                                                                                                                                                           & \multicolumn{1}{c|}{}                              &                               \\ \hline
distorted helix                                                                                   & \multicolumn{1}{c|}{$(a,b,c,d)$}                                       & P2$_1$ (4.7)                     &                                                                      $1$, $\{\left. C_{2z}\right|00\frac{3c}{2}\}$                                                                                                                    & \multicolumn{1}{c|}{$\checkmark$}                  & $\checkmark$                  \\ \hline
\end{tabular}
}
\label{tabSM:table1}
\end{table*}

\section{Representations of the order parameter}\label{secSM:op}

In this Section, we address equivalent ways of representing the order parameters $\mathbf{M}_{\alpha}(\mathbf{Q}_1)$ and $\mathbf{M}_{\alpha}(\mathbf{Q}_2)$ for Phases I and II of \eia. $\mathbf{M}_{\alpha}(\mathbf{Q})$ are the Fourier components of the magnetic structure $\mathbf{M}_{\alpha}(z_i)$ in each phase and, therefore, is a pseudovector in the Euclidean space. $\mathbf{M}_{\alpha}(z_i)$ transforms as one or more Irreps of the parent gray space group. Within each Irrep of the space group, the spatial operations in the group $g\in G\oplus\mathcal{T}G$ can be represented by a matrix. These matrices are defined in a vector space called \textit{representation space} and have a dimension given by the product of the dimension of the corresponding Irrep of the little group of $\mathbf{Q}$ and the number of legs in the star of $\mathbf{Q}$. For the Irreps relevant for \eia, $m\Delta_6$ of $P6_3/mmc1'$ associated with the ordering vector $\mathbf{Q}_1$ is four-dimensional, and the Irrep $m\Gamma_5^+$ associated with $\mathbf{Q}_2$ is two-dimensional. 

It is useful to represent $M_{\alpha}(\mathbf{Q}_1)$ and $M_{\alpha}(\mathbf{Q}_2)$ in the representation space of $m\Delta_6$ and $m\Gamma_5^+$ respectively. We start with $M_{\alpha}(\mathbf{Q}_1)$ in the next section. 

\subsection{Order parameter for Phase I}\label{secSM:helical}

The order parameter for Phase I takes the form of a four-component vector in the representation space of $m\Delta_6$, 
\begin{equation}
\bm{\Psi}_{I}=\left(\Delta_1,\Delta_2,\Delta_3,\Delta_4\right) \text{ ,}
\label{eqSM:psiI}
\end{equation}

\noindent where $\Delta_{j}$ are real numbers. The basis of the representation space is defined in terms of four basis functions. The transformation of the basis functions upon the action of the symmetry operations of the space group is then used to construct the matrix representation of each operation within the Irrep. The choice of basis function is arbitrary since there are many possible choices of basis vectors for a vector space. The order parameters $\bm{\Psi}_I$ obtained in Isotropy \cite{stokes2017isotropy} are written in a basis that we call \textit{helical basis}. This means that the basis functions correspond to four independent helical magnetic structures that we denote by $\mathbf{f}_{x\pm}(z_i)$ and $\mathbf{f}_{y\pm}(z_i)$: \begin{align}
&(1,0,0,0)\rightarrow \mathbf{f}_{x-}(z_i) \text{ ,}\\
&(0,1,0,0)\rightarrow 
\mathbf{f}_{x+}(z_i) \text{ ,}\\
&(0,0,1,0)\rightarrow 
\mathbf{f}_{y-}(z_i) \text{ ,}\\
&(0,0,0,1)\rightarrow 
\mathbf{f}_{y+}(z_i) \text{ ,}
\end{align}

\noindent where
\begin{align}
    &\mathbf{f}_{x\mp}(z_i)=\left(
\pm\cos\left(\frac{2\pi z_i}{3c} \right ), 
-\sin\left(\frac{2\pi z_i}{3c} \right ), 
0
\right)^T\text{ ,}\label{eqSM:fc}\\[0.3cm]
&\mathbf{f}_{y\mp}(z_i)=
\left(
\mp\sin\left(\frac{2\pi z_i}{3c} \right ), 
-\cos\left(\frac{2\pi z_i}{3c} \right ), 
0\right)^T\text{ .}
\label{eqSM:fv}
\end{align}

\begin{table*}[t!]\centering
\caption{Candidate magnetic structures for Phase II that are consistent with the experimentally observed Bragg peaks, $\mathbf{Q}_{1}=(0,0,1/3)$ and $\mathbf{Q}_{2}=(0,0,1)$. All of the structures are allowed by symmetry to have finite T-mod and H-mod optical responses. For each structure, we specify the form of the order parameter $\bm{\Psi}_I$ and the antiferromagnetic component $\bm{\Psi}_{L}$ written in the helical basis (see Secs. \ref{secSM:helical} and \ref{secSM:basis}). The last column shows the magnetic space group (MSG) for each structure, whose generators can be found in Table \ref{tabSM:table1}.}
\label{tabSM:table2}
\resizebox{0.7\textwidth}{!}{
\begin{tabular}{|c|c|c|c|}
\hline
\multirow{2}{*}{Magnetic structure}                                                               & \multirow{2}{*}{$\bm{\Psi}_{L}$}                             &\multirow{2}{*}{$\bm{\Psi}_{I}$}                                                                     & \multirow{2}{*}{MSG}    \\ 
& \multicolumn{1}{c|}{}             & \multicolumn{1}{c|}{}                                                      &       
                                                                                    \\  \hline
\multirow{4}{*}{\begin{tabular}[c]{@{}c@{}}colinear nodeless \\ amplitude-modulated\end{tabular}} & $\left(0,a\right)$       & $\left(0,0,b,b\right)$                                                             & Cm'c'm' (No. 63.462)    \\ \cline{2-4} 
                                                                                                  & $\left(a,\sqrt{3}a\right)$            & $\left(-\frac{b}{2},-b,-\frac{\sqrt{3}b}{2},0\right)$                           & Cmcm (No. 63.457)       \\ \cline{2-4} 
                                                                                                  & $\left(a,\sqrt{3}a\right)$               & $\left(\frac{-b+\sqrt{3}c}{2},-b,-\frac{\sqrt{3}b+c}{2},-c\right)$                & Cm'c'2$_1$ (No. 36.176) \\ \cline{2-4} 
                                                                                                  & $\left(0,a\right)$                         & $\left(b,b,c,c\right)$                                                      & Cmc2$_1$ (No. 36.172)   \\ \hline
\multirow{4}{*}{broken helix}                                                                     & $\left(0,a\right)$                        & $\left(0,0,b,c\right)$                                                           & C2'2'2$_1$ (No. 20.33)  \\ \cline{2-4} 
                                                                                                  & $\left(a,\sqrt{3}a\right)$             & $\left(\frac{-b}{2},c,-\frac{\sqrt{3}b}{2},0\right)$              & C222$_1$ (No. 20.31)    \\ \cline{2-4} 
                                                                                                & $\left(a,b\right)$                        & $\left(c,-c,d,d\right)$                                                            & P2$_1$/m (No. 11.50)    \\ \cline{2-4} 
                                                                                                  & $\left(a,b\right)$                        & $\left(c,d,e,f\right)$                                                             & P2$_1$ (No. 4.7)        \\ \hline
\end{tabular}
}
\end{table*}

\noindent Here, the superscript $T$ denotes the transpose of the vector. $\mathbf{f}_{x\pm}(z_i)$ is a vector function in Euclidian space and  describes 60\degree-helix with clockwise ($-$) or counterclockwise (+) helicity and moments parallel to $\hat{x}$ every third Eu layer. Recall that $z_i=0,c/2,c,\cdots$. Similarly, $\mathbf{f}_{y\pm}(z_i)$ describes 60\degree-helices with opposite helicities with moments parallel to $\hat{y}$ in every third Eu-layer. A generic magnetic structure associated with the ordering vector $\bm{\Psi}_I$ in Eq.(\ref{eqSM:psiI}) can thus be written as 
\begin{equation}
    \mathbf{M}(z_{i}^{\alpha})=\Delta_1\mathbf{f}_{x-}(z_{i}^{\alpha})+\Delta_2\mathbf{f}_{x+}(z_{i}^{\alpha})+\Delta_3\mathbf{f}_{y-}(z_{i}^{\alpha})+\Delta_4\mathbf{f}_{y+}(z_{i}^{\alpha}) \text{ .}
    \label{eqSM:MDeltas}
\end{equation}

\noindent Here, we define $\bm{M}_{\alpha}(z_i)\equiv \bm{M}(z_i^\alpha)$, where $z_{i}^{\alpha}$ denotes the position of the Eu layer belonging to the sublattice $\alpha$. Note that $z_{i}^{\alpha}=0,c,2c,\cdots$ for $\alpha=1$ and $z_{i}^{\alpha}=\frac{c}{2},\frac{3c}{2},\frac{5c}{2},\cdots$ for $\alpha=2$.

We now have all the ingredients needed to relate $\bm{\Psi}_{I}$ and $\mathbf{M}_{\alpha}(\mathbf{Q}_1)$. Inverting Eq.(\ref{eqSM:M}),  substituting into it Eq.(\ref{eqSM:MDeltas}), and recalling that $z_i=0,c,2c,\cdots$ for sublattice $\alpha=1$ and $z_i=c/2,3c/3,5c/c,\cdots$ for sublattice $\alpha=2$, we obtain

\begin{equation}
    \frac{\mathbf{M}_{1}(\mathbf{Q}_1)}{N_m}=\frac{\mathbf{M}_{2}(\mathbf{Q}_1)}{N_m}=\frac{3}{2}\begin{pmatrix}
        \Delta_1-\Delta_2+i\left(\Delta_3-\Delta_4\right)\\
        -\Delta_3-\Delta_4+i\left(\Delta_1+\Delta_2\right)\\
        0
    \end{pmatrix} \text{ ,}
    \label{eqSM:MQ1final}
\end{equation}

\noindent where $N_{m}$ denotes the number of magnetic unit cells.

In Sec.\ref{secSM:basis}, we define another basis which is constructed from orthogonal collinear structures rather than helical. We call it \textit{collinear basis} and this is the basis adopted in the discussion carried out in the main text. The helical to colinear basis transformation is provided in Sec.\ref{secSM:basis}.

\subsection{Order parameter for Phase II}

We now repeat the analysis for Phase II, which is characterized by both $\bm{\Psi}_I$ and by a \Neel~component to the magnetic order. The later takes the form of a two-component vector in the representation space of $m\Gamma_5^+$, 
\begin{equation}
\bm{\Psi}_{L}=\left(l\cos\theta,l\sin\theta\right) \text{ .}
\label{eqSM:psiL}
\end{equation}

\noindent $l$ and $\theta$ are real numbers and their meaning will become clear shortly. As before, we need to specify the basis functions for $m\Gamma_5^+$. We adopt the same basis used as Isotropy \cite{stokes2017isotropy}, which consists of two orthogonal in-plane AFM configurations: 
\begin{align}
    &\left(1,0\right)\rightarrow \mathbf{g}_{x}(z_i)\text{ ,}\\
    &\left(0,1\right)\rightarrow \mathbf{g}_{y}(z_i)\text{ ,}
\end{align}

\noindent with
\begin{align}
    &\mathbf{g}_{x}(z_i)=(-1)^{\frac{2z_i}{c}}\left(1,0,0
    \right)^T\text{ ,}\\
    &\mathbf{g}_{y}(z_i)=(-1)^{\frac{2z_i}{c}}\left(0,1,0\right)^T \text{ .}
\end{align} 

\noindent $\mathbf{g}_{x}(z_i)$ and $\mathbf{g}_{y}(z_i)$ are thus vector functions in Euclidean space that describe an AFM order with moments pointing along $\hat{x}$ and $\hat{y}$, respectively. A generic N\'eel order with order parameter $\bm{\Psi}_{L}$ in Eq.(\ref{eqSM:psiL}) can thus be written as \begin{equation}
    \mathbf{M}(z_{i}^{\alpha})=l\cos\theta \,\mathbf{g}_{x}(z_{i}^{\alpha})+l\sin\theta \,\mathbf{g}_y(z_{i}^{\alpha}) \text{ .}
    \label{eqSM:Ml}
\end{equation}   

\noindent Recall that $z_{i}^{\alpha}=0,c,2c,\cdots$ for $\alpha=1$ and $z_{i}^{\alpha}=\frac{c}{2},\frac{3c}{2},\frac{5c}{2},\cdots$ for $\alpha=2$. From this equation we can readily see that $l$ and $\theta$ denote, respectively, the amplitude and orientation with respect to the lattice axes of the moments in the \Neel~state. 

The relation between $\mathbf{M}_{\alpha}(\mathbf{Q}_2)$ and $\bm{\Psi}_{L}$ is obtained by inverting Eq.(\ref{eqSM:M}) and using Eq.(\ref{eqSM:Ml}): 
\begin{align}
    &\frac{\mathbf{M}_{1}(\mathbf{Q}_2)}{N_m}=-\frac{\mathbf{M}_{2}(\mathbf{Q}_2)}{N_m}=\begin{pmatrix}
        l\cos\theta\\
        l\sin\theta\\
        0
    \end{pmatrix}
\end{align}

\subsection{Basis choice for the representation space}\label{secSM:basis}

A more convenient choice of basis functions for the magnetic order $\mathbf{M}_{\alpha}(\mathbf{Q}_1)$ consists of using the following four orthogonal collinear magnetic structures obtained by linear combinations of the helical basis functions [Eqs.(\ref{eqSM:fc})-(\ref{eqSM:fv})]:
\begin{align}
&\mathbf{f}_{A_1,x}(z_i)=\frac{1}{\sqrt{2}}\left(\mathbf{f}_{x+}(z_i)-\mathbf{f}_{x-}(z_i)\right)\text{ ,}\\
&\mathbf{f}_{A_1,y}(z_i)=\frac{1}{\sqrt{2}}\left(\mathbf{f}_{y-}(z_i)+\mathbf{f}_{y+}(z_i)\right)\text{ ,}\\
&\mathbf{f}_{A_2,x}(z_i)=\frac{1}{\sqrt{2}}\left(\mathbf{f}_{y+}(z_i)-\mathbf{f}_{y-}(z_i)\right)\text{ ,}\\
&\mathbf{f}_{A_2,y}(z_i)=-\frac{1}{\sqrt{2}}\left(\mathbf{f}_{x+}(z_i)+\mathbf{f}_{x-}(z_i)\right) \text{ .}
\end{align}

\noindent These combinations result in amplitude modulated structures, where the moments in all Eu layers point along the same direction, while the norm of the moments vary sinusoidally from layer to layer. In the structure $\mathbf{f}_{A_1,x}(z_i)$ ($\mathbf{f}_{A_1,y}(z_i)$), referred as $A_1$, the moments point along $\hat{x}$ ($\hat{y}$) and the node of the sinusoidal amplitude modulation lies in between Eu planes. These are the nodeless amplitude-modulated structures defined in the main text. Similarly, $\mathbf{f}_{A_2,x}(z_i)$ ($\mathbf{f}_{A_2,y}(z_i)$) are nodal amplitude modulated structures, denoted $A_2$, with moments pointing along $x$ ($y$) and nodes at every third Eu layer. 

The order parameter for Phase I [Eq.(\ref{eqSM:psiI})] in this colinear basis takes the form shown in the main text:
\begin{equation}
\bm{\Psi}_{I}=\left(a_1\cos\phi_1,a_1\sin\phi_1,a_2\cos\phi_2,a_2\sin\phi_2\right)\text{ .}
\label{eqSM:psiI_new}
\end{equation}

\noindent Here, $a_{1,2}$ and $\phi_{1,2}$ sets the amplitude and in-plane orientation of the moments in $A_{1,2}$. The transformation between the helical and the colinear basis is given by
\begin{equation}
  \begin{pmatrix}
a_1\cos\phi_1\\
a_1\sin\phi_1\\
a_2\cos\phi_2\\
a_2\sin\phi_2
\end{pmatrix}=\frac{1}{\sqrt{2}}\begin{pmatrix}
-1 & 1 & 0 & 0\\ 
0 & 0 & 1 & 1\\ 
0 & 0 & -1 & 1\\ 
-1 & -1 & 0 & 0
\end{pmatrix}\begin{pmatrix}
\Delta_1\\
\Delta_2\\
\Delta_3\\
\Delta_4
\end{pmatrix}  \text{ .}
\end{equation}

As a result, Eq. (\ref{eqSM:MQ1final}) can be re-expressed as 

\begin{equation}
    \frac{\mathbf{M}_{1}(\mathbf{Q}_1)}{N_m}=\frac{\mathbf{M}_{2}(\mathbf{Q}_1)}{N_m}=\frac{-3}{2}\begin{pmatrix}
        a_1 \cos \phi_1 +i\, a_2 \cos \phi_2\\
       a_1 \sin \phi_1+i\,a_2 \cos \phi_2\\
        0
    \end{pmatrix}
    \label{eqSM:MQ1final_mod}
\end{equation}

\section{Nematic director for Phase II}\label{secSM:nematic}

In Phase II, we have the coexistence of two Potts-nematic order parameters, $\bm{\eta}$ which is related to $\bm{\Psi}_I$, and $\bm{\xi}$ that is related to the \Neel~ state that emerges in Phase II. Therefore, the general form of the nematic order parameter associated with the magnetic structure in Phase II is $\bm{\eta}+\kappa\bm{\xi}$, where $\kappa$ is a real number. In this section we show that for the forms of $\bm{\eta}$ and $\bm{\xi}$ adopted in the main text [Eq.(\ref{eq:nematic_psi}) and Eq.(\ref{eq:nematic_xi})], $\kappa=1$ is the most natural choice. 

The nematic director can be calculated directly from the magnetic structure $\mathbf{M}(z_i)$ by
\begin{equation}
    \bm{n}=\frac{1}{N_l}\sum\limits_{z_i=0}^{5/2}\begin{pmatrix}
        M_{x}^2(z_i)-M_{y}^2(z_i)\\[0.2cm]
        2M_{x}(z_i)M_{y}(z_i)
    \end{pmatrix} \text{ .}
    \label{eqSM:nematicM}
\end{equation}

\noindent Here $x,y$ denotes the Cartesian components of $\bm{M}(z_i)$. Besides, $N_l=6$ corresponds to the number of Eu layers in the magnetic unit cell. Finite birefringence requires a non-zero $\bm{n}$.

In Phase II, $\bm{M}(z_i)$ takes the following form in the collinear basis 
\begin{align}
\bm{M}(z_i)=&\sum\limits_{j=1}^2\left[a_j\cos{\phi_j}\,\bm{f}_{A_j,x}(z_i)+a_j\sin{\phi_j}\,\bm{f}_{A_j,y}(z_i)\right]\nonumber\\
&+l\cos\theta\,\bm{g}_x(z_i)+l\sin\theta\,\bm{g}_y(z_i) \text{ ,}
    \label{eqSM:genericM}
\end{align}

\noindent where $\bm{g}_{\mu}(z_i)$, $\bm{f}_{A1,\mu}(z_i)$ and $
\bm{f}_{A2,\mu}(z_i)$ (with $\mu=x,y$) are the basis functions defined in Sec.\ref{secSM:op}. Substituting Eq.(\ref{eqSM:genericM}) into Eq.(\ref{eqSM:nematicM}), we obtain
\begin{equation}
    \bm{n}=2\left(\bm{\eta}+\bm{\xi}\right) \text{ .}
\end{equation}

\section{Free energy functional}\label{secSM:freeE}

In this section, we derive the analytic form of the Landau functional discussed in the main text. The generic Landau function for \eia~that captures a sequence of phase transitions where $\mathbf{Q}_1$ and $\mathbf{Q}_2$ emerge takes the form of a power series on the components of the order parameters $\bm{\Psi}_I$ and $\bm{\Psi}_L$ defined in the main text (see also Sec. \ref{secSM:op}).  $\bm{\Psi}_I$ is the order parameter for Phase I [Eq.(\ref{eq:psiI})] and $\bm{\Psi}_L$ is the order parameter corresponding to the AFM component of the magnetic state that develops in Phase II [Eq.(\ref{eqSM:psiL})]. The terms in the power series are constrained by the symmetries of the paramagnetic space group $P6_3/mmc1'$ (No. 194.264) of \eia and can be readily obtained using the tool \textsc{invariants} in the Isotropy software~\cite{stokes2017isotropy} with the condition of coupled $m\Delta_6$ and $m\Gamma_5^+$ Irreps. They can be grouped into three classes of terms, 
\begin{equation}
F\left(\bm{\Psi}_{I},\bm{\Psi}_{L}\right)=F_A\left(\bm{\Psi}_{I}\right)+F_L\left(\bm{\Psi}_{L}\right) + F_{AL}\left(\bm{\Psi}_{I},\bm{\Psi}_{L}\right).
\label{eqSM:FreeE}
\end{equation}

\noindent We now show the expressions for each of them.

\subsection{The form of $F_A$}

We start with $F_A$, which involves only the components of $\bm{\Psi}_{I}$. In the helical basis [Eq.{\ref{eqSM:psiI}}] and up to order $\mathcal{O}(\Delta_j^6)$, it reads
\begin{widetext}
    \begin{align}
 F_{A}= & \alpha_1 \left(\sum_j\Delta_j^2\right)+\frac{\beta_1}{2} \left(\sum_j\Delta_j^2\right)^2+\alpha_2 \bm{\eta}\cdot \bm{\eta}-\frac{\beta_2}{2}\eta_1(\eta_1^2-3 \eta_2^2)+\frac{\beta_4}{2} \left(\sum_j\Delta_j^2\right)^3+\gamma_3 \left(\sum_j\Delta_j^2\right) \bm{\eta}\cdot \bm{\eta} \notag \\[0.2cm]
 &+ \gamma_1 \left[\Delta_3^2 (\Delta_3^2 -3 \Delta_1^2 )^2+\Delta_4^2 (\Delta_4^2 -3 \Delta_2^2 )^2\right]-\frac{\gamma_2 }{2} (\Delta_1 \Delta_2-\Delta_3 \Delta_4) \left[\Delta_1^2 \left(\Delta_2^2-3 \Delta_4^2\right)-8 \Delta_1 \Delta_2 \Delta_3 \Delta_4+\Delta_3^2 \left(\Delta_4^2-3 \Delta_2^2\right)\right]
\end{align}
\label{eqSM:FDelta}
\end{widetext}

\noindent where 
\begin{equation}
\boldsymbol{\eta}=\begin{pmatrix}
    \eta_1\\
    \eta_2
\end{pmatrix}=\begin{pmatrix}
    -\Delta_1\Delta_2-\Delta_3\Delta_4
        \\
        \Delta_2\Delta_3-\Delta_1\Delta_4
    \end{pmatrix}
\end{equation}

\noindent is a 3-state Potts nematic composite order parameter \cite{little_three-state_2020} allowed to develop in Phase I. In the main text, $\boldsymbol{\eta}$ is expressed in the collinear basis (see Sec. \ref{secSM:basis}). We emphasize that the sixth-order terms are essential to remove the degeneracy between different amplitude-modulated phases.

Note that $\boldsymbol{\eta}$ vanishes for choices of $\Delta_j$ resulting in magnetic structures that preserve the three-fold rotation symmetry around $\hat{z}$ ($C_{3z}$). For instance, $\boldsymbol{\eta}=0$ if $\Delta_2=\Delta_3=\Delta_4=0$ in Eq.(\ref{eqSM:psiI}), which is the order parameter for a 60\degree-helix. Importantly, $\eta\neq 0$ is required for a finite T-mod signal. Therefore all of the magnetic structures listed in Tables \ref{tabSM:table1} and \ref{tabSM:table2} with an allowed T-mod optical signal have a non-zero nematic component.

Although the values for the parameters in Eq.(\ref{eqSM:FDelta}) could be determined through a microscopic theory for magnetism in \eia, this is beyond the scope of this work and a phenomenological approach suffices to capture the sequence of observed phase transitions in this material. The parameter $\alpha_1=\alpha_0(T-T_{N1})$, with $\alpha_0>0$, changes sign at the onset of Phase I, at temperature $T_{N1}$. Moreover, $\beta_1>0$ and $\beta_4>0$ guarantee that $F_{A}$ is bounded. The parameters $\alpha_2$ and $\gamma_3$, when chosen to be negative, energetically favor nematic magnetic structures. 

Rewriting $F_A$ in the collinear basis (see Sec. \ref{secSM:basis}), we find that $\beta_2$ and $\gamma_1$ are the coefficients of terms involving combinations of $\sin\phi_{1,2}$ and $\cos\phi_{1,2}$ and, therefore they favor specific orientations of the magnetic moments within the Eu planes. Throughout this work, motivated by the experimental observation that the magnetic moments are not pinned to the crystal axes, we focus on the limit where $\beta_2$ and $\gamma_1$ are much smaller than the energy scale associated with built-in strain. In practice, to simplify the analysis, we will set these coefficients to zero and choose an arbitrary direction for the moment in Phase I. Of course, these coefficients are not identically zero, and the direction is set by the local strain, as discussed in the main text.

The coefficient $\gamma_2$ distinguishes between the nodeless ($a_1\neq 0$) and nodal ($a_2\neq 0$ and $a_1=0$) amplitude-modulated structures. This can be readily seen in the collinear basis, where the term proportional to $\gamma_2$ in Eq.(\ref{eqSM:FDelta}) simplifies to 
\begin{equation}
    \frac{1}{16}\left(a_1^2-a_2^2\right)\left[\left(a_1^2-a_2^2\right)^2-12 a_1^2 a_2^2 \cos^2\left(\phi_1-\phi_2\right)\right] \text{ .}
\end{equation}

\subsection{The form of $F_L$}

The second term in the right-hand side of Eq. (\ref{eqSM:FreeE}) involves only components of the emergent AFM order parameter in Phase II. Up to order $\mathcal{O}(l^6)$, we find
\begin{equation}
F_L(\bm{\Psi}_L)=\alpha_3l^2+\frac{1}{2}\beta_3l^4 + \frac{1}{3}\,\gamma_3 l^6 \cos6\theta  \text{ .}
\end{equation}

\noindent We emphasize that $\alpha_3$ \textit{does not} need to change the sign for the AFM component to become non-zero. Having two order parameters with independent transition temperatures requires more fine-tuning of the model parameters. Here, instead, we argue that $\alpha_3>0$ and that the non-zer0 $\bm{\Psi}_{L}$ is induced by $\bm{\Psi}_{I}$ through the coupling term $F_{AL}(\bm{\Psi}_{I},\bm{\Psi}_{L})$, as we show in the next section. Note that $\gamma_3$ enforces the moments to point along high-symmetry directions of the lattice. Similarly to what we discussed above, we set this term to zero to capture the fact that the moment direction is selected locally by the strain. 

\subsection{The form of $F_{AL}$} \label{sec:SM_FAL}

The last term in Eq. (\ref{eqSM:FreeE}) includes both linear and quadratic couplings between $\bm{\Psi}_{I}$ and $\bm{\Psi}_{L}$: 
\begin{equation}
F_{AL}=\delta_1F_{AL}^{(1)}+\delta_2F_{AL}^{(2a)}+\delta_3F_{AL}^{(2b)} \text{ .}
\end{equation}

\noindent We omitted the explicit dependence on $\bm{\Psi}_{I}$ and $\bm{\Psi}_{L}$ to shorten the notation. In the collinear basis, the coupling that is linear in $l$ is given by:
\begin{equation}
   F_{AL}^{(1)} =\frac{a_{1}l}{\sqrt{2}}\left[(a_{1}^{2} - 2a_{2}^{2})\cos(\theta - \phi_{1}) - a_{2}^{2}\cos(\theta + \phi_{1} - 2\phi_{2})\right] \text{ .}
   \label{eqSM:Flinear}
\end{equation}

\noindent In the presence of a non-zero $F^{(1)}_{AL}$, when $\bm{\Psi}_{I}$ becomes non-zero it immediately triggers the development of $\bm{\Psi}_{L}$ and the two transitions happen at the same temperature -- except for fine-tuned values of the three angles. We know from the scattering experiments that this is not the case in \eia~where Phases I and II are separated in temperature. From Eq. (\ref{eqSM:Flinear}), we note that $F_{AL}^{(1)}$ vanishes identically if $a_1=0$. This is an important result as it tells us that the nodal collinear amplitude-modulated phase ($a_2\neq0$ and $a_1=0$, as well as the symmetry-related structures) is the only candidate magnetic structure for Phase I that allows for the development of $\bm{\Psi}_I$ and $\bm{\Psi}_{L}$ at distinct temperatures. This is in agreement with the results of Table \ref{tabSM:table1} for the structures (expressed in the helical basis) that display a $T$-mod signal but not an $H$-mod signal. 

The quadratic couplings, also in the collinear basis, take the form
\begin{align}
&F_{AL}^{2a} =-\frac{l^2}{2} \left[a_{1}^2\cos\left(2(\theta - \phi_{1})\right) + a_{2}^2\cos\left(2(\theta - \phi_{2})\right)\right]\\
&F_{AL}^{2b} =\left(a_1^2+a_2^2\right)l^2.
\end{align}

\noindent These are the terms that can trigger the second phase transition at a temperature $T_{N,2}$ lower than $T_{N,1}$ depending on the signs of the coefficients $\delta_2$ and $\delta_3$. Indeed, $\delta_3 < 0$ favors a coexistence between $a_2$ and $l$, whereas the sign of $\delta_2$ sets the relative angle between $\theta$ and $\phi_2$, with $\delta_2>0$ favoring collinear directions and $\delta_2<0$, orthogonal directions. To summarize, the free energy analysis gives us the following picture for the phase transitions in \eia: at $T_{N1}$, a transition into a nodal collinear amplitude-modulated state takes place. As the temperature is further lowered, $a_2$ increases and through the couplings $F_{AL}^{(2a)}$ and $F_{AL}^{(2b)}$ it can renormalize the coefficient $\alpha_3>0$ in $F_{L}$, reducing its value. When the renormalized coefficient of $l^2$  eventually changes sign, a finite $\bm{\Psi}_{L}$ develops, triggering the transition to Phase II. A non-zero $\bm{\Psi}_{L}$ makes $F_{AL}$ also non-zero, and together with the quadratic couplings, also change the form of $\bm{\Psi}_{I}$.

\subsection{Equal moment condition}\label{secSM:equal}

\begin{figure*}[t!]
 \centering
\includegraphics[width=0.6\linewidth]{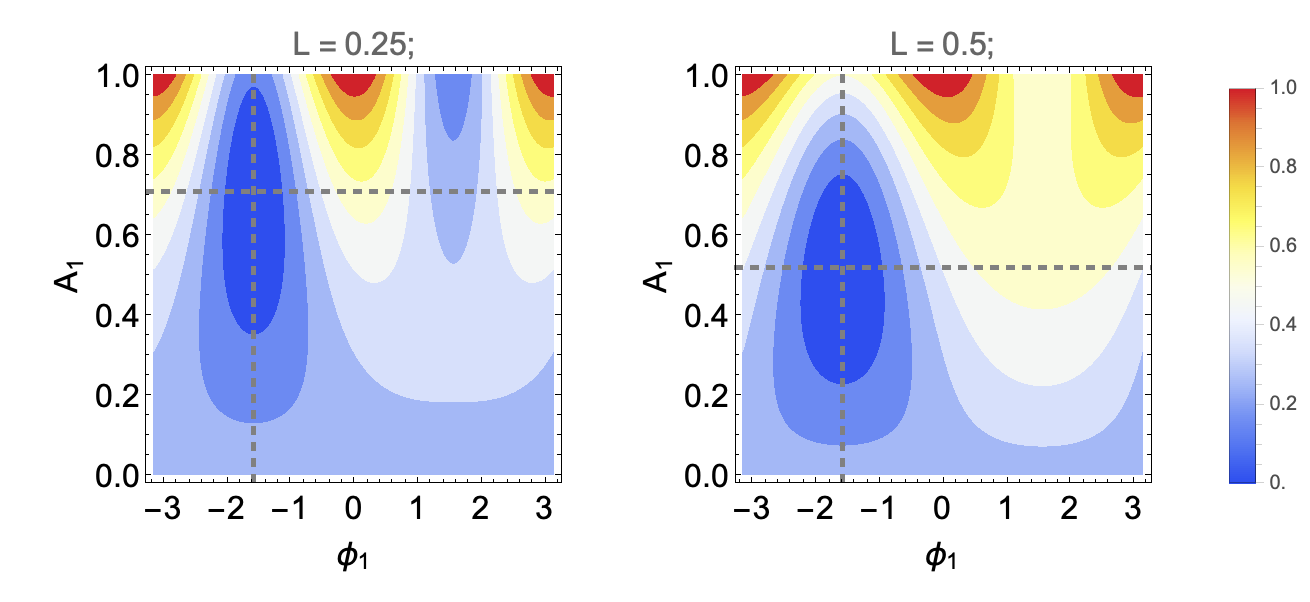}
 \caption{The free energy landscape given by Eq.~\ref{eq:FN_total}, for $A_2 = 1$, $\phi_2 = 0$, $\theta= \pi/2$, and two values of $L$, as noted on the figures. The dashed lines denote the equal spin length condition (eq.~\ref{eqSM:equalM}), and clearly coincide with the minima of this free energy term. }
 \label{fig:S_EnergyLandscapes}
\end{figure*}

The qualitative analysis of the free energy presented above demonstrates that the phenomenology of the magnetic phases of \eia{} can be captured by the Landau free-energy introduced here, provided that the coefficients satisfy certain conditions. In order to be able to perform a more quantitative analysis, additional information about the Landau coefficients is necessary, otherwise the parameter space is intractably large. Ultimately, these coefficients should be derived from a microscopic model which, as we discussed in the main text, is itself challenging. 
To achieve some progress, we instead restrict the Landau coefficients such that configurations with equal-amplitude moments are favored by minimization of the free energy (away from the transition). 

To motivate our choice of parameters, we start by considering a magnetic structure in Phase II with order parameters given by Eqs. (\ref{eqSM:psiL}) and (\ref{eqSM:psiI_new}).

Calculating $\left|\bm{M}_{\alpha}(z_i)\right|$ for this generic structure, we find that the norm of the magnetic moments are layer-independent if the following conditions are fulfilled: 
\begin{equation}
   a_{1} =  \sqrt{a_{2}^2 + 2l^2}-\sqrt{2}l , \hspace{5pt} \phi_1 = \phi_2\pm\frac{\pi}{2},\hspace{5pt} \theta=\phi_2\mp\frac{\pi}{2}. 
  \label{eqSM:equalM}
\end{equation}

The form of Eq.~\ref{eqSM:equalM} suggests that for angles $\phi_1=\phi_2\pm\pi/2$ and $\theta=\phi_2\mp\pi/2$, the equal moment condition would be favored by a term of the Landau functional of the form
\begin{equation} \label{eq:S_norm}
 F_{N0} = \left((a_{1}+\sqrt{2}l )^2-(a_{2}^2 + 2l^2)\right)^2, 
\end{equation}

\noindent We find by inspection that such a term is allowed by the symmetries of the paramagnetic group and, in fact, can be obtained directly from the action by constraining the Landau coefficients. In particular, by choosing $\delta_2=\delta_1$, $\delta_3=-\frac{\delta_1}{2}$ in $F_{AL}$,  $\alpha_2=-\frac{\delta_1}{2}$ in $F_{A}$, and enforcing the relationships between $\phi_1$, $\phi_2$ and $\theta$ given in Eq.~\ref{eqSM:equalM}, we reproduce a term proportional to $F_{N0}$.
This motivates us to keep the angles $\phi_1$, $\phi_2$ and $\theta$ arbitrary and enforce the same relationship between the parameters $\delta_1$, $\delta_2$, $\delta_3$ and $\alpha_2$. We obtain:

\begin{widetext}
\begin{align}
F_N & = -\frac{\delta_1}{8} (a_1^4 + a_2^4 + 4a_1^2 l^2 \cos(2 (\theta - \phi_1)) + 4\sqrt{2} a_1 l \left(-(a_1^2 - 2a_2^2) \cos(\theta - \phi_1) + \right. \nonumber \\
& \left. + a_2^2 \cos(\theta + \phi_1 - 2\phi_2)\right) + 4l^2 (a_1^2 + a_2^2 + a_2^2 \cos(2 (\theta - \phi_2))) + 2a_1^2 a_2^2 \cos(2 (\phi_1 - \phi_2))
\label{eq:FN_total}
\end{align}
\end{widetext}
It turns out that $F_N$ is minimized when condition \ref{eqSM:equalM} is met. To show that, we plot $F_{N}$ as a function of $a_1$ and $\phi_1$, for two different values of $l$. In both cases, we find minima of $F_N$ when the equal-moment condition is satisfied.

Thus, to perform the quantitative analysis shown in Fig.\ref{fig:EIA_Fig9} of the main text, we minimize the following free energy:

\begin{widetext}\label{eqSM:constF}
\begin{align}
\Tilde{F}&= \left(T-T_{N1}\right) a^2 +\frac{\beta_1}{2}a^4+\frac{\beta_4}{2}a^6 -\frac{\delta_1}{8}  \left(a_1^4 + 2a_1^2 a_2^2 \cos(2 (\phi_1 - \phi_2)) + a_2^4\right)+\nonumber \\
&+\frac{\gamma_2 }{16}\left(a_1^2-a_2^2\right)\left[\left(a_1^2-a_2^2\right)^2-12 a_1^2 a_2^2 \cos\left(\phi_1-\phi_2\right)^2\right] +\alpha_3l^2+\frac{1}{2}\beta_3l^4 -\frac{1}{2} \, \delta_1 \left[a_1^2 l^2 \cos\left(2 (\theta - \phi_1)\right) + \right. \nonumber \\
&\left. \sqrt{2} a_1 l \left(-\left(a_1^2 - 2 a_2^2\right) \cos\left(\theta - \phi_1\right) + a_2^2 \cos\left(\theta + \phi_1 - 2 \phi_2\right)\right) + l^2 \left(a^2 + a_2^2 \cos\left(2 (\theta - \phi_2)\right)\right)\right],
\end{align}
\end{widetext}
where we introduced the notation $a^2=a_1^2+a_2^2$.

For the results shown in Fig.\ref{fig:EIA_Fig9}, we introduced a temperature dependence for $\delta_1$, such that $\delta_1=\chi_0-\chi_1\left(T_{N1}-T\right)\Theta\left(T_{N1} - T\right)$, with $\chi_1=0.075$ and $\chi_0=0.05$. Here, $\Theta(x)$ is the Heaviside step function, which is $0$ for $x\leq0$ and $1$ for $x>0$. Such a temperature dependent term  plays a similar role as higher-order terms in the free energy that are only relevant farther from the transition, since the order parameter scales as $a^2 \sim \left(T_{N1}-T\right)\Theta\left(T_{N1} - T\right)$. 

At $T_{N1}$, the amplitude-modulated phase is favored by $\alpha_2 < 0$. Note that this condition is satisfied by our parameters, since we set $\alpha_2 = -\delta_1/2$ and $\delta_1 = \chi_0 > 0$ at $T_{N1}$. As temperature is lowered, the assumption is that higher-order terms (i.e. beyond sixth-order) effectively renormalize $\delta_1$ and make it switch sign below $T_{N1}$, which favors the equal-moment condition.

The other parameters were set to 
$\alpha_3= 0.025$, $\beta_1 = 0.5$, $\beta_3 = 0.01$, $\beta_4 = 0.1$, $\gamma_2 = 0.1$. Recall that, as explained above, $\beta_2$, $\gamma_1$ and $\gamma_3$ were set to zero to model the dominant role played by the local strain in setting the moments direction as compared to the intrinsic crystalline anisotropy.   

The starting point for energy minimization in Fig.~\ref{fig:EIA_Fig9}(a) at the lowest temperature was a series of random moment configurations. Since all relative orientations of the moments and the lattice are degenerate when setting $\beta_2=\gamma_1=\gamma_3=0$, the moment orientation in the optimized structure was determined by the choice of the random initial conditions (and could be changed by a change of the random number generator). At every temperature $T$ the structure from the last temperature step ($T-dT$) was used as the initial condition, mimicking the experimental evolution of the structure.

\section {Spin Hamiltonian}

\subsection{Heisenberg term}\label{sec:S_Heis}
In the following we construct the exchange Hamiltonian that captures the tendency towards the order at two wave vectors, as observed in \eia. Let us suppose the Heisenberg exchange between $n$-th neighbors is:
\begin{equation}\label{eq:interaction}
 \mathcal{J}_n\left(J_0, k, n, d \right) = J_0 \frac{\cos{\left( kn\right)}}{\left(kn\right)^d}.
\end{equation}
 For $d=0$ the interaction does not decay at all, and is represented by a function in $k$ space. $d = 1,2,3$ correspond to leading distance - dependent terms of the Ruderman–Kittel–Kasuya–Yosida (RKKY) interaction in one, two and three dimensions~\cite{roth_generalization_1966}. Let us assume a single-$q$ structure of the form:

 \begin{equation}\label{eq:S_spiral}
 \mathbf{S_n}=S_q\left(\cos{\left(q n\right)\mathbf{\hat{x}}}+\sin{\left(q n\right)\mathbf{\hat{y}}}\right),
\end{equation}
and calculate the total energy of such a state, assuming Heisenberg interactions, with the distance-dependent exchange given by Eq.~\ref{eq:interaction}. The total exchange energy is: 

\begin{equation}\label{eq:S_oscillExchange}
 E_{ex}=J_0 S_q^2 \sum_{n=1}^N \frac{\cos{\left( kn\right)}}{\left(kn\right)^d} \cos{nq},
\end{equation}
where $N$ corresponds to the number of neighbors considered in the interaction. 

\begin{figure}[h]
\centering
\includegraphics[width=0.4\textwidth]{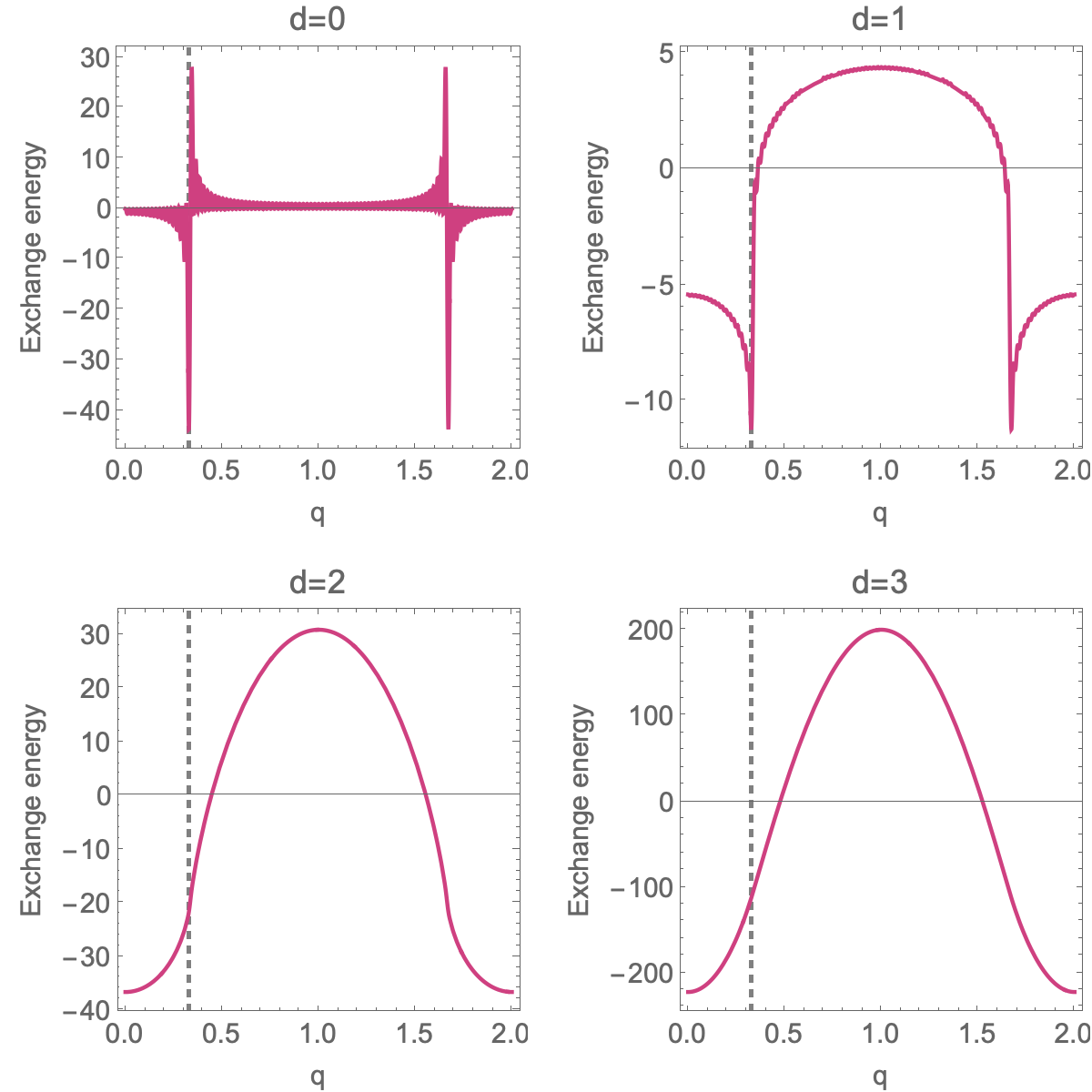}
\caption{The exchange energy (Eq.~\ref{eq:S_oscillExchange}) for $k=\pi/3$, $J=-1$, $N=100$ and $S_q=1$ as a function of $q$, for $d = 0,1,2,3$. The dashed line marks $\bm{Q_1}$.}
\label{fig:S_OscillatoryDep}
\end{figure}

In Fig.~\ref{fig:S_OscillatoryDep} we plot the exchange energy (Eq.~\ref{eq:S_oscillExchange}) for $k=\pi/3$, $J=-1$, $N=100$ and $S_q=1$ as a function of $q$, for $d = 0,1,2,3$. It is interesting to see that only for $d=0$ and $d=1$ the exchange energy is minimized for a $q\neq 0$; in other cases the interaction drops off too quickly to overcome the nearest-neighbor ferromagnetic tendency. 

\begin{figure}[h]
 \centering
 \includegraphics[width=1\linewidth]{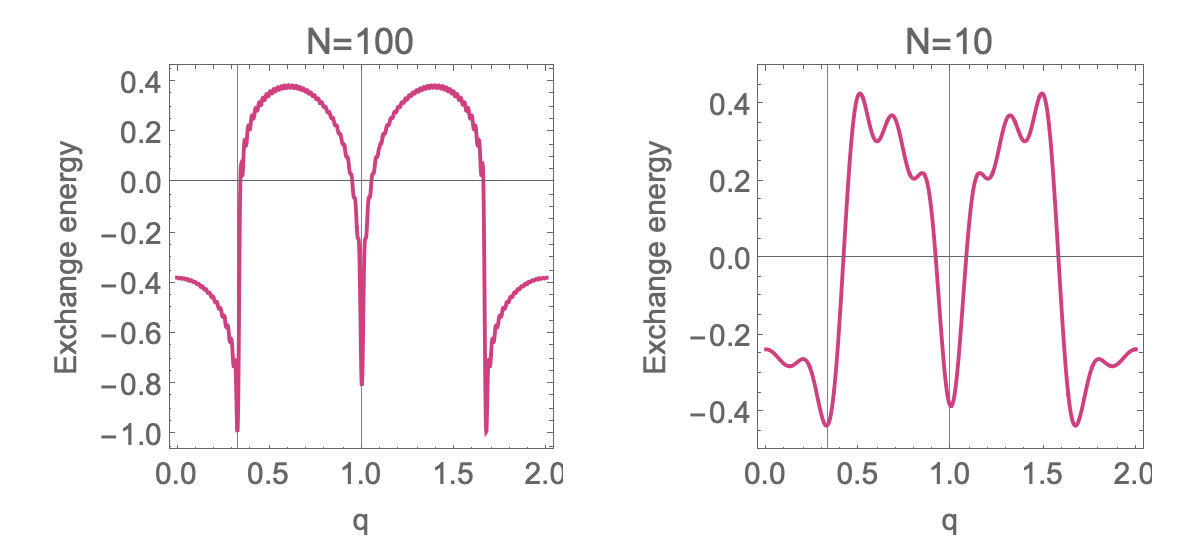}
 \caption{Exchange energy of a state with periodicity $q$, if the bi-linear terms are given by Eq.~\ref{eq:S_interactionBoth}, for: (a) $J_1=-0.1, k_1=\pi/3,J_2=0.12, k_2=\pi, N=100$;(b) $J_1=-0.1, k_1=0.4\pi,J_2=0.125, k_2=\pi, N=10$.}
 \label{fig:S_Exchanges}
\end{figure}

We now have all the tools needed to choose the distance-dependent exchange interaction, which will favor both the $\bm{Q_2}$ and the $\bm{Q_1}$ order. Our Heisenberg term will be constructed as a superposition of terms favoring the $\bm{Q_2}$ and $\bm{Q_1}$ orders:

\begin{equation}\label{eq:S_interactionBoth}
 \mathcal{J}_n = J_1 \frac{\cos{\left( k_1 n\right)}}{k_1 n}+ J_2 \frac{\cos{\left( k_2 n\right)}}{k_2 n},
\end{equation}
where $k_1=\pi/3$ and $k_2=\pi$. In Fig.~\ref{fig:S_Exchanges} (a,b) we plot the total exchange energy as a function of $q$ for two sets of parameters. In panel (a) we show the exchange energy for $N=100$ neighbors, while in Fig.~\ref{fig:S_Exchanges} (b) we modify the parameters to obtain the two minima at $q=1, 1/3$ for $N=10$. By controlling $J_2/J_1$, we can choose the relative depth of the two energy minima, and therefore pick the lowest energy $q$; for the rest of our modeling we will use $J_2/J_1=-1.25$, which ensures that the lowest energy states is found for $\bm{Q_1}$, but also that the $\bm{Q_2}$ state is close to it in energy. 

To summarize, we find that a long-range RKKY-like exchange is needed to capture the susceptibility towards order at two values of $q$. However, the result of energy minimization of such a model is always a single $q$ state: the deepest minimum in Fig.~\ref{fig:S_Exchanges} is chosen. 

\subsection{Atomistic simulations}\label{sec:SI_atom}
Atomistic simulations shown in Fig.~\ref{fig:EIA_Fig12} were performed with the Spirit code \cite{muller_spirit_2019}. For simplicity we simulated a cubic lattice with $6 \times 6\times 12$ sites. The Hamiltonian used for simulations is:

\begin{equation}\label{eq:SI_hamilt}
 H=\sum_{ij} \mathcal{J}_{ij}\bm{S}_i\cdot \bm{S}_j + \sum_{ijkl} \mathcal{J}_{ijkl}\left(\bm{S}_i\cdot \bm{S}_j\right) \left(\bm{S}_k\cdot \bm{S}_l\right),
\end{equation}
with the Heisenberg parameters given in Table~\ref{tab:Heisenberg}, and the fourth order terms in Table~\ref{tab:fourthTerms}. All other parameters needed to reproduce the simulations are given in Table~\ref{tab:parameteres}.

\begin{table}[h]
\centering
\begin{tabular}{cccc}
$da$ & $db$ & $dc$ & $-\mathcal{J}_{ij}$ \\
\hline
1 & 0 & 0 & 1 \\
0 & 1 & 0 & 1 \\
0 & 0 & 1 & 0.25 \\
0 & 0 & 2 & 0.202254 \\
0 & 0 & 3 & -0.218169 \\
0 & 0 & 4 & -0.0386271 \\
0 & 0 & 5 & -0.0190983 \\
0 & 0 & 6 & 0.125 \\
0 & 0 & 7 & -0.0136416 \\
0 & 0 & 8 & -0.0193136 \\
0 & 0 & 9 & -0.0727232 \\
0 & 0 & 10 & -0.0404508 \\
\hline
\end{tabular}
\caption{Heisenberg terms used for atomistic simulations. $da$,  $db$ and $dc$ denote the spin-spin distances along the three orthogonal directions in units of lattice length. All exchanges are given in $\unit{meV}$. The signs in the table are consistent with the Spirit convention~\cite{muller_spirit_2019}.}
\label{tab:Heisenberg}
\end{table}

\begin{table}[h]
\centering
\begin{tabular}{cccccccccc}
$da_j$ & $db_j$ & $dc_j$ & $da_k$ & $db_k$ & $dc_k$  & $da_l$ & $db_l$ & $dc_l$ & $-\mathcal{J}_{ijkl}$ \\
\hline
0 & 0 & 1 & 0 & 0 & 0 & 0  & 0 & 1 & -0.2 \\
0 & 0 & 1 & 0 &  0 & 1 & 0  & 0 & 2 & -0.2 \\
\hline
\end{tabular}
\caption{Fourth order exchange terms used for atomistic simulations. $da_j$,  $db_j$ and $dc_j$ denote the distances between spins $i$ and $j$ along the three orthogonal directions in units of lattice length. All exchanges are given in $\unit{meV}$. The signs in the table are consistent with the Spirit convention~\cite{muller_spirit_2019}.}
\label{tab:fourthTerms}
\end{table}

\begin{table}[h]
\centering

\begin{tabular}{ll}
Parameter & Value \\
\hline
llg\_seed & 20006 \\
llg\_n\_iterations & 5000000 \\
llg\_n\_iterations\_log & 5000 \\
llg\_damping & 0.3 \\
llg\_beta & 0.1 \\
llg\_dt & 1.0E-5 \\
llg\_force\_convergence & 10e-9 \\
\hline
\end{tabular}
\caption{LLG Parameters}
\label{tab:parameteres}
\end{table}

\end{document}